# Recent progress on the characterization of the high-pressure behaviour of $AVO_4$ orthovanadates

Daniel Errandonea[1,*] and Alka B. Garg[2]

[1]Departamento de Física Aplicada - ICMUV - MALTA Consolider Team, Universitat de València, c/Dr. Moliner 50, 46100 Burjassot (Valencia), Spain

[2]High Pressure and Synchrotron Radiation Physics Division, Bhabha Atomic Research Centre, Mumbai 400 085, India

**Abstract**

$AVO_4$ orthovanadates are materials of fundamental and technological importance due to the large variety of functional properties exhibited by them. These materials have potential applications such as scintillators, thermophosphors, photocatalysts, and cathodoluminescence materials among others. They are also used as laser-host crystals. Studies at high pressures and temperatures are helpful for understanding the physical properties of the solid state, in particular, the phase behaviour of $AVO_4$ materials. For instance, they have contributed to the understanding of the macroscopic properties of orthovanadates in terms of microscopic mechanisms. A great progress has been made in the last decade towards the study of the pressure-effects on the structural, vibrational, and electronic properties of $AVO_4$ compounds. Thanks to the combination of experimental and theoretical studies, novel metastable structures with interesting

*Corresponding author, Email: daniel.errandonea@uv.es, Fax: (34) 96 3543146, Tel.: (34) 96 354 4475

physical properties have been discovered and the high-pressure structural sequence followed by $AVO_4$ oxides has been understood. In this article, we will review high-pressure studies carried out on the phase behaviour of different $AVO_4$ compounds. The studied materials include rare-earth orthovanadates and other compounds; for example, $BiVO_4$, $FeVO_4$, $CrVO_4$, and $InVO_4$. In particular, we will focus on discussing the results obtained by different research groups, who have extensively studied orthovanadates up to pressures exceeding 50 GPa. We will make a systematic presentation and discussion of the different results reported in the literature. In addition, with the aim of contributing to the improvement of the actual understanding of the high-pressure properties of ternary oxides, the high-pressure behaviour of orhovanadates will be compared with related compounds; including phosphates, chromates, and arsenates. The behaviour of nanomaterials under compression will also be briefly described and compared with their bulk counterpart. Finally, the implications of the reported studies on technological developments and geophysics will be commented and possible directions for the future studies will be proposed.

**Contents**

## 1. Introduction

The human quest for new and exotic materials has given thrust to the research in materials science. All normal matter in the universe consists of either the pure elements in the Periodic Table or their combination. However, researchers are trying to create materials with exotic properties by disturbing their original electronic structure by subjecting them to various external parameters such as pressure and/or temperature. When the temperature of certain materials is reduced, many exotic phenomena happen. In particular, after the observation of a large drop in resistivity at low temperatures, the branch of superconductivity was born. On the other hand, pressure can change the inter-atomic distances in solids by an order of magnitude, dramatically altering the electronic properties, breaking the existing bonds, or forming new chemical bonds. These facts in turn lead to a variety of fascinating phenomena such as pressure-induced metallization, amorphization, and superconductivity among others. Hence, compression provides a unique possibility to control the structure and properties of materials without altering their chemical composition. In fact, pressure-induced irreversible phase transitions can lead to the discovery of new phases of materials, which could remain metastable at ambient conditions. Many of these new phase have novel and interesting properties, being therefore potentially useful for technological developments.

Research under high pressure (HP) was pioneered by Prof. P. Bridgman in the early years of $20^{th}$ century [1], for which he was awarded the Nobel prize in physics in 1946. One decade later, high-pressure research was revolutionized with the invention of a hand-held device called diamond-anvil cell (DAC) [2]. This high-pressure device is capable of squeezing materials up to pressures larger than 500 gigapascals (GPa). Recently with the use of nanocrystalline diamonds, pressures of the order of terapascals have been achieved [3]. Another major breakthrough was the development and application of various synchrotron sources across the globe for a variety of high-pressure measurements [4]. With these resources in hand, high-pressure research has matured exponentially with investigations covering several research fields, which ranges from inorganic/organic solids to liquids and gases.

With this brief introduction to the subject we would like to focus on the main topic of this article, in which we will review the results of various high-pressure studies on $AVO_4$-type orthovanadates ($A$ = trivalent element), in general, and rare-earth orthovanadates, $RVO_4$ ($R$ = rare-earth atom), in particular. These vanadates are a subset of the large family of $AXO_4$-type compounds. Most of the compounds belonging to this class possess many interesting properties, and indeed many of them are being employed in the industry [5]. For instance, tungstates are currently being used to fabricate cryogenic phonon-scintillation detectors [6], phosphates and arsenates are widely employed in the nuclear-waste management industry [7], and orthovanadates have applications in green technologies as photocatalyst materials for degradation of propane and hydrogen sulphide [8]. The $AVO_4$ orthovanadates, which are the focus of this report, can also be used as thermophosphors, ionic conductors, and non-linear optical materials [9, 10]. On top of that, their doping with trivalent rare-earth cations makes them extremely useful as laser-host materials [11]. In fact, $YVO_4$, which can be grown as a single crystal using different techniques [12] and can also be found in nature as the mineral wakefieldite-(Y), is one of the most widely used materials in the laser industry. In addition, compounds like $LuVO_4$ and $YbVO_4$ have been found to be promising materials for self-Raman lasers [13].

High-pressure studies not only have contributed to the understanding of the physical properties of orthovanadates, but also to the synthesis of novel metastable polymorphs. One of them is the metastable scheelite phase, which can be quenched from HP conditions, and has the potential to be used as negative electrode material in place of graphite, presently being used in Li-ion batteries [14]. With the plethora of potential applications of *A*$VO_4$-type orthovanadates, in order to understand the basic physics behind their properties, significant amount of research work has been carried out by various laboratories across the globe. Such studies were performed at ambient conditions as well as under high-pressure. Since most of the research articles published on these compounds span over five decades and are dispersed and fragmented in various journals, consolidating the HP studies on all the *A*$VO_4$-type compounds studied till date in the form of a single review would be a significant contribution to the scientific community in general and high-pressure community in particular.

The focus of the present article is to review the effect of compression on rare-earth orthovanadates along with a few other metal orthovanadates such as $BiVO_4$, $FeVO_4$, $InVO_4$, and $CrVO_4$, to a pressure of about half a megabar (50 GPa). We will comment on the results of HP investigations obtained on almost all the rare-earth orthovanadates including recent studies on nanoparticles. Though the most widely and routinely employed techniques for the characterization of the various phases of these compounds are powder x-ray diffraction and Raman spectroscopic techniques, in a few cases optical absorption, photoluminescence, electrical resistance, and other techniques have also been used. Results of first-principles calculations coupled with experimental outcomes indicate a definite systematic in the pressure-induced phase transition sequence in *A*$VO_4$ compounds, hence, theoretical results will also be included in the present review.

In the next paragraph, we will give an outline of the article. Starting with a brief historical background about the high-pressure studies on *A*$VO_4$ compounds, we will then describe the details of crystal structures of the various phases adopted by these compounds at ambient pressure along with the expected Raman and infrared (IR) vibrational modes associated with them. High-pressure investigations on zircon-structured $AVO_4$ vanadates and related compounds will be presented in the following section with a description of relevant experimental/theoretical studies. In Section 5, the observed phase transitions and the mechanisms of phase transformations will be described. A brief discussion will be made on the pressure-induced phase-transition sequence on nanoparticles of $AVO_4$ compounds. Their behaviour will be compared with those of their bulk counterparts. The systematic of phase transitions along with various equations of state parameters, coordination changes around the different polyhedral units, axial, bulk and polyhedral compressibility will be followed. Finally, a summary and impact of the review on a wider group of research fields in general and high-pressure research in particular will be presented. This will be preceded by a discussion of the technological and geophysical implications of the summarized research and the possible future directions for the high-pressure study of orthovanadates.

## 2. Historical background

Young and Schwartz reported the earliest high-pressure study on metal orthovanadates in the year 1962 [15]. These authors studied chromium and iron orthovanadates with the aim of finding whether the high-pressure modifications of these compounds have any resemblance with the high-pressure phases of $SiO_2$, an important geophysical material. This work was followed by a short communication on the HP behaviour of $FeVO_4$ by Laves *et al.* in 1964 [16]. In both studies, a pressure of around 6 GPa and a temperature of approximately 1000 K were applied to the compounds in a belt-type apparatus [17] for about 24 hours. The quenched products were analysed by powder x-ray diffraction (XRD). These studies provided evidence of the existence of previously unknown polymorphs, but the complete crystal structure of those polymorphs was not determined. A follow up of these studies took place a decade later in which more accurate information on the crystal structure of the HP forms of $FeVO_4$ was reported [18, 19].

One of the first *in-situ* high-pressure studies on $AVO_4$ orthovanadates was carried out in the early 80s. In this work, the lattice parameters of $BiVO_4$ were determined under different pressure and temperature conditions up to 5 GPa [20] providing accurate information on a reversible ferroelastic phase transition. The first *in-situ* HP investigation on rare-earth orthovanadates was published by Jayaraman *et al.* in 1987 [21]. These authors analysed the vibrational and optical-absorption behaviour of $YVO_4$ up to a pressure of 15 GPa, detecting a pressure-induced irreversible phase transition in the material beyond 7.5 GPa. Based on ambient-pressure x-ray diffraction measurements on a pressure-cycled sample, the HP phase was assigned to a scheelite-type structure [21]. Subsequent works on $TbVO_4$ and $DyVO_4$ found changes in the Raman spectra which were compatible with the zircon-scheelite transition reported in $YVO_4$ [22]. Luminescence measurements in rare-earth-doped $YVO_4$ confirmed the existence of a pressure-induced phase transition in $YVO_4$ [23].

An *in-situ* confirmation by x-ray diffraction of the non-reversible zircon-to-scheelite transformation in $YVO_4$ had to wait until the 21st century and the development of synchrotron powder x-ray diffraction in a high-resolution angle-dispersive mode [24]. After this study, a number of HP experiments followed on rare-earth orthovanadates. Most of the members of this family have been investigated covering different pressure ranges [25 - 29], with the exception of radioactive $PmVO_4$. These studies reveal the existence of a zircon-scheelite phase transition in the compounds with the rare-earth cations having ionic radii smaller than that of promethium (Pm) and another post-scheelite phase at around 20 GPa or higher pressure. $ScVO_4$ was found to follow a similar behaviour [30]. Compounds with larger rare-earth cationic radii were found to undergo a phase transition from zircon to the monazite structure [31]. In these compounds (e.g. $CeVO_4$), evidence of a second pressure-driven transition has been reported around 20 GPa [32, 33].

Substantial amount of the progress achieved in understanding the experimentally observed high-pressure behaviour of orthovanadates was attained with the contribution of theoretical studies based on well-refined computational models. Experimental findings are well corroborated with theoretical results in terms of transition pressures, equation of state parameters, identification of crystal structures, and the vibrational behaviour [26, 28 - 30]. Calculations have also predicted the HP behaviour of infrared modes [28], and helped to understand the influence of pressure on the electronic structure of $AVO_4$ compounds [34]. One of the major contributions of theory is the identification of a phonon softening, responsible for the pressure-driven instability of the zircon phase and the possible relation of it with the pressure-induced delocalization of f-electrons [30]. Interestingly, it has been found that the silent $B_{1u}$ mode softens in all the zircon structured orthovanadates studied hitherto. The softening of this mode is associated to the violation of one of the Born's stability criteria, which explains why the zircon structure becomes unstable under compression [35].

In addition to the characterization of HP structural and vibrational behaviour of $AVO_4$ compounds, recently, some progress has also been made on the characterization of their optical properties. It has been established that in the zircon phase, compression has a negligible influence on the electronic band gap of these wide band-gap materials [34]. In addition, a narrowing down of the band-gap is observed across the zircon-scheelite transition. The most interesting consequence of these studies is the possibility of obtaining a metastable phase at ambient pressure with a band gap smaller than 3 eV.

Related with the changes induced by compression in the electronic structure of orthovanadates, recently a pressure-enhanced light emission in Er-doped $GdVO_4$ has been reported [36]. In spite of the recent progress, research on the influence of pressure on electronic and optical properties of $AVO_4$ compounds is in a nascent stage.

Other interesting topics that have just started to be explored are the influence of non-hydrostaticity on the HP behaviour of orthovanadates [37] and the phase transition mechanisms [38]. Non-hydrostatic compression has been found to strongly influence the HP behaviour of $ScVO_4$ [37] and $HoVO_4$ [28]. In case of $HoVO_4$, a different structural sequence was found compared to the results under hydrostatic conditions; and in ScVO4, an unexpected metastable polymorph at ambient conditions was obtained after the pressure cycling of the sample. The deformation behaviour of $GdVO_4$ and the zircon-scheelite transition mechanism have been recently explored using radial x-ray diffraction [38]. The studies mentioned in this paragraph have opened up new avenues to explore the HP behaviour of $AVO_4$ orthovanadates.

Among rare-earth orthovanadates, $LaVO_4$ is the only compound, which can be crystallized in both the monazite and zircon structures at ambient pressure [39]. In particular, it has been found that when monoclinic monazite-type $LaVO_4$ is squeezed up to 12 GPa, a phase transition to another monoclinic phase takes place. The structure of the HP phase of $LaVO_4$ has been identified, and interestingly, after the transition, there is an increase in coordination number around V and La from four to six and nine to ten, respectively [40]. These results are relevant for understanding the HP behaviour of monazite-type oxides (e.g. phosphates, chromates, and selenates) and could contribute to the characterization of post-monazite phases observed in other vanadates such as $NdVO_4$, $PrVO_4$, and $CeVO_4$.

Out of many compounds belonging to $AVO_4$ series and excluding the rare-earth orthovanadates, the most recent *in-situ* HP studies has been reported on $InVO_4$ [41]. The interest of this compound comes from its potential application as a photo-anode for water splitting [42]. The crystal structure of $InVO_4$, which will be described in the next section, differs mainly from the zircon and monazite structures in terms of the low coordination of the indium ion. A new wolframite-type polymorph has been identified near 7 GPa by HP Raman spectroscopic and XRD studies and confirmed by *ab-initio* calculations [43]. The phase transition involves drastic changes in the physical properties of $InVO_4$ and the increase of coordination number of vanadium. In addition, it has been proposed that the new phase of $InVO_4$ could be suitable for the development of green-technology applications like photocatalytic hydrogen production [42]. Recent theoretical studies predicted two additional phase transitions in $InVO_4$ [44], which are yet to be confirmed experimentally.

Before the end of this section, we would like to make a brief comment on high-pressure investigations on nanoparticles of $AVO_4$ compounds. Nowadays these nanoparticles can be routinely prepared using soft-chemical methods [45]. They have been widely studied at ambient conditions because of their potential applications in medicine and in biological fields [46]. Of particular interest are their potential uses for the inactivation of the growth of different carcinoma [47] and the encapsulation of biomolecules [48]. In spite of the fact that the structural behaviour of nanomaterials under HP might be different from that of the bulk, due to the important role played by the surface in the structure stabilization [49], little efforts have been made to explore the HP behaviour of $AVO_4$ nanoparticles. To the best of our knowledge only zircon-type nanorods [50], zircon/monazite-type nanoparticles of $LaVO_4$ [51], and nanoboxes of $YVO_4$ [52] have been studied under compression. Details of these studies will be presented in a separate section dedicated to $AVO_4$ nanoparticles.

**3. Crystal structure**

In this section, we will describe in detail the crystal structure of commonly observed phases in $AVO_4$ compounds at ambient conditions. All the rare-earth orthovanadates (except for $LaVO_4$, which being dimorphic, can also be stabilized in the monazite structure) and many other ternary oxides crystallize in the so-called zircon structure [53]. Zircon is the name of the mineral $ZrSiO_4$, which is used to describe the family of all isomorphic compounds. Compounds adopting the zircon-type structure have a tetragonal symmetry in which the atomic positions of both cations are fixed by the crystal symmetry. High-quality large single crystals can be grown for zircon-type orthovanadates. The crystalline structure of rare-earth orthovanadates was solved by Milligan and Vernon in 1950s by means of neutron diffraction measurements [54]. These experiments allowed the determination of the unit-cell parameters and the atomic positions of the oxygen atoms (the only positions not fixed by the symmetry of the zircon crystal structure). Much more precise refinements of the zircon structure of rare-earth orthovanadates evolved from 1960s to 1990s [55, 56].

The zircon crystal structure is shown in Fig.1a. As it has been mentioned earlier, it is characterized by a tetragonal symmetry with space group $I4_1/amd$ (space group No. 14, $Z$ = 4). It is usually described in the literature using setting 2 (with origin at 2/$m$). In $AVO_4$ compounds, the $A$ and V cations are located at high-symmetry positions, which correspond to Wyckoff positions 4a (0, 3/4, 1/8) and 4b (0, 1/4, 3/8), respectively. On the other hand, the oxygen atoms are located at the 16h Wyckoff position, which is described by (0, $y$, $z$). In Table I, we summarize the oxygen positions of different compounds taken from the literature [55 - 64] along with the unit-cell parameters. As seen in the table, that there is direct correlation between the size of the trivalent metal atom *A*, the unit-cell parameters, and oxygen positions. In fact, Chakoumakos *et al.* have determined empirical relations correlating these parameters with the ionic radius of the trivalent atom *A* [56]. We have used these relations to estimate the structural parameters of $PmVO_4$, which has not been experimentally studied due to the short half-life of promethium (Pm) isotopes. The relationships given in Ref. 56 can also be used to estimate the structural parameters of vanadate solid solutions by using the Vegard´s law [65].

As shown in Fig.1a, the zircon-type structure of the orthovanadates can be described as formed by isolated $VO_4$ tetrahedral units, which surround the trivalent metal atom *A* forming eight-vertex $AO_8$ bisdisphenoids (triangular dodecahedra). Since each V atom is surrounded by four equivalent oxygen atoms, the $VO_4$ tetrahedron is regular with V-O lengths of approximately 1.7 Å which are quite rigid even under compression [24, 25, 28, 33]. The isolation of the $VO_4$ tetrahedron allows the description of lattice vibrations as internal and external modes of this unit [66]. In the $AO_8$ dodecahedron, the trivalent atom *A* is coordinated by eight oxygen atoms, with identical four short bond lengths and four long bond lengths. Depending on the compound, the average *A*-O distance ranges from 2.2 to 2.5 Å. The typical distortion index of the bond length is 0.06 [67]. Because of the structural characteristics described above, the principal structural unit in zircon can be described as a succession of alternating $VO_4$ and $AO_8$ polyhedral units running parallel to the crystallographic *c*-axis. Each of the chains is joined laterally by edge-sharing $AO_8$ dodecahedra, being this structural characteristic responsible for the zircon's prismatic habit and (100) cleavage as well as for the extreme birefringence of zircon-type crystals.

Another common structure observed in orthovanadates is the $CrVO_4$-type structure, which is shown in Fig 1b. This crystal structure is observed in $CrVO_4$, $InVO_4$, $TlVO_4$, and $FeVO_4$ ($FeVO_4$-II type polymorph) [68]. For the first time, this structure was accurately determined by Frazer and Brown [69]. The crystal structure is orthorhombic and belongs to the space group *Cmcm* (No. 63) with Z = 4. It consists of chains of nearly regular edge-sharing octahedral units of $CrO_6$, which are linked together by tetrahedral $VO_4$ groups. Again, these units can be considered isolated from each other. As an example of this structure, we include in Table II the structural information of the $CrVO_4$-type polymorph of $InVO_4$ (known as $InVO_4$-III) [41]. This information is representative of the four vanadates that are isomorphic and has been obtained using synchrotron x-ray powder diffraction. The unit-cell parameters of isomorphic compounds are given in Table III. Unlike the zircon structure, in the $CrVO_4$-type structure, the $VO_4$ tetrahedron is slightly distorted, with two long V-O distances and two short V-O distances with their values as 1.6579 Å and 1.7983 Å for $InVO_4$ at 0.8 GPa [41]. The $AO_6$ octahedron is also slightly distorted having four long equatorial distances and two short axial bonds with their values as 2.1623 Å and 2.1483 Å for $InVO_4$ [41]. We would like to remark here that a monoclinic distortion of the $CrVO_4$-type structure has been reported for $CrVO_4$ and $InVO_4$, which is obtained by heating a precipitated amorphous phase [70]. Its crystal structure belongs to space group *C*2*/m* (No. 12) with Z = 8 and similarly to the $CrVO_4$-type structure, it consists of chains of $VO_4$ tetrahedral and $CrO_6$ ($InO_6$) octahedral units.

Other than, the two commonly observed structures described above, a few $AVO_4$ orthovanadates have other metastable or stable crystal structures at ambient conditions. Two of them are $AlVO_4$ and $FeVO_4$ ($FeVO_4$-I type polymorph) which have a triclinic crystal structure shown in Fig. 1c. Although the $AlVO_4$-type structure is triclinic (space group *P*-1, No. 2, Z = 6) and has six formula units per unit cell, it can be considered as a distortion of the $CrVO_4$-type structure. Detailed information on the $FeVO_4$-I type structure can be found in the literature [71]. The structure consists of isolated $VO_4$ tetrahedral units, which are connected by $FeO_6$ (or $AlO_6$) octahedral units.

$BiVO_4$ is the compound belonging to the *A*$VO_4$-type series, which shows the richest polymorphism. Other than the zircon structure of $BiVO_4$, the compound also adopts two other polymorphs, the monoclinic fergusonite-type (described with space groups *I*2/*a* or *C*2/*c*, No. 15, Z = 4) and the orthorhombic pucherite-type (space group *Pbcn*, No. 60, Z = 4). Both structures are similar and are intimately connected to the tetragonal scheelite-type structure (space group $I4_1/a$, No. 88, Z = 4), which has been observed at high-temperature in $BiVO_4$. The fergusonite- and pucherite-type structures are shown in Figs. 1d and 1e, respectively. Detailed crystallographic information about these structures can be found in the literature [61, 72]. Interestingly, the scheelite-type structure has been found as a high-pressure polymorph in many orthovanadates. The first structural description of such scheelite phase was reported for $ErVO_4$ by Range and Meister [73]. This structure is discussed in more detail in the section dedicated to HP phase transitions. Here we will only state that in this structure, the V atom is coordinated by four oxygen atoms forming isolated tetrahedral units and the trivalent metal atom *A* is coordinated by eight oxygen atoms forming similar dodecahedra as in the zircon-type structure. The fergusonite structure is a monoclinic distortion of scheelite related to it by means of a group-subgroup relationship [74]. On the other hand, the pucherite structure has very similar unit-cell parameters as fergusonite. It can be considered as a high-symmetry version of it, which becomes the scheelite structure if the *a*- and *b*-axes, which differs by 6% in pucherite, are identical [75].

Another relevant structure for this review is the monazite structure (space group $P2_1/n$, No. 14, Z = 4) shown in Fig. 1f. This is one of the crystal structures adopted by $LaVO_4$ at ambient conditions [40] and the HP structure of several $AVO_4$ orthovanadates [31]. In the monazite structure, the V and trivalent *A* atoms are four- and nine-fold coordinated, respectively, and the $VO_4$ polyhedral units only share corners and edges with the $AO_9$ units.

It is interesting to note that the presence of isolated $VO_4$ tetrahedra is a common feature in most of the crystal structures of $AVO_4$ orthovanadates, which have been described in this section. The same feature is also observed in the low-temperature phases of a few rare-earth orthovanadates which has an orthorhombic crystal structure (space group *Fddd*, No. 70, Z = 8). The orthorhombic structure can be depicted as a distortion of zircon, where the symmetry is lowered by the cooperative Jahn-Teller effect [76], which causes a distortion in the *ab*-plane of zircon along [110] and [1-10] axes. In particular, the orthorhombic *a*- and *b*-axes, which are different by 2%, are the [110] and [1-10] diagonals of the zircon unit cell.

**4. Raman and IR spectra**

Among the $AVO_4$ orthovanadates, the vibrational properties under HP have been studied mainly for zircon-, monazite-, and $CrVO_4$-type compounds. Therefore, we will concentrate on the discussion of the vibrational modes of the compounds adopting these structures at ambient pressure. The Raman and IR modes of the HP structures will be discussed in the section where pressure-induced phase transitions are described. We will start with the zircon-type structure. Different authors have performed a detailed group-theoretical analysis of phonons in the zircon-type structure [66, 77]. In order to describe this structure a primitive cell can be chosen with only two formula units. The twelve atoms contained by this primitive cell give rise to thirty-six phonon branches at the center of the Brillouin zone (BZ), the Γ point. They decompose in the irreducible representations of $D_{4h}$ symmetry as follow. $\Gamma = (2\ A_{1g} + B_{1u}) + (4\ B_{1g} + 4\ A_{2u}) + (A_{2g} + 2\ B_{2u}) + (B_{2g} + A_{2u}) + (5\ E_g + 5\ E_u)$ [78]. The A and B modes are non-degenerate, whereas the E modes are doubly degenerate. Among these modes there are two acoustic modes ($\Gamma_{acoustic} = A_{2u} + E_u$) and five silent modes ($\Gamma_{silent} = A_{2g} + A_{1u} + B_{1u} + 2\ B_{2u}$). Consequently, there is a total of twelve Raman-active modes ($\Gamma_{Raman} = 2\ A_{1g} + 4\ B_{1g} + B_{2g} + 5\ E_g$) and seven IR-active modes ($\Gamma_{IR} = 3\ A_{2u} + 4\ E_u$). These modes can be classified either as internal or external vibrations of the $VO_4$ tetrahedron. The external modes are associated with either pure translations (T) or pure rotations (R) of the tetrahedron. The internal modes can be decomposed into symmetric and asymmetric stretching ($\nu_1$ and $\nu_3$) and bending motions ($\nu_2$ and $\nu_4$) of the vanadate unit [79].

The Raman spectra of zircon-type rare-earth orthovanadates have been measured at ambient conditions by several authors [66, 80 - 82]. *Ab-initio* calculations of the Raman spectrum have also been carried out [81]. Usually, in the experiments a maximum of eleven Raman-active modes are detected. The absence of some of the expected Raman modes in the experiments is due to their weak Raman-scattering cross-section. Polarized Raman experiments and *ab-initio* calculations have allowed a precise assignment of the Raman modes. The agreement between theory and experiment is excellent. In Table IV, as an example, we present the frequencies of the Raman modes of $TbVO_4$ [81] and $NdVO_4$ [82]. One can see that the differences between calculated and measured frequencies are less than 5%. For most of the modes, calculations tend to slightly under-estimate the frequency of the phonons. In Table IV, one can also see that the Raman modes can be divided in two frequency regions with a clear phonon gap between them. The low-frequency region is between 100 - 480 $cm^{-1}$, and the high-frequency region is between 750 – 900 $cm^{-1}$. The modes in the high-frequency region are usually the most intense modes and correspond to the symmetric-stretching internal mode $\nu_1(A_{1g})$, and two asymmetric-stretching modes $\nu_3(E_g)$ and $\nu_3(B_{1g})$. The rest of the rare-earth orthovanadates have a similar spectral distribution of Raman modes. An interesting feature to highlight on the Raman spectrum of zircon-type $AVO_4$ compounds is that the frequency of the internal $VO_4$ phonons increased with increasing the atomic number as a consequence of increase of mass of the trivalent metal *A* [65]. This is a due to the fact that, the *A*-O distances decrease as the atomic number increases (due to the increase of the atomic mass of the lanthanide atom) going from La to Lu. The decrease of the *A*-O bond distance leads to an enhancement of the energy of the $VO_4$ tetrahedron, which results in the hardening of the internal $VO_4$ frequencies [66].

Regarding the IR-active modes, much less efforts have been made in comparison to the Raman measurements. The compound for which maximum information is available so far is $YVO_4$, where the seven expected IR modes have been characterized [83 - 85]. As a first approximation, the IR spectrum is qualitatively similar to the Raman spectrum, with internal modes in the high-frequency region and external modes lying in low-frequency region, and a phonon gap between both groups of modes. According to the most recent mode assignment [85], the IR modes associated to internal vibrations of $VO_4$ tetrahedron are located in the high-frequency region around 778 and 930 $cm^{-1}$. These modes correspond to one $A_{2u}$ and one $E_u$ mode, which exhibit an LO–TO splitting due to the coupling of the atomic displacement with the long-range electric field [85]. The low-frequency modes are located from 190 to 315 $cm^{-1}$, and have been assigned to two $A_{2u}$ and three $E_u$ modes. These five modes correspond to lattice vibrations and show the LO–TO splitting. In contrast with the Raman vibrations, apparently anharmonic effects cannot be neglected in the IR modes [83].

We will comment now on the lattice vibrations of $CrVO_4$-type vanadates. For these compounds, group theoretical analysis led to the following vibrational representation at the Γ point of the BZ in standard notation: $\Gamma = 5\ A_g + 4\ B_{1g} + 6\ B_{1u} + 3\ A_u + 2\ B_{2g} + 7\ B_{2u} + 4\ B_{3g} + 5\ B_{3u}$ [68, 86]. Among these modes, there are three acoustic modes ($B_{1u} + B_{2u} + B_{3u}$), three silent modes ($A_u$), fifteen Raman-active modes ($5\ A_g + 4\ B_{1g} + 2\ B_{2g} + 4\ B_{3g}$), and fifteen IR-active modes ($5\ B_{1u} + 6\ B_{2u} + 4\ B_{3u}$). Studies on the Raman and IR modes of $CrVO_4$-type vanadates are sparse compared to the zircon-type vanadates. The IR spectrum was investigated by Olivier [87] and Baran [88], who tentatively assigned the observed modes. More recently, all the expected fifteen Raman modes of $InVO_4$ have been measured, identified, and calculated [44]. The results obtained for $InVO_4$ are summarized in Table V. The agreement between calculations and experiments is excellent as can be seen in the table. Calculations tend to give frequencies slightly smaller than experiments, but differences are comparable to those found in zircon-type vanadates. The computer simulations also provide a full description of the IR modes of $InVO_4$ [44]. Isomorphic vanadates are expected to have similar Raman and IR spectra. The accurate predictions obtained from *ab-initio* calculations for the Raman spectra of $InVO_4$ suggest that this methodology can be a powerful tool for guiding the future IR experiments. As in the case of zircon-type compounds, for $CrVO_4$-type compounds also, the vibrational spectra can be interpreted in terms of modes of the $VO_4$ tetrahedron, which can be considered as an independent and isolated unit. Thus the modes can be classified either as internal (the $VO_4$ center of mass does not move) or external (movements of $VO_4$ tetrahedron as rigid unit and the trivalent atom). As can be seen in Table V, the internal stretching modes are the ones with the highest frequency and the external translational modes lie at the lower frequency. This occurs for both Raman and IR modes.

Before we conclude this section, we will comment on lattice vibrations of the monazite-type structure. According to group theory, the monazite structure has seventy-two vibrational modes. The symmetry decomposition of the zone center phonons is dictated by the factor group 2/*m*; being Γ = 18 $A_g$ + 18 $B_g$ + 18 $A_u$ + 18 $B_u$. Three of these modes ($A_u$ + 2 $B_u$) are acoustic vibrations. The rest of the vibrational modes correspond to thirty-three IR-active modes (17$A_u$ + 16 $B_u$) and thirty-six Raman-active modes (18 $A_g$ + 18 $B_g$). All the Raman modes of monazite-type $LaVO_4$ have been measured and identified which are summarized in Table VI [40]. The mode assignment is based upon polarized Raman measurements, which allow the separation of $A_g$ and $B_g$ modes. A good agreement is observed between experiments and calculations. As in the other two family of compounds discussed before, for monazite-type oxides the vibrations can also be classified in terms of stretching, bending and rotational-translational modes of the $VO_4$ tetrahedron [89]. In $LaVO_4$, there is an isolated group of Raman-active phonons in the wavelength range from 768 to 862 $cm^{-1}$ corresponding to stretching modes (see Table VI) with the highest frequency mode corresponds to the stretching of the shortest V-O bond. On the other hand, the $A_g$ mode at 859 $cm^{-1}$, which is the most intense mode [40], is related to the $\nu_1$ breathing mode. The bending modes are observed in 309 - 440 $cm^{-1}$ range. A typical example is the $A_g$ mode at 374 $cm^{-1}$ in which the opposed O-V-O bonds in the $VO_4$ tetrahedron bend in phase, as in $\nu_2$ vibrations. Finally, the rotational-translational lattice modes display wavenumbers from 64 to 252 $cm^{-1}$. Here, the most remarkable fact is the increase in the La vibrational amplitude as the wavenumber of the phonon decreases.

The frequencies of the IR-active modes have also been calculated and are given in Table VI. The IR modes show a similar frequency distribution as the Raman modes. Different authors agree about the calculated frequencies of the IR-mode assignment [40, 90]. Calculations also agree within 1.5 % with the experimentally measured stretching modes [90], which are the only IR modes reported up to now.

## 5. High-pressure structural studies on zircon/monazite-type orthovanadates

### 5.1 X-ray diffraction experiments

Powder x-ray diffraction is the most versatile technique to obtain the structural details of a material and is generally performed in two geometries; namely: angle dispersive x-ray diffraction (ADXRD) or energy dispersive x-ray diffraction (EDXRD). In these configurations a monochromatic (ADXRD) or polychromatic (EDXRD) x-ray beam interacts with the sample and the diffracted beam is collected on two dimensional area detectors such as an image plate or and charge-coupled device (ADXRD) or a high-purity germanium point detector (EDXRD). After the data collection, the raw data is processed through various methods and the structural refinement is carried out to extract the crystallographic details. More frequently, a sample in the form of powder is employed for XRD measurements under HP, however, with the availability of intense x-ray beam from fourth generation synchrotron sources, studies on single crystals are picking up [91 - 95]. Fig. 2 depicts the ambient-pressure lattice parameters for all the $RVO_4$ compounds adopting the zircon structure. One can clearly see the increase in the tetragonal distortion with the decrease in the ionic radii of the rare earth, which is also reflected in the *c/a* axial ratio. The overall volume of the unit cell decreases by ~ 18 % as one moves from La to Lu, which is a consequence of well-known rare-earth contraction effect [94].

The first *in-situ* HP x-ray diffraction study in a zircon-type vanadate was carried out in the year 2004. This study was performed by Wang *et al.* [24] on the widely used laser-host material $YVO_4$ up to 26 GPa at room temperature. In their experiments, a fine powder of $YVO_4$ was loaded in a DAC using nitrogen as pressure-transmitting medium and ruby was used as the optical pressure marker. A micron-sized monochromatic x-ray beam, with wavelength 0.4176 Å, from a synchrotron source along with an image-plate area detector were employed for high-resolution data collection. Based on these *in-situ* ADXRD experiments, Wang *et al.* reported an irreversible phase transition in $YVO_4$ from the low-density zircon structure to the high-density scheelite structure at around 8.5 GPa, with nearly 12 % volume discontinuity [24]. On further compression, these authors mentioned the signature of another HP phase around 26 GPa, which can be clearly identified by the broadening of the Bragg peak identified as (112) in Fig. 3. However, they did not give any detail about the crystal structure of the second HP phase. Subsequent to this study, high-pressure XRD measurements were reported by Mittal *et al.* [95] in the year 2008 on $LuVO_4$, the end-member of the rare-earth orthovanadate series. These authors observed a zircon to scheelite phase transition similar to the one observed for $YVO_4$. In addition, they reported a second HP phase in the compound and for the first time, they proposed the crystal structure of this phase as fergusonite-type. In the year 2009, high-pressure XRD studies were reported by Garg *et al.* [26] on $YbVO_4$. These experiments also showed the same pressure-induced phase-transition sequence observed in $YVO_4$ and $LuVO_4$. After these reports, systematic HP investigations followed on $RVO_4$ compounds by various researchers, including the authors of this review [25, 26, 28, 29, 33, 36, 59, 96, 97]. Measurements on these compounds now have established beyond doubt that the $RVO_4$ compounds with smaller ionic radii (Sm-Lu) transform from the zircon to the scheelite phase below 10 GPa with nearly 10 % volume discontinuity. In many of these compounds, a reversible second HP phase transition to a fergusonite structure has been reported (e.g. $EuVO_4$ [59] and $HoVO_4$ [28]). In case of $TbVO_4$, experiments carried out under quasi-hydrostatic conditions have found the fergusonite structure to remain stable up to nearly 50 GPa [33]. The systematic behaviour observed in the HP studies of the vanadates described in this paragraph has also been found in $YV_{1-x}P_xO_4$ solid solutions [98] in which the increase of the phosphor content shifts the transition pressure towards higher pressures.

In Fig. 3 we show the pressure evolution of XRD patterns in $YVO_4$ up to 26.6 GPa [24], the highest pressure reached in the measurements. Up to 7.8 GPa, all the diffraction peaks could be assigned to the ambient-pressure zircon phase and the data collected at 8.5 GPa shows an additional peak indicating the onset of a pressure-induced phase transition in the compound. On further compression, the intensity of the new peaks increases and that of the zircon-phase peaks decreases, which clearly indicates the coexistence of the two phases until 12.5 GPa. Beyond this pressure, diffraction peaks from only the HP phase are observed. On pressure release, the HP could be retained indicating the irreversible nature of the transition. The structural details of scheelite-type $YVO_4$ at ambient pressure are given in Table VII. The same behaviour has been observed in the XRD patterns measured under compression in $ScVO_4$ and most of the $R$VO$_4$ compounds going from $SmVO_4$ to $LuVO_4$. In Fig. 4, we show the Rietveld refinement of the zircon and scheelite phases observed in $YVO_4$ along with the residual plot to illustrate the unequivocal assignment of the HP phase to the scheelite structure [24]. The appearance of a strong peak in the HP phase near $2\theta = 8^o$ is a finger print of the presence of the scheelite structure. This peak corresponds to the overlap of (112) and (103) Bragg reflections of the scheelite structure.

In order to illustrate the pressure-induced zircon-scheelite-fergusonite structural sequence, in Fig. 5 we show a selection of XRD patterns measured in $TbVO_4$ up to nearly 50 GPa [33]. Since fergusonite is a monoclinic distortion of scheelite, the main feature of the scheelite-fergusonite transition, found at 33.9 GPa in $TbVO_4$, are the appearance of extra weak peaks and the splitting and broadening of some of Bragg peaks of the scheelite structure. In particular, there is a Bragg peak of the new structure appearing at low angle, which is a fingerprint of the fergusonite structure. In addition, the (112) and (103) peaks of scheelite, located near $2\theta = 8^o$, gradually split as a consequence of the monoclinic distortion of the fergusonite structure. As mentioned before, this phase has been found in other members of the rare-earth vanadate compounds as a post-scheelite phase. One of them is $EuVO_4$, for which the crystal structure of fergusonite has been accurately determined. This structure at 26.4 GPa is given in Table VIII.

It is important to note that, due to the irreversible nature of zircon-scheelite transition, the scheelite-type polymorph can be recovered as a metastable phase at ambient conditions. However, the transition from scheelite to fergusonite is reversible, leading to the recovery of the scheelite phase as a metastable phase on pressure release. This is shown in Fig. 5 by the XRD pattern measured on $TbVO_4$ at 0.3 GPa after pressure release, which can be undoubtedly assigned to the scheelite structure. The ambient-pressure scheelite structure has been determined for a few of the studied vanadates. The results are summarized in Table VII. Interestingly the scheelite structure can also be prepared in rare-earth vanadates by mechanochemical synthesis from $R_2O_3$ and $V_2O_5$ powders [99], which opens the door to prepare these compounds in large quantity required for industrial applications.

As shown in Fig. 6a, the scheelite structure consists of $VO_4$ tetrahedral and $RVO_4$ dodecahedral units similar to the zircon structure. However, the scheelite structure is much more compact than the zircon structure. In particular, at ambient pressure, for a given compound, the volume of scheelite-type structure is approximately 12 % lower than that of the zircon-type phase. On the other hand, in spite of the large volume drop observed across the zircon-scheelite transition, there is no change in the coordination number around *R* and V cations. Regarding the bond distances, the average *R*−O and V−O bond distances are similar in both polymorphs. This is a consequence of the fact that the transition from zircon to scheelite mainly involves reorientation of $VO_4$ tetrahedral and $RO_8$ dodecahedral units. In addition, the $RO_8$ dodecahedra become slightly more regular in the HP scheelite polymorph. Notice that the scheelite-structure is a symmetrized version of the fergusonite structure (shown in Fig. 1d), in which two unit-cell parameters are identical and the monoclinic $\beta$ angle is equal to 90º.

From Rietveld refinements of HP x-ray diffraction experiments, the pressure dependence of unit-cell parameters for different compound has been obtained for the zircon and scheelite phases. The same systematic has been obtained for all the studied compounds. In order to illustrate how pressure affects the unit-cell parameters, in Fig. 7 we show the results obtained for $TbVO_4$ [33, 63]. The observed non-isotropic axial compression results in gradual increase in the axial ratio (*c/a*) from 0.881 at ambient pressure to 0.891 at 6.4 GPa in zircon-type $TbVO_4$. The opposite behaviour is found for the scheelite phase, in which the *c*-axis is the most compressible axis instead of *a*. In particular, in scheelite the axial ratio *c/a* decreases from 2.231 at ambient pressure to 2.185 at 30.3 GPa.

An important fact to consider in HP studies is the influence of non-hydrostatic stresses [100], which depend on the pressure-transmitting medium (PTM) used in the experiments. The presence of such stresses can influence not only the transition pressure, but also the structural sequence and other physical properties. For instance, in case of $HoVO_4$, non-hydrostaticity reduces the zircon-scheelite transition pressure from 8.2 to 4 GPa [28] and triggers the second transition from scheelite to fergusonite at 20.4

GPa. The same transition is expected to occur under quasi-hydrostatic conditions beyond 28 GPa. More recently, it has been shown than when highly non-hydrostatic (no PTM) conditions are applied to $ScVO_4$ [37], the zircon structure can be transformed directly into a metastable fergusonite structure at 6 GPa, whereas, the same compound compressed quasi-hydrostatically transforms to the metastable scheelite phase at 8.2 GPa [30].

HPXRD measurements on $R$VO$_4$ compounds with larger ionic radii (La - Pr) show a different behaviour than other members of the family. In $CeVO_4$, $PrVO_4$, and $NdVO_4$ an irreversible zircon to monazite phase transition, with a large volume discontinuity, takes place beyond 4 GPa. In Fig. 8 we show the pressure evolution of XRD patterns collected for $CeVO_4$ up to 45 GPa under quasi-hydrostatic conditions using neon as pressure-transmitting medium [33]. As one can see, in the diffraction patterns collected at pressures lower than 3.9 GPa, all the observed peaks could be fitted with the ambient-pressure zircon structure. However, in the data collected at 3.9 GPa, in addition to the diffraction peaks from the ambient-pressure phase, a few additional peaks are seen which is a clear indication of the onset of a pressure-induced structural transformation in the compound. Under further compression, the change of the Bragg peaks intensity indicates that the phase fraction of the zircon phase reduces and that of the HP phase increases. In the patterns collected at 5.9 GPa and beyond, all the observed Bragg peaks could be assigned to the HP phase. The structure of this HP phase is assigned as monoclinic monazite (space group $P2_1/n$, No. 15, Z = 8).

On compressing $CeVO_4$ further, another phase transition is detected around 14.5 GPa. An orthorhombic structure was suggested for this phase based on a Lebail analysis [33]. This orthorhombic phase remains stable up to 45 GPa, the highest pressure reached in the experiments [33]. On pressure release, the monoclinic monazite phase was retained at ambient pressure indicating the metastable nature of the first HP phase. Similar results were reported by Garg *et al.* [32] when $CeVO_4$ was compressed with a methanol-ethanol mixture as pressure-transmitting medium. However, based on a Rietveld analysis, these authors proposed that the second HP phase is isostructural to monazite, but having a much smaller volume. Nevertheless, on pressure release the monazite phase was recovered as observed for the measurements made with neon as pressure-transmitting medium. As it has been mentioned in an earlier paragraph, the two different post-monazite structures observed in these studies [32, 33] could be due the presence of different non-hydrostatic stresses as a result of different pressure-transmitting medium used in the experiments.

Based on *in-situ* x-ray diffraction studies in $PrVO_4$, the compound next to $CeVO_4$ in the series, the zircon-monazite phase transition has been found to occur at 6 GPa [101]. Additionally, XRD experiment carried out on a pressure-recycled sample allowed the refinement of the monazite-type structure [31]. The details of this structure are given in Table IX. Note that the monazite structure involves an increase in the coordination number of Pr from eight to nine and a distortion of the $VO_4$ tetrahedron, which in monazite has four slightly different V-O distances: 1.699(4) Å, 1.698(4) Å, 1.671(4) Å, and 1.666(5) Å. *In-situ* XRD measurements have been carried out on $NdVO_4$, up to 12 GPa. This compound also shows an irreversible HP phase-transition to a monazite-type structure [96]. The compression of the HP monazite phase in all the studied compounds and ambient phase of monazite-type $LaVO_4$, will be discussed in the next paragraph.

$LaVO_4$ is the only rare-earth orthovanadate, adopting the monazite structure at ambient conditions. Three XRD studies have been reported on this compound. In one of the investigations, carried out up to 9.1 GPa, only the compressibility behaviour was reported, with no observation of any structural phase transition [102]. In another study, an isostructural phase transition was proposed to occur near 8.6 GPa [96]. However, most recent studies combining single-crystal and powder XRD, performed under hydrostatic conditions, located the transition pressure at 12.2 GPa [40]. In Fig. 9 we show powder XRD patterns illustrating this transition. In the figure, we also include the refined XRD profile and the residual to illustrate the quality of the structure determination. The crystal structure of the HP phase has been determined as a monoclinic structure, which can be described with the same monoclinic space group ($P2_1/n$) as the low-pressure phase but with a doubled unit-cell volume. The HP phase of $LaVO_4$ (shown in Fig. 6b) is isomorphic to $BaWO_4$–II, the HP post-scheelite structure of $BaWO_4$ [103]. Details on the HP structure of $LaVO_4$ are given in Table X. The HP phase remains stable up to 21 GPa and upon decompression the phase transition is reversible with a large hysteresis. The phase transition results in an increase of density by 8 % in $LaVO_4$. The volume difference between the two phases indicates a first-order phase transition. In addition, the phase transition implies an increase of coordination number of V (La) from 4 to 6 (9 to 10) [40].

A last comment on the behaviour of $LaVO_4$ under compression is dedicated to its compressibility, which is illustrative of the compressibility of different monazite-type polymorphs found in $AVO_4$ compounds. It has been found that the compression of $LaVO_4$ is highly non-isotropic. The determined values for the different linear compressibility coefficients are $\kappa_a = 4.39 \times 10^{-3}$ GPa$^{-1}$, $\kappa_b = 4.68 \times 10^{-3}$ GPa$^{-1}$, and $\kappa_c = 0.45 \times 10^{-3}$ GPa$^{-1}$, which means that the volume compression of monazite $LaVO_4$ is mainly accommodated within the (001) plane. The pressure variation of monoclinic $\beta$ angle is given by $\partial\beta/\partial P =$ -0.18 º/GPa. From these parameters, the four independent components of the isothermal compressibility tensor [104] of monazite $LaVO_4$ can be determined. The calculate values are $\beta_{11}$ = 4.86 × $10^{-3}$ $GPa^{-1}$, $\beta_{22}$ = 4.68 × $10^{-3}$ $GPa^{-1}$, $\beta_{33}$ = 0.45 × $10^{-3}$ $GPa^{-1}$, and $\beta_{13}$ = -0.32 × $10^{-3}$ $GPa^{-1}$.

An additional topic that deserves to be commented is the possible decomposition of rare-earth orthovanadates under high pressure. Such phenomenon was not observed up to 50 GPa in most of the experiments carried out. However, when using large-wavelength x-rays for the experiments, after long exposure a partial decomposition was detected [28, 59]. This fact is apparently triggered by x-ray absorption, which induces photoelectric processes leading to the dissociation of $V_2O_5$ units from the vanadates. This is a common phenomenon in ternary oxides when x-rays with wavelengths larger than 0.6 Å are used [105].

### 5.2 Raman spectroscopy

Raman spectroscopy is a powerful experimental tool, which provides information about the vibrational properties of materials. The technique probes elementary excitations in the material by utilizing the inelastic scattering of a monochromatic light source. Since the Raman spectra of phonons (lattice and molecular vibrations) have a very high selectivity, it permits a finger-print analysis of the materials phase that includes its chemical composition and structural state. High-pressure Raman spectroscopy [106] is complementary to XRD measurements for analysing structural phase transitions because it is a subtle local probe capable of distinguishing small traces of various local phases coexisting in a compound.

The first HP Raman investigation on a compound belonging to $AVO_4$ series was reported by Jayaraman *et al.* nearly three decades ago on pure and doped crystals of $YVO_4$ up to 15 GPa [11]. In their measurements, the researchers found the appearance of new Raman peaks at around 7.5 GPa indicating the presence of phase transition in the compound. Subsequently, HP Raman experiments were reported on single crystals of $TbVO_4$ and $DyVO_4$ up to 18 and 10 GPa, respectively [22]. Both compounds showed a phase transition around 6.5 GPa. On pressure release, the HP phase was found to be metastable. Nearly two decades later, Rao *et al.* published HP Raman investigations on $LuVO_4$ up to 26 GPa [107]. The authors of this work reported changes in the Raman spectra at around 8 GPa across the phase transition previously reported by x-ray diffraction [11, 22]. Around 16 GPa subtle indications of a second phase transition were observed. On pressure release, the Raman spectra resemble with that of first HP phase. HP Raman measurements followed on other compounds of the $R$VO4 series [30, 31, 51, 81, 82, 108]. As observed in XRD studies, for the compounds with lower ionic radii (Sm - Lu including Y and Sc), Raman measurements supported the zircon-to-scheelite transition. On the other hand, for larger ionic-radii compounds (Ce - Nd) they substantiated the zircon-to-monazite transition.

In Figs. 10 and 11 we show HP Raman spectra measured on $PrVO_4$ and $SmVO_4$, respectively [31] where one can clearly see that for both compounds under compression, the Raman spectra resembles with that of zircon phase up to 6 GPa. The experimentally determined pressure dependences of the Raman modes are shown in Fig. 10 and 11. Most of the modes harden under compression. Because of the dissimilar pressure dependence of different modes, some of them merge or cross. In particular, the crossover of the $E_g$ and $B_{1g}$ mode with wavenumbers near 250 $cm^{-1}$ is a typical feature of zircon-type vanadates. In Figs. 10 and 11 it can be seen that noticeable changes occur in the Raman spectra of $PrVO_4$ and $SmVO_4$ beyond 6 GPa indicating the onset of a phase transition. In $PrVO_4$ ($SmVO_4$), the Raman spectra of the HP phase can be identified with the monazite (scheelite) structure [31]. As commented previously in section 4, thirty-six Raman-active modes are allowed in the monazite structure. In contrast, the scheelite structure (space group $I4_1/a$ and point group $C_{4h}^6$) has only thirteen Raman-active modes: Γ = 3 $A_g$ + 5 $B_g$ + 5 $E_g$ [109]. The difference in the number of Raman modes between both structures is quite evident in Figs. 10 and 11.

As mentioned in the beginning of this section, Raman spectroscopy has been useful to detect two HP phase transitions in zircon-type $A$$VO_4$ compounds. This is illustrated in Fig. 12 where we show the Raman spectra measured for $TbVO_4$ up to 36 GPa [81]. The differences between the spectra measured at 6.7 GPa and 8.3 GPa clearly indicates the transition from zircon to scheelite phase. The scheelite structure is found to be stable up to 26.3 GPa. However, beyond 26.7 GPa some of the Raman peaks split into two and new peaks appear. Both facts are a consequence of the scheelite-fergusonite transition. Remember that scheelite has thirteen Raman-active modes, but fergusonite has eighteen Raman-active modes. In particular, the group theoretical considerations indicate that the fergusonite structure, which is centrosymmetric, has the following Raman-active modes: Γ = 8 $A_g$ + 10 $B_g$. The modes of fergusonite are derived from the reduction in the tetragonal $C_{4h}$ symmetry of scheelite to the monoclinic $C_{2h}$ symmetry. Therefore, there is a direct relationship among the Raman modes of the scheelite and the fergusonite phases, which justify the subtle difference between Raman spectra of two phases. Notice that the second-order nature of this transition in $YVO_4$ has been proved using Landau's theory [26, 110, 111], which implies small changes between the Raman spectrum of scheelite and fergusonite. This interpretation is consistent with two facts: there is a low-frequency mode in scheelite (an external translational $B_g$ mode) which softens under compression [26] and at the phase transition there is a change in sign in the pressure coefficient from negative to positive for the corresponding $A_g$ mode in fergusonite [26]. This feature supports the assignation of the HP scheelite-fergusonite transition as a classical displacive transition [112].

In Fig. 12, it can be seen that the fergusonite phase of $TbVO_4$ is stable up to 31.4 GPa. At 34.4 GPa an additional phase transition is found. This transition was previously not detected by XRD [33]. The difference between Raman and XRD results are attributed to the use of a less hydrostatic pressure medium in the Raman experiments [81]. An orthorhombic structure has been proposed for the phase detected at 34.4 GPa based upon density-functional theoretical calculations [81]. Details on this structure will be provided in the section devoted to theoretical calculations. However, this structural assignment is yet to be confirmed by XRD experiments. The new HP phase has been observed up to 36.6 GPa with no evidence of chemical decomposition or amorphization of $TbVO_4$. Upon pressure release from 36.6 GPa the scheelite phase is recovered as a metastable phase (see Fig. 12).

The knowledge of the pressure dependence of phonons is useful to estimate the pressure dependence of thermodynamic, elastic, and transport properties of materials. Quantitative information has been obtained for the pressure dependence of the Raman modes of different phases of $R\mathrm{VO}_4$ vanadates and similar behaviour has been found in most of them. As a representative case of $R\mathrm{VO}_4$ series [26], we will present here the frequencies (ω) and pressure coefficients (dω/dP) that have been determined for the zircon, scheelite, and fergusonite phases of $YVO_4$. An excellent agreement has been found between experiments and calculations carried out using the general-gradient approximation (GGA) and local-density approximation (LDA) [26]. The results are summarized in Table XI.

In the zircon phase, the frequencies of most of the Raman modes increases under compression, except for two low-frequency external modes assigned as $E_g$ and $B_{2g}$ modes (see Table XI). These modes show negative pressure coefficients. Their existence has also been observed in other zircon-type vanadates [22, 27, 31, 44, 81] and

it is typical of the destabilization induced by pressure in the zircon structure. For instance, this can be seen for $PrVO_4$ and $SmVO_4$ in Figs. 10 and 11. Regarding the modes that harden under HP, one of the most noticeable facts is that the high-frequency internal modes are those with the largest pressure coefficients (see Table XI), which is a direct consequence of the incompressibility of the $VO_4$ tetrahedron. Another characteristic of the Raman spectrum of the zircon phase is that there are pairs of modes, which progressively merge under compression due to their different pressure dependence. In the case of $YVO_4$, these are $B_g$ and $E_g$ modes which at ambient pressure have wavenumbers 156.8 and 163.2 $cm^{-1}$, respectively. We would like to remark here that the pressure coefficients of equivalent Raman modes are comparable among all studied zircon-type orthovanadates. One example of this is the highest-frequency symmetric stretching internal vibrations of $VO_4$. The frequency and pressure coefficient of these modes in different compounds are: 891.1 $cm^{-1}$ and 5.99 $cm^{-1}$/GPa for $YVO_4$, 885 $cm^{-1}$ and 5.6 $cm^{-1}$/GPa for $TbVO_4$, 905.6 $cm^{-1}$ and 5.33 $cm^{-1}$/GPa for $ScVO_4$, 871.1 $cm^{-1}$ and 5.63 $cm^{-1}$/GPa for $NdVO_4$, 899 $cm^{-1}$ and 6.31 $cm^{-1}$/GPa for $LuVO_4$, and 898 $cm^{-1}$ and 5.4 $cm^{-1}$/GPa for $YbVO_4$.

In the scheelite structure, again the high-frequency modes are the ones which get maximum affected by compression. In this phase, there is only one mode with a negative pressure coefficient which is a low-frequency $B_g$ mode (with $\omega$ = 166.2 $cm^{-1}$ in $YVO_4$). The existence of this mode has been attributed by Manjon *et al.* [26] to mechanical instabilities in the scheelite structure, which trigger the scheelite-fergusonite transition [26]. In contrast with the zircon and scheelite structures, in fergusonite all the Raman modes have been found to have positive pressure coefficients. Another distinctive feature of this phase is the increase of number of high-frequency modes from three to four (see Table XI). This is a consequence of the distortion of the

$VO_4$ tetrahedron, which is regular in zircon and scheelite, but becomes irregular (with two short and two long V-O bonds) in fergusonite. A conclusion that can be drawn from Table XI is the tendency to decrease the pressure coefficient of the phonons following the zircon-scheelite-fergusonite sequence because of the decrease in the compressibility of the structures. Therefore, following the structural sequence, the effect of pressure is reduced on bond distances and the relevant restoring force of the different lattice vibrations.

To conclude this section, we would like to comment on the influence of pressure in monazite-type $LaVO_4$. Raman spectra as a function of pressure are shown in Fig. 13 [40]. The changes observed around 12.2 GPa are consistent with the occurrence of the phase transition described in the section devoted to XRD measurements. The main changes caused by the phase transition in the Raman spectrum are the proliferation of Raman modes, the decrease in the wavenumber of the highest frequency mode, and the appearance of new phonons in the region from 500 to 700 $cm^{-1}$. Upon decompression, the changes are reversible, with large hysteresis. As mentioned earlier, the HP phase has a monoclinic $BaWO_4$–II-type structure. This structure has seventy-two Raman-active modes: $\Gamma = 36\ A_g + 36\ B_g$; however, in the experiments only thirty-one modes have been detected. Possible reasons for measuring fewer modes than theoretically expected might be due to the broadening and overlapping of Raman modes and the presence of modes with intensities below the noise threshold. The same phenomenon is observed in isostructural phases to $BaWO_4$-II in other materials; e.g. $BaWO_4$ [113]. An interesting observation is the enlargement of the number of high-frequency modes in the HP phase, a fact that it is coherent with the enlargement of the coordination number of the V atom [40].

## 5.3 Optical absorption, luminescence, and transport measurements

Although the most widely and routinely used techniques for the study of pressure-induced phase transitions, are powder XRD and Raman spectroscopy, there are a few other techniques, such as optical absorption, photoluminescence, and transport measurements, which have been used by researchers. Optical absorption is a simple technique providing information about the changes in the electronic band gap of the material [114]. The use of type IIA diamonds in the DAC allows the study of compounds with a band-gap energy ($E_{gap}$) smaller than 5.5 eV. Therefore, HP optical-absorption measurements can be accurately performed in zircon-type vanadates, for which $E_{gap} < 4$ eV [36, 115].

Jayaraman *et al.* reported the first HP optical-absorption study on $YVO_4$, $TbVO_4$, and $DyVO_4$ [21, 22]. These authors reported the decrease in the band gap of the HP phase under compression with slopes of -37 meV/GPa and -33meV/GPa for $TbVO_4$, and $DyVO_4$, respectively. More recently HP optical-absorption experiments on four $R$VO$_4$ orthovanadates ($R$ = Lu, Nd, Y, Yb) were reported by Panchal *et al.* [36]. These measurements led to the determination of the pressure dependence of $E_{gap}$ in the different polymorphs of the studied compounds. The zircon phase of the different compounds is found to have a direct band gap, with the top of the valence band and the bottom of the conduction band at the Γ point of the BZ [36]. All the compounds have a band gap between 3.7 and 3.8 eV. This is a consequence of the fact that in zircon-type rare-earth vanadates the band structure near the Fermi level is dominated by electronic states of the vanadate ion. In particular, the top of the valence band mainly consists of non-bonding O 2p states and the bottom of the conduction band has an antibonding V 3d - O 2p character.

Pressure dependence of $E_{gap}$ has been determined and is shown in Fig. 14 [36]. In the zircon phase, there is a small linear increase in $E_{gap}$ under compression for all compounds. This is a consequence of the fact that under pressure the V 3d states move faster towards higher energies than O 2p states, leading to the observed opening of the band gap. At the transition pressure, a collapse of the band gap takes place, being the change in $E_{gap}$ larger for the zircon-scheelite transition than for the zircon-monazite transition [36]. The collapse of $E_{gap}$ is associated with the contraction of the unit-cell volume. Additionally, in the HP phase the band gap closes when pressure increases. The pressure coefficients of the HP phases are of the order of -20 meV/GPa; i.e. comparable to the values previously found for $TbVO_4$ and $DyVO_4$ [22]. In case of $NdVO_4$, two discontinuities are found in $E_{gap}$ (se Fig. 14), suggesting the possible existence of a second HP phase transition. The HP scheelite phase is a direct-gap semiconductor with the valence-band maxima and conduction-band minima located at the Γ point of the BZ. Monazite $NdVO_4$ is an indirect-gap material with the maximum of the valence band at the Z point of the BZ and the minimum of the conduction band at the Y point. In the HP phases, there is a partial delocalization of 4f electrons of the rare-earth atoms. In addition, some hybridization between orbitals occurs. The observed decrease of $E_{gap}$ under compressioncan is related with both phenomena.

Photoluminescence (PL) spectroscopy is another technique to explore the effect of pressure in the electronic properties of materials. In these measurements, a luminescent ion is used as a probe to investigate the consequences of structural changes induced by pressure on electronic properties [116]. The first HP luminescence study on a zircon-type vanadate was reported by Chen *et al.* [117] on $Eu^{+3}$ doped $YVO_4$. Abrupt changes in the $Eu^{+3}$ ion's luminescent spectra at around 7 GPa indicated the presence of the pressure-driven phase transition. The same authors published another report on the

study of lifetime measurements of luminescent peaks of $Eu^{+3}$ in $YVO_4$ [118]. Beyond the transition pressure, the appearance of new luminescence peaks and an abrupt reduction in the intensity of the peaks are clear evidences of the pressure-induced phase transition. Similar results have been obtained from $Nd^{3+}$ and $Pr^{+3}$ doped $YVO_4$ [119, 120] and $Eu^{+3}$ doped $GdVO_4$ [110]. Results for $Eu^{+3}$ doped $GdVO_4$ are shown in Fig. 15. The appearance of new luminescence peaks at around 6.3 GPa indicates the pressure induced phase transition. Group theory predicts four luminescence peaks for the Eu ion from $^5D_0$ to $^7F_{1,2}$ electronic transitions in the zircon-type structure. Changes at around 2.8 GPa, shows the distortion in the local symmetry around $Eu^{+3}$ and the lifting of degeneracy of $^5D_0$ and $^7F_1$ transitions. The drastic changes observed in the luminescence spectra indicate the changes of the local symmetry of $Eu^{3+}$ ions and the crystal field experienced by $Eu^{3+}$ ions in the host material with increasing pressure. Additionally, subtle changes observed in $Er^{3+}$ doped $GdVO_4$ at higher pressures [108] indicate the occurrence of the scheelite-fergusonite transition. The changes are less obvious than those changes induced in the zircon-scheelite transition because the scheelite-fergusonite transition involves minor structural changes. However, it has been demonstrated that in the fergusonite phase, pressure helps to enhance simultaneously the red and green light emission in $Er^{3+}$ doped $GdVO_4$ [108].

Resistivity measurements have been used in HP research since the foundational period and have been applied to semiconductor materials to characterize the effect of pressure in their electronic properties [121]. HP resistivity measurements have been reported on $CeVO_4$ and $LuVO_4$ up to 30 and 10 GPa, respectively [32, 122]. Resistance measurements made for the first compound up to 33 GPa are shown in Fig. 16. In the figure, it can be seen that there are two abrupt changes in the behaviour of the resistance, which can be correlated with changes of the volume associated to phase transitions.

Resistance measurements confirm the occurrence of phase transitions around 5 and 12 GPa. Near ambient pressure, in the zircon phase, $CeVO_4$ behaves as an insulator with a resistance in the range of GΩ. As pressure increase the resistance decrease reaching values in the range of MΩ before the phase transition occurs. Such decrease of several orders of magnitudes in the resistance cannot be expected in samples exhibiting an intrinsic behaviour because it will involve an anomalous decrease of 1 eV in $E_{gap}$ for an applied pressure of 5 GPa. The observed decrease of the resistivity is more likely due to the presence of deep donor levels, which under compression transform into shallow donor levels. At the phase transition, an abrupt decrease is found in the resistance in $CeVO_4$. The same occurs for $LuVO_4$ [122]. Such decrease of the resistance can be related with the band gap collapse observed in optical experiments [34]. The pressure influence in the resistivity of scheelite-type $LuVO_4$ is minor. However, in $CeVO_4$, the resistance decreases with pressure in the HP monazite phase, as observed in other monazite-structured oxides [123]. In addition, an order of magnitude fall is detected in the resistance of $CeVO_4$ around 12 GPa, indicating the substantial narrowing down of the band gap. This is the same pressure region where the second-pressure induced phase transformation is observed. The drop of the resistivity can be originated by the creation of donor levels proposed to be associated to the formation of oxygen vacancies. The values observed for the resistance at the highest pressure are compatible with a low-resistance semiconductor but not with a metal.

### 5.4 Brillouin spectroscopy and ultrasonic measurements

The elastic constants of zircon-type vanadates have not been systematically investigated yet. In particular, Brillouin-scattering measurements have been carried out only on a few compounds at ambient pressure [124 - 128]. Such studies provide quite valuable information to constrain the independent six elastic constants ($C_{11}$, $C_{33}$, $C_{44}$,

$C_{66}$, $C_{12}$, and $C_{13}$) of a tetragonal crystal like zircon [129], which can be used to determine its bulk and shear modulus [130]. In addition, in compounds like $CeVO_4$ and $SmVO_4$ the elastic constants have been calculated via *ab-initio* calculations [58, 131]. The reported results for the elastic constants are summarized in Table XII. Note that similar values for the elastic constants are obtained from experiments and calculations, giving computing simulations slightly smaller values than experiments. However, for $CeVO_4$, calculations give an unusually large value of $C_{66}$ [131]. On the other hand, the elastic constants reported for the vanadates resemble very much those reported for other zircon-type oxides [35, 132 - 134]. However, it is to be noticed that these constants decrease in going from silicates to phosphates and to vanadates [124]. This fact is consistent with the decrease of the phonon frequency distribution in the same order in these materials [124].

Among the six elastic constant of zircon, $C_{11}$ and $C_{33}$ are related with the longitudinal compression along the *a*-axis and *c*-axis, respectively. $C_{12}$ and $C_{13}$ are related to transverse compression, and $C_{44}$ and $C_{66}$ to shear. For zircon-type materials, the elastic constants should satisfy the Born criteria of mechanical stability for tetragonal crystals [135]: $C_{11} > 0$, $C_{11} - C_{12} > 0$, $C_{11} (C_{12} + C_{33}) - 2 C_{13}^2 > 0$; $C_{44} > 0$, $C_{66} > 0$. From Table XII it can be easily verified that the elastic constants of zircon-type vanadates fulfil the Born criteria. The value of $C_{33} > C_{11}$ indicates that the bonding strength along [100] and [010] direction is stronger than the bonding along [001] and $C_{44} > C_{66}$ suggesting that [100] (001) shear is harder than [100] (010) shear.

The elastic constants summarized in Table XII also allow us to obtain macroscopic information for the different vanadates listed in the table. In particular, the $C_{33}/C_{11}$ ratio describes the longitudinal elastic anisotropy. The results reported for $HoVO_4$, $ErVO_4$, $YVO_4$, and $SmVO_4$ indicate that the stiffness of these four compounds

along the *c*-axis is approximately 30 % greater than perpendicular to it. This result is in good agreement with the non-isotropic compression of zircon described in section 5.1. In fact, $C_{33}/C_{11}$ is similar to the ratio of the axial compressibilities determined from XRD for the *c*-axis and *a*-axis in different zircon-type vanadates [24, 25, 28, 29, 33, 58, 59, 97]. In case of $CeVO_4$, calculations give $C_{33}/C_{11} = 1$ [131], which is in contradiction with the known axial compressibilities of $CeVO_4$, which in highly non isotropic [32, 33]. This and the unusual large $C_{66}$ obtained from the same calculations suggests that Brillouin experiments and new elastic constant calculations are needed for $CeVO_4$.

Macroscopic elastic constants can also be determined from the elastic constants. In particular, within the Voigt approximation [136], the bulk modulus can be calculated as follows: $B = ( 2 C_{11} + 2 C_{12} + C_{33} + 4 C_{13}) / 9$, and the shear modulus is given by the equation $G = (2 C_{11} + C_{33} - C_{12} – 2 C_{13} + 6 C_{44} + 3 C_{66}) / 15$. Using these equations, the calculated values of the bulk (shear) moduli for $ErVO_4$, $YVO_4$, $CeVO_4$, and $SmVO_4$ are 138.7 GPa (64.6 GPa), 136.1 GPa (62.7 GPa), 128.5 GPa (55.9 GPa), and 122.6 GPa (52.3 GPa), respectively. The fact that $B > G$ indicates that the parameter limiting the mechanical stability of zircon-type vanadates is the shear modulus G. Besides, since G represents the resistance to fracture, the Pugh's criterion can be used to determine if zircon-type vanadates are ductile or brittle [137]. According to this criterion, if $B/G > 1.75$ the material is ductile and if $B/G < 1.75$ the material is brittle. For $ErVO_4$, $YVO_4$, $CeVO_4$, and $SmVO_4$ $B/G > 2.14$, therefore they are ductile.

### 5.5 Deformation studies

Radial x-ray diffraction in a DAC is a technique that has been recently employed to study under pressure the influence of induced dislocations on the crystal structure of $GdVO_4$ across its phase transitions to the HP polymorphs [38]. These studies not only provide precise information about transition mechanisms but also on the pressure

effects on plastic deformation of materials, which is of considerable importance to both Earth science and technological applications. The experiments carried out for $GdVO_4$ allowed to obtain, as a function of pressure, the unit-cell parameters, lattice strain, and preferred orientations of both zircon and scheelite polymorphs. An interesting fact is the developing of a (001) compression texture associated with a dominant slip plane along <100>{001} in the zircon-type structure starting from 5 GPa. After the phase transition, the (001) texture transforms into a (110) texture in the HP scheelite polymorph. This information supports the conclusions extracted from a transmission-electron microscopy study of shock-deformed zircon ($ZrSiO_4$) [138]. It also confirms the first-order character previously associated to the zircon-scheelite transition [24, 25, 28, 29, 33, 58, 59, 97] and it is consistent with the non-reversibility of this transition. In addition, it is more consistent with the displacive mechanism proposed for the zircon-scheelite transformation by Kusaba *et al.* [139] than with the bond-switching reconstructive mechanism proposed for the transition by Smirnov *et al.* [140].

A qualitative characterization of the deformation of zircon-type $GdVO_4$ under compression has been obtained by analysing the pressure dependence of the ratio of differential stress $t$ and the shear modulus G [141]. The magnitude $t$ is defined as the difference between the greatest and the smallest compressive stresses. In the zircon phase of $GdVO_4$ this magnitude ($t$/G) is found to increase rapidly from 0 at ambient pressure to 0.0412 at 5 GPa. After reaching the onset of the transition, the value of $t$/G begins to decrease gradually [38]. This observation indicates that the lattice strains of the zircon structure first enhances under compression but decreases with the emergence of the scheelite structure. This can be a consequence of the fact that the grains with higher deviatoric strains transform to the scheelite phase at lower pressure. This conclusion is in agreement with observation made in non-hydrostatic experiments [28,

37]. In case of scheelite structure, *t*/G is found to increase from 0.036 at 5 GPa to 0.096 at 25.6 GPa. The larger values of *t*/G in the scheelite phase are a consequence of the fact that this phase inherits the lattice strain of the zircon phase during the phase transformation. Finally, the radial XRD measurements found evidence of the appearance of the monoclinic fergusonite structure (space group: I*2/a*) above 31.2 GPa [38]. However, a full conversion to this phase was not achieved and thus the deformation behaviour of the fergusonite phase has not yet been analysed in detail.

**5.6 Theoretical studies**

With the advancement in computational techniques, *ab-initio* calculations have become almost an indispensable tool for the investigation of materials under extreme conditions. They usually serve as a complimentary tool to the experimental observations. In addition, they can predict properties of materials in the pressure range not easily accessible for the experiments. A*b-initio* calculations have been reported on a number of $AVO_4$ compounds. In collaboration with experiments, these calculations have helped to identify HP structures in different compounds [26, 44, 81, 82]. The reliability of the theoretical studies is supported by the consistency with the experimental results [38]. Thus, *ab-initio* calculations can be considered as quite powerful instrument for making predictions to guide future experiments [142].

In total-energy calculations, which are usually carried out at 0 K (temperature effects can also be included in the calculations if required), the Gibbs free-energy or the enthalpy as a function of pressure are calculated and the thermodynamically most stable phase is obtained by energy minimization [143]. A comprehensive description of the technical details of the methods of calculation is beyond the scope of this review since there are many review articles on HP *ab-initio* calculations in the literature. For

instance, readers can refer the paper published by Mujica *et al.* [144]. Most of the progress on the theoretical study of $AVO_4$ compounds under compression has been done by A. Muñoz *et al.* [26, 30, 40, 44, 58, 77, 81, 82, 102]; however, the contributions from other researchers cannot be ignored [43, 131, 145 – 147].

In order to illustrate the outcome of total-energy calculations we include in Fig. 17 the results of calculations carried out for $TbVO_4$ [81]. Both the total energy and the Gibbs free energy (the figure shows the energy difference between phases) indicate a zircon-scheelite-fergusonite structural sequence in agreement with the experiments. In addition, calculations predict the existence of a post-fergusonite phase not yet discovered experimentally [81]. The structural information of the post-fergusonite phase is included in Table XIII, which is an orthorhombic structure (space group *Cmca*) with an increased coordination number for both cations. In the HP orthorhombic polymorph, Tb and V atoms are coordinated to eleven and seven oxygen atoms, respectively. The structure is shown in Fig. 6c. It is a potential HP structure not only for $TbVO_4$, but also for the members of the zircon-type $AVO_4$ in which the trivalent cation has a smaller ionic radius as Nd; e.g. $HoVO_4$ [28]. The transition from fergusonite to this structure implies a large volume collapse.

For the compounds from $NdVO_4$ to $LaVO_4$, calculations have not only described the experimental observation properly but also predicted the existence of a post-monazite structure. Results of enthalpy calculations are shown in Fig. 18. They support the existence of zircon – monazite – $BaWO_4$-II transitions [81]. In the case of $LaVO_4$ this structure ($BaWO_4$-II-type) has been confirmed by powder XRD and Raman experiments [40]. The same structure has been theoretically predicted for other compounds. The structural information for the post-monazite structure of $NdVO_4$ is given in Table XIV. This structure has been described in the previous section (section

5.1) while discussing the phase transitions in $LaVO_4$. The crystal structure is shown in Fig. 6b. The transition from monazite to the post-monazite phase implies a collapse of the volume and an increase of the coordination number for both V and Nd atoms.

Calculations have not only helped in identifying the crystal structure of HP phases but also in the assignation of Raman-modes. In the case of IR modes, which have not yet been measured under compression, *ab-initio* calculations have already predicted the pressure behaviour of these modes. Experimental studies should be carried out to confirm these theoretical findings. A notable result determined from calculations is the behaviour of the $B_{1u}$ silent mode of the zircon phase. This mode has been found to have a strong non-linear pressure dependence. The behaviour of this mode in $TbVO_4$ is shown in Fig. 19. This mode is associated to rotations of the rigid vanadate ion. At 7.2 GPa, its frequency becomes imaginary indicating that the zircon structure becomes dynamically unstable. The same behaviour has been observed in other zircon-type vanadates [58]. The behaviour of $B_{1u}$ silent mode and the presence of two Raman-active soft modes in the zircon structure (see section 5.2) are related to mechanical instabilities of the zircon structure and could be the trigger to the observed phase transition. The pressure at which the $B_{1u}$ mode becomes imaginary agrees with the pressure at which the Born stability criteria is violated in the zircon structure.

Density-functional theory (DFT) based calculations not only help to ascertain the crystal structure of HP phases, but also to understand the influence of pressure on phonons and many other physical properties. As can be seen in Tables IV to VI, *ab-initio* calculations give an accurate description of Raman and IR modes of different vanadates, contributing to the mode assignment. This is particularly useful for the identification of the Raman modes of the HP phases; especially in those cases where the unpolarizing effect of the pressurized diamonds of the DAC make the mode

assignment difficult. Specially, computing simulations have been quite useful for assigning the modes of scheelite-type, monazite-type, and other HP phases of different orthovanadates [26, 29, 40, 44, 81, 82]. The results provided by calculations not only describe the frequency of the phonons properly but also their pressure coefficients. Additionally, calculations have also predicted the Raman modes of post-fergusonite and post-monazite phases. In case of $TbVO_4$, the post-fergusonite orthorhombic structure [81] (space group *Cmca* and point group symmetry $D_{2h}$) has thirty-six Raman modes: $\Gamma = 9\ B_{1g} + 7\ B_{2g} + 11\ B_{3g} + 9\ A_g$. The frequency and pressure coefficient for all these modes have been already calculated and are waiting for the experimental confirmation. Clearly, the duplication of the number of Raman modes (since fergusonite-type $TbVO_4$ has eighteen Raman modes) can be the cause of the appearance of broad bands in the Raman spectra measured for the post-fergusonite phase of $TbVO_4$. More details on these theoretical results can be found in Ref. 81. A theoretical assignation and study of HP behaviour of the Raman modes of the post-monazite phase of $LaVO_4$ has also been performed [40]. For this compound, the computer simulations also anticipate that the HP post-monazite phase has considerably more phonons than the low-pressure monazite phase. This is a consequence of the doubling of the unit cell. In particular, the monoclinic HP $BaWO_4$–II-type phase of $LaVO_4$ (space group $P2_1/n$ and point group symmetry $C_{2h}$) has seventy-two Raman-active modes ($\Gamma = 36\ A_g + 36\ B_g$). The excellent agreement found between theory and experiments for the behaviour of the Raman modes of the low-pressure phase [40] suggests that calculations can be an excellent guide for future experiments. Details on the frequency and pressure coefficient of the Raman modes of $BaWO_4$–II-type $LaVO_4$ can be found in the literature [40]. Predictions made by the DFT based calculations for the unknown HP phases of vanadates have established a rational basis for the future systematic exploration of high-

pressure behaviour of these compounds by means of Raman experiments. Calculations have also provided the information about the IR modes of different polymorphs of orthovanadates which have not yet been characterized under compression by experiments.

In this paragraph, we will describe the contribution made by *ab-initio* calculations to the understanding of HP behaviour of the band structure of $AVO_4$ vanadates. Calculations have not only helped to grasp the influence of pressure on the electronic band gap of different polymorphs, but have also contributed to reach a systematic understanding of $AVO_4$ vanadates [34, 37, 43, 44, 145, 147, 148, 149]. The most relevant conclusions obtained from calculations are that the electronic structure of orthovanadates near the Fermi level originates largely from the molecular orbitals of the vanadate ion. However, cation substitution influences these electronic states becoming quite important in compounds like $CeVO_4$. Compounds in which the band structure is mainly dominated by V 3d and O 2p states, the energy of the band gap is marginally affected by compression [34]. On the other hand, the atomic rearrangement induced by the phase transition in the crystal structure causes a collapse of the band gap, which in some compounds can be of the order of 1 eV [34]. Basically, the observed structural changes cause a reduction of the Coulomb attraction among the O 2p and V 3p states, causing a reduction of their splitting, which lead to a reduction of band-gap energy ($E_{gap}$). A charge transfer from O to V or from O to the trivalent metal might also contribute to the reduction of $E_{gap}$ [22]. However, the last phenomenon should lead to a pressure-induced metallization, which has not been observed yet in the pressure range reached in the experiments. Notice that ambient-pressure fergusonite-type $BiVO_4$ (a distorted version of scheelite-type) has a much smaller $E_{gap}$ than zircon-type $BiVO_4$ [150], which is consistent with the conclusion that the pressure-induced collapse of $E_{gap}$

is intrinsic to the structural changes triggered by the phase transition. The conclusions obtained from calculations on the influence of HP on the band structure of $AVO_4$ compounds have been shown to be useful for understanding the effect of chemical pressure (induced by cation substitution) in other vanadates; for instance, spinel-type vanadates [151].

To conclude this section, we would like to mention briefly two relevant facts. The first one is that the *ab-initio* calculations have contributed to understand how deviatoric stresses influence the HP behaviour of $AVO_4$ vanadates [28, 37]. In particular, in $HoVO_4$ calculations found that deviatoric stresses of 1.2 GPa will be enough to trigger the transition from scheelite to fergusonite at 20 GPa. However, under complete hydrostatic conditions the scheelite polymorphs is predicted to transform into a polymorph different that fergusonite, the orthorhombic HP polymorph (space group C*mca*), at a pressure larger than 30 GPa. On the other hand, in the case of $ScVO_4$, calculations have shown that a direct transformation from zircon to fergusonite can be induced under non-hydrostatic conditions. In particular, a uniaxial stress of 1.2 GPa is sufficient to trigger the transition from zircon to fergusonite at 2.5 GPa [28]. The second relevant fact is that up to now only calculations have provided information on the influence of pressure in the elastic constants [58]. Calculations of the elastic constant have given useful information on the mechanical stability of $AVO_4$ compounds under compression. They are also helpful for estimating macroscopic elastic moduli and other properties like hardness.

## 6. Pressure-induced phase transitions

### 6.1 Zircon-scheelite-fergusonite sequence

Zircon, the mineral $ZrSiO_4$, under compression has been shown to transform from its high-symmetry tetragonal structure ($I4_1/amd$) to a lower-symmetry scheelite-

type ($CaWO_4$) tetragonal structure ($I4_1/a$) [152]. Both structures have been observed in a number of $AXO_4$ compounds and are composed of $AO_8$ dodecahedra and $XO_4$ tetrahedra (See Figs. 1a and 6a). The HP zircon-scheelite transition has been observed in many different oxides, including silicates [152], vanadates [28, 58, 59], phosphates [153], chromates [154], and arsenates [155], among others. Between the rare-earth orthovanadates, almost all $RVO_4$ compounds have been investigated under HP. By the combination of various experimental techniques and theoretical methods it has been established beyond doubt, that the compounds with smaller ionic radii (Sm - Lu, including Y and Sc) follow the zircon to scheelite phase transition.

A remarkable fact is that the zircon and scheelite structures are related via a group-subgroup relationship ($I4_1/a \subset I4_1/amd$). In addition, both structures are related via symmetry operations. Indeed, twinning zircon subsequently on (200), (020), and (002) generates the scheelite structure [156]. Because of these symmetry relations, the $c/a$ ratio in zircon ($\approx 0.9$ as can be seen from Table I) is roughly equal to $2a/c$ in scheelite ($\approx 2.2$ as can be seen from Table VII). Apparently, the transformation of zircon into scheelite implies the slip of planes [38]. Consequently, the transition involves small displacements on the trivalent cations and the V atoms. On the other hand, the oxygen atoms arrange themselves to retain similar shapes for the $VO_4$ tetrahedron and $AO_8$ dodecahedron. The structural rearrangement caused by the phase transition includes rotations of the $VO_4$ tetrahedral units. This can be seen in Fig. 20 where the zircon and scheelite structures are compared. In spite of the structural relationship between zircon and scheelite, the transition mechanisms associated to the transition make it a first-order transition. Significant aspects of it are the large volume collapse involved in the transition and the non-reversibility of the transition, which favour the recovery of the scheelite-type phase as a metastable polymorph at ambient conditions.

The post-scheelite phase observed in many $AVO_4$ compounds beyond 20 GPa has a fergusonite-type structure. The zircon-scheelite-fergusonite sequence has been observed from $LuVO_4$ [95] to $EuVO_4$ [59]. The monoclinic fergusonite structure (space group $I2/a$) is obtained from a slight monoclinic distortion of the scheelite structure. This can be seen in Fig. 20 where both structures are compared. Fergusonite is the HP phase of many compounds isomorphic to scheelite. It was first found under compression in $CaWO_4$ [157], and after that, in $SrWO_4$ and other compounds [158 - 161]. Fundamentally, the transformation of scheelite into fergusonite consists of small translations of the different cations (e.g. Ca and W in $CaWO_4$ or Y and V in $YVO_4$) from their high-symmetry positions. These structural changes lead to a polyhedral distortion. Nevertheless, the building blocks of fergusonite in vanadates are also $VO_4$ tetrahedra and $AO_8$ dodecahedra as for zircon and scheelite.

After the phase transition, in fergusonite the $\beta$ angle becomes slightly different that 90º and the $c$-axis becomes different that the $a$-axis. However, in general, there are no discontinuities detected at the transition in the pressure dependence of the unit-cell parameters and volume. The structural changes associated to the phase transformation cause a reduction of the symmetry from tetragonal to monoclinic and a lowering of the point-group symmetry from 4/m (in scheelite) to 2/m (in fergusonite). A detailed discussion of the transition mechanism can be found in the literature [161]. Notice that the scheelite-fergusonite transition has also been observed at high temperature in compounds like $LaNbO_4$ [162], being characterized as a second-order ferroelastic transition. This characterization is consistent with the group-subgroup relationship between fergusonite and scheelite ($I2/a \subset I4_1/a$). However, it is still under debate whether the equivalent HP transition is first- or second-order [158].

After the transition, the increase of pressure gradually induces an enhancement of the distortion of the fergusonite structure, which produces an intensification of the polyhedral distortion. It has been argued that as a consequence of this fact, in $AXO_4$ fergusonites the coordination of the $X$ cations (e.g. V or W) gradually changes from 4 to 4 + 2 [163]. Therefore, the fergusonite structure operates as a bridge phase between the scheelite structure, containing $XO_4$ tetrahedra, and a HP structure containing $XO_6$ octahedra, like for instances the orthorhombic HP structure proposed for $TbVO_4$ [81]. To conclude this section, we would like to add that upon decompression the fergusonite phase back transforms to the scheelite phase with little or no hysteresis. This is a consequence of the displacive mechanism, characteristic of the scheelite-fergusonite transition. Since the zircon-scheelite transition is non-reversible the structural sequence observed in vanadates under compression is zircon-scheelite-fergusonite, but under decompression only the fergusonite-scheelite transition is observed, with scheelite remaining as a metastable phase at ambient conditions, as we described above. Post-fergusonites phases have not been recovered yet at ambient conditions and fergsuonite is only recovered in extreme non-hydrostatic conditions applied to the studied compound [33]. This quenching of the scheelite-type polymorph as a metastable phase is of technological interest given the reduction of the band gap in the scheelite polymorphs, which absorb visible light while the zircon polymorphs absorb only ultraviolet light [33].

**6.2 Zircon-monazite-post-monazite**

Zircon-type $AVO_4$ compounds with trivalent metals with large ionic radii (La-Nd) shows an irreversible zircon to monazite phase transition around 5 GPa with a large volume discontinuity. The transition has also been found in phosphates [153]. It involves a change in the coordination number around the trivalent cation from 8 to 9 as

can be seen by comparing Figs. 1a and 1f. In contrast, the coordination around V atom remains unchanged. The irreversible character of the transition, which involves a large volume discontinuity, along with the change in the coordination number, qualifies this transition to be first order in nature. As we discuss above for scheelite and zircon, there are also structural similarities between monazite and zircon. The first structure is similar to the second in many ways. In monazite the isolated $VO_4$ polyhedra share edges and corners with the $AO_9$ polyhedra, which bear a resemblance to $AO_8$ polyhedra in zircon. The chains of the trivalent metal that are the framework of the crystal structure are analogous in monazite and zircon; despite the fact that polyhedral chains in monazite are twisted compared to those in zircon. This twisting is what favours the connection of ninth oxygen to the trivalent metal. Consequently, the $VO_4$ tetrahedra, which are isolated in zircon, are no longer totally independent in monazite. Hence, the monazite structure has a greater structural connectivity than zircon. Monazite is more densely packed than zircon (as also scheelite is), lacking the interstitial voids and channels present in zircon. It is known than in different zircon-type ternary oxides the substitution of the *A* cation by a larger cation (which can be considered as a chemical pressure) causes the transformation of zircon into monazite. Therefore, the finding of the zircon-monazite transition under compression is expected from the crystal-chemistry point of view. In the case of vanadates, the transition mechanism involves a rotation of the $VO_4$ tetrahedra and a lateral shift of the (100) plane of zircon. This way the symmetry of the structure is reduced from $I4_1/amd$ (zircon) to $P2_1/n$ (monazite). In addition, the structural rearrangements produce the formation in monazite of an extra bond in the equatorial plane of the coordination polyhedron of *A*. This bond occupies the empty space among the $AO_8$ polyhedra of zircon. Therefore, the HP monazite phase is slightly less compressible than the low-pressure zircon phase.

On further compression, several post-monazite phases have been found in the orthovanadates [32, 33, 40]. The crystal structure of these phase depend on the amount of hydrostaticity present in the experiment. Notice than non-hydrostaticity can also induce a zircon-scheelite transition instead of the zircon-monazite transition as found in $CeVO_4$ [33]. However, the monazite-scheelite transition predicted by calculations in some vanadates [40] has never been found experimentally under hydrostatic conditions. The common feature of all proposed post-monazite structures for vanadates is that they look like a highly distorted monazite or a distorted barite-type structure. All of them have been described either with space group $P2_12_12_1$ or $P2_1/n$ (the space group of monazite) which are translationengleiche subgroups of the *Pnma* space group of barite, which suggest a possible zircon-monazite-post-monazite-barite sequence in vanadates with large trivalent cations. Additionally, the post-monazite structures share a tendency to increase the coordination of both cations. The post-monazite structures remain stable at least up to a pressure close to 50 GPa [33, 40].

### 6.3 Crystal chemistry of zircon-type vanadates

The zircon, scheelite, and fergusonite structures of $AVO_4$ compounds are formed by $VO_4$ tetrahedra and $AO_8$ eight-coordinated polyhedra, which can be described as two interpenetrating tetrahedral units. As described above, monazite shares many similarities with these structures, being the main difference the introduction of a ninth oxygen in the coordination sphere of the trivalent metal to form $AO_9$ polyhedra. We have described previously that depending on the cation *A*, zircon-type vanadates will transform either to scheelite (and next to fergusonite) or monazite. In those compounds where the trivalent cation has an ionic radius close to that determining the frontier between one structural transition or the other, either the zircon-scheelite or the zircon-monazite transition can occur depending on the hydrostaticity or non-

hydrostaticity of the experiments. For instance, in $CeVO_4$ the zircon-scheelite transition can be observed instead of the zircon-monazite transition if non-hydrostatic compression is applied [167, 168]. There are a few compounds where post-monazite or post-fergusonite phases have been found, but in all of them these new phases has six-coordinated V atoms and a trivalent metal with a coordination number larger than eight.

In Table XV, we summarize the different transition sequences and pressures reported for different $AVO_4$ compounds. We include only results obtained from experiments carried out starting from the ambient-pressure stable structure. Cases in which more than one transition pressure have been reported we list only the lowest and highest pressure reported in the literature [21– 38, 40, 58, 59, 62, 81, 95, 96, 97, 101, 163 –171]. These results are presented in Fig. 21 and for simplicity; we have only included lanthanide vanadates. The different compounds are plotted using the ionic radii at the horizontal axis. There are a few facts to be highlighted in the figure. For the compounds undergoing the zircon-monazite transition there is a clear decrease of the transition pressure from Nd to La, with $LaVO_4$ being stable in the monazite phase. A similar phenomenon has been observed for the transition to the post-monazite phase. Indeed, the existing results suggest that the post-monazite phase should be found in $PrVO_4$ at pressure slightly higher than 15 GPa. For the zircon-scheelite transition the influence of cation substitution in the transition pressure is not so evident, however, apparently it decreases as the ionic radii of the lanthanide increase from 8 GPa in $LuVO_4$ to 6.5 GPa in $SmVO_4$. For the scheelite-fergusonite transition pressure apparently there is a more important dependence on the ionic radii of the lanthanide, however, definite conclusion cannot be made yet because the published experimental data have been obtained using different pressure transmitting medium, generating

different quasi-hydrostatic or non-hydrostatic conditions, which could strongly affect the results [100].

Crystal-chemistry arguments are usually used for the systematic interpretation of the experimental evidence already accumulated on the HP behaviour of $AVO_4$ oxides. In particular, the rules proposed by Bastide [172] have been effectively applied to improve the understanding of pressure-induced structural transitions in $AVO_4$ compounds [5, 89]. According to Bastide's ideas, the thermodynamic pressure can be assimilated to the chemical pressure that is induced in a given compound when substituting a given atom by another with the same valence but a larger ionic radius. Thus, $AXO_4$ ternary oxides are organized in a diagram using the ionic radii of the two cations. High-pressure phase transitions are probable to take place from the structure of a particular oxide to that of a larger-cation-hosting material [5].

A clear example of the arguments presented above is the zircon-schelite transition. Scheelite is the crystal structure of $CaWO_4$. Ca has an ionic radius of 1.02 Å and W a radius of 0.42 Å. Thus, it is fully compatible with Bastide's arguments the transition from a zircon-type structure to scheelite oserved in experiments. This is because in zircon ($ZrSiO_4$), Zr has an ionic radius of 0.78 Å and Si a radius of 0.22 Å. Therefore, substituting Zr by Ca and Si by W will be equivalent to applying an external pressure. This is indeed what occurs in $ZrSiO_4$ which undergoes the zircon-scheelite transition around 23 GPa [152]. In the case of $AVO_4$ vanadates, the compounds with trivalent cations with radii equal or smaller than Sm are located in Bastide's diagram close to the zircon-scheelite boundary [5] and therefore this is the most reasonable transition from a crystal-chemistry point of view, being in fact the sequence found by the experiments. When the radius of the trivalent metal is larger than that of Sm, the $AVO_4$ compound will be located in Bastide's diagram closer to the zircon-monazite

boundary than to the zircon-monazite boundary (see Fig. 21). Therefore, the first transition will be more likely, as observed in the experiments. This situation occurs not only in vanadates but also in lanthanide phosphates [173], being in the phosphates Tb (and not Sm) the cation who determines the boundary between one transition sequence and the other. Notice the fact that $LaVO_4$, the vanadate hosting the largest lanthanide, crystallizes as monazite instead of zircon is fully consistent with these arguments.

Crystal-chemistry arguments also warrant the existence of the pressure-induced scheelite–fergusonite transition. Notice that fergusonite is the structure of $LaNbO_4$. La has an ionic radius of 1.16 Å (larger than that of Ca) and Nb a radius of 0.48 Å (larger than that of W). Therefore, fergusonite, which indeed is the HP phase of $CaWO_4$ [157], is a quite expected HP post-scheelite structure in the orthovanadates. The same reasons support the finding of post-fergusonite and post-monazite structures, as those observed in $TbVO_4$ [81] and $LaVO_4$ [40]. See that such HP structures are observed at ambient conditions in compounds with large cations, having a larger cation coordination number than zircon, scheelite, fergusonite, and monazite (e.g. $SrUO_4$ and $CsIO_4$). Therefore, it is not surprising that an increase in the V and rare-earth cation coordination is associated to the post-fergusonite and post-monazite transitions in $TbVO_4$ and $LaVO_4$, respectively. The facts described above show that crystal-chemistry arguments can be used to predict new HP phases, quite successfully as has been done in the past. The predictions made using these arguments indicate that there are still exciting roads to be surveyed for deep understanding of the behaviour of $AVO_4$ vanadates under compression.

## 7. High-pressure studies on other $AVO_4$ compounds

There are at least six compounds belonging to $AVO_4$ family other than rare-earth orthovanadates, $ScVO_4$, and $YVO_4$, whose synthesis and crystal structure has

been well documented. They are, $CrVO_4$, $InVO_4$, $TlVO_4$, $FeVO_4$, $AlVO_4$, and $BiVO_4$. Their crystal structure has been described in section 3. All of these compounds show excellent photocatalytic activities and most of the research carried out on these compounds is mainly focused on their photocatalytic behaviour, except for $BiVO_4$, which in addition to its photocatalytic behaviour also shows ferroelasticity [174 - 176]. Except $TlVO_4$ and $AlVO_4$, the other four compounds *i.e.* $CrVO_4$, $InVO_4$, $FeVO_4$ and $BiVO_4$ have been investigated under high pressure. The most recent works are reported on $InVO_4$ [41, 43, 44]. In this section, we will be discussing the high-pressure behaviour of these four compounds.

### 7.1 Different polymorphs of $BiVO_4$

Bismuth vanadate, $BiVO_4$, has received considerable attention since the discovery of a ferroelastic-paraelastic phase transition at 255 °C in this compound by Bierlein and Sleight in 1975 [177]. The low-temperature ferroelastic phase has the monoclinic fergusonite structure ($I2_1/a$) while the high temperature paraelastic phase adopts the tetragonal scheelite structure ($I4/a$). The transition is a second-order displacive transformation with complete reversibility. The monoclinic phase is a slight distortion of the scheelite phase (as we commented for other vanadates) where Bi and O atoms are shifted by approximately 0.1 Å from their ideal tetragonal positions. The monoclinic $\beta$ angle gradually approaches 90° as the temperature of the transition approaches. The vanadium tetrahedron remains rigid with no significant change in size. The same transition can be induced by pressure. There are two permissible orientations of the ferroelastic state, resulting in twinning in the low-symmetry form, making it ferroelastic.

Detailed high-pressure high-temperature crystallographic studies of the ferroelastic-paraelastic transition on single crystals by x-ray diffraction have been

reported [20, 178] and it has been found that in both phases the reduction of unit-cell volume is principally a result of Bi-O polyhedral compression. This behaviour is similar to that observed for other scheelite-type compounds, in which tetrahedral units are rigid and compression occurs in the eight-coordinated polyhedron. It has also been determined that the stress-strain hysteresis curves of $BiVO_4$ single crystals studied by thermal mechanical analysis shows a decrease in the ferroelastic to paraelastic phase transition temperature as the stress is increased [179]. High-pressure Raman measurements show the same transition at 1.5 GPa at ambient temperature [180]. In this study, the authors found a softening of a zone-center optical phonon with $B_g$ symmetry in the high-temperature phase and one with $A_g$ symmetry in the low-temperature phase. Most recently, the determination of the pressure-temperature phase diagram of $BiVO_4$ has been extended using this technique [181].

Raman measurements have also been carried in the zircon-type polymorph of $BiVO_4$, which behaves similarly as the rare-earth vanadates with a small lanthanide cation [181], but transforming directly from the zircon to the fergusonite polymorph. This is consistent with the finding of Bhattacharya *et al.* [182], who have reported that mechanical grinding of zircon-type $BiVO_4$ at room temperature also results in an irreversible transformation to the monoclinic fergusonite-type form. The amount transformed depends on the duration of grinding. The same transformation is also observed when zircon-type $BiVO_4$ is heated up to 350 – 400 °C.

The properties of $BiVO_4$ under high pressure have been explored using first-principles calculations [147]. These studies focused on the mechanical properties. According to calculations, the compound shows four polymorphs with ductility in all the phases. In spite of this progress, calculations on phase transitions and transformation mechanisms have not been performed yet.

### 7.2 Compounds adopting the $CrVO_4$ structure

$CrVO_4$, $InVO_4$, and $FeVO_4$-II adopt the same crystal structure, known as $CrVO_4$-type structure (C*mcm*) [68]. $CrVO_4$ is known to transform to a rutile-type structure at 750 °C and 6 GPa [184]. A similar polymorph and a wolframite-type polymorph have also been obtained under high-pressure and high-temperature for $FeVO_4$ [16]. However, none of these two compounds has been studied systematically under compression yet. Surprisingly, $InVO_4$ is the only compound adopting the $CrVO_4$–type structure (the $InVO_4$−III polymorph), which has been studied under high pressure by *in-situ* x-ray diffraction [41]. The compound has also been investigated by Raman spectroscopic measurements and first-principles calculations [41, 43, 44]. Though a few of the results of these studies have been mentioned in the previous sections of this article, here we will elaborate more on them. Different preparation and temperature conditions can lead to three different polymorphic forms of $InVO_4$ i.e. monoclinic $InVO_4$-I (S.G. *C*2/*m*), orthorhombic $InVO_4$−III (space group C*mcm*), and an undetermined structure InVO4−II. In spite of the ambiguities about the structure of InVO4−II, it is believed that the three known polymorphs of $InVO_4$ have $InO_6$ octahedral and $VO_4$ tetrahedral units as building blocks of the crystal structure (see Fig. 1b for $InVO_4$−III.) The orthorhombic form ($InVO_4$−III) has been reported to be the most stable phase in the $In_2O_3$−$V_2O_5$ system.

High-pressure powder XRD measurements carried out in $InVO_4$−III found a phase transition from this orthorhombic polymorph to monoclinic wolframite-type $InVO_4$-V (space group *P*2/*c*) near 7 GPa [41]. The structure of new HP polymorph is shown in Fig. 6d. In the small pressure region from 6.2 to 7.2 GPa, another phase whose structure could not be identified was also reported. $InVO_4$-V is stable until 23.9

GPa, the highest pressure reached in the measurements. On fast decompression, the XRD pattern at ambient pressure was found to correspond to the coexistence of both $InVO_4$-III and $InVO_4$-V. Similar results were reported from Raman investigations [41]. In Fig. 22 we show the pressure evolution of XRD data illustrating the structural transformation of $InVO_4$.

*Ab-initio* calculations confirmed the structural sequence determined from experiments [43] and predicted the physical properties of the HP phase, in particular the behaviour of the electronic band gap. More recent calculations [43] have predicted new polymorphs of $InVO_4$ with the transition sequence as $CrVO_4$-type (phase III) → wolframite (phase V at 4.4 GPa) → raspite (phase VI at 28.1 GPa) → $AgMnO_4$-type (phase VII at 44 GPa). The first transition is in agreement with the experimental findings. The subsequent transition pressures predicted from theory are higher than those previously covered by experiments. Therefore, this prediction is calling for new experiments. In Fig. 23, we show the pressure dependence of the unit-cell volume in the different polymorphs of $InVO_4$ [44]. Experiments and calculations agree quite well for the $CrVO_4$-type and wolframite-type polymorphs [41, 44]. The transition from $CrVO_4$-type to wolframite is associated with a large volume drop of approximately 17.5 %. This volume collapse is associated with a change in the coordination of the polyhedron around V, from $VO_4$ to $VO_6$. The subsequent transitions also imply collapses in the volume and an increase of the coordination number of In and V. In the $AgMnO_4$-type phase, In and V are eight coordinated by oxygen atoms. The lattice vibrations and electronic properties of the new polymorphs have been calculated [44]. However, their discussion is beyond the scope of this review. We refer to those interested in them to the recent work by Lopez-Moreno *et al.* [44]. We will only mention here the most remarkable effects are the drastic changes in the interatomic

bond distances and the increase in the coordination of In and V, proposed to occur at elevated pressures, in the phonons. Structural changes have an important effect not only in the Raman and infrared phonon frequencies at the Γ point, but also in the branches of the phonon dispersion in other points of the Brillouin zone [44].

## 8. Nanoparticles of *A*$VO_4$ compounds

It is a well-known fact that nanomaterials under high pressure do not necessarily mimic the high-pressure behaviour of their bulk counterpart [183 - 185]. Because of it, it is essential to carry out specific HP measurements on these samples. It particular, the compressibility and structural sequence can be modified in nanomaterials and in some cases; a pressure-induced amorphization can be induced. The behaviour of nanoparticles under HP depends on the particle size [186, 187] but also the surface state and defect density in the nanoparticles could strongly modify it [188].

There exists a variety of methods to grow lanthanide vanadate nanoparticles [189]. These techniques provide scientists the flexibility for the selective synthesis of nanocrystals of different polymorphs of the same compound [190] and doping of the nanocrystals with different dopants [191]. Therefore, the properties of $AVO_4$ nanocrystals can be tailored for making them optimum for different technological applications [192]. However, in spite of the relevance of HP studies for these applications, little efforts have been dedicated to the study on $AVO_4$ nanomaterials under compression. To the best of our knowledge there are only three articles published on the HP properties on $AVO_4$ compounds. Out of them, two are on $LaVO_4$ and one is on $YVO_4$ [50, 52, 98].

Zircon-type $LaVO_4$ nano-rods synthesized by a soft chemical rout [50] have been studied up to 17 GPa by means of XRD. The irreversible zircon to monazite phase transition has been detected at around 4 GPa with large pressure range of phase

coexistence. This result has been confirmed by Raman measurements [50]. There are no reports on HP investigations on bulk $LaVO_4$ with the zircon structure. Nevertheless, the observed HP structural sequence for the nano-rods of $LaVO_4$ is similar to the HP behaviour of bulk $CeVO_4$, $PrVO_4$, and $NdVO_4$. The only difference is that the phase transition observed in the nano-rods is more sluggish than in bulk materials. This fact might be associated with the specific morphology of $LaVO_4$ nano-rods. This morphology might modify the surface energy, which eventually influences the phase stability of the compound. Future studies are needed to explore this hypothesis. On further compression, a second phase transition was detected in the $LaVO_4$ nano-rods at around 12 GPa. The structure of the second HP phase, a post-monazite phase, has not been identified, but the phase transition coincides with that reported for the monazite-to-post-monazite transition in bulk $LaVO_4$ [40].

The second report on nanocrystals of $LaVO_4$ describes HP Raman measurements on both monazite- and zircon-type pure $LaVO_4$ and Eu-doped $LaVO_4$ [98]. The study on monazite-type $LaVO_4$ shows subtle changes at around 11.2 GPa, indicating an isostructural phase transition in the compound. These results are similar with those observed in bulk $LaVO_4$ under non-hydrostatic conditions [96]. Measurements on zircon-structured $LaVO_4$ show the expected phase transition to the monazite phase beyond 5.9 GPa. In addition, the pressure evolution of the luminescence spectra of $Eu^{3+}$ doped zircon-type $LaVO_4$ shows drastic changes across the zircon to monazite transition. These studies have established a correlation between the microstructure and luminescence behaviour on nanocrystals of $LaVO_4$.

The third HP study on nanocrystals is performed on $Eu^{3+}$-doped $YVO_4$ nanoboxes [52]. In this material a very different behaviour than in its bulk counterpart was found [52]. A combination of Raman, XRD, and luminescence measurements

clearly show an irreversible amorphization in the compound beyond 12.7 GPa in contrast to the zircon to scheelite phase transition observed in the bulk sample at around 7.5 GPa. However, small humps present in the XRD patterns indicate the presence of a short-range order in the amorphous phase. The reported results also support the occurrence of an inverse Hall-Petch effect [193] in nanocrystalline $YVO_4$, which is more compressible than bulk $YVO_4$. In particular, the bulk modulus is considerably reduced by 23 % in the nanocrystal [52]. Systematic studies in nanoparticles of different sizes are needed to confirm this behaviour.

Results of these three works described here, indicates that $AVO_4$ nanocrystals are waiting for future studies to explore systematically the effect of particle size and morphology in its HP behaviour. The importance of the interactions between the nanomaterial and the pressure-transmitting medium used for HP studies also deserves to be studied. A particular issue of interest is the reduction of the transition pressure and the possibility to create metastable polymorphs by the application of low-pressures [183]. Specially, reduction of zircon-scheelite transition pressure will facilitate the preparation of scheelite-type nanoparticles by pressure cycling for technological applications.

## 9. Systematic comparison with related oxides

### 9.1 Common trends of pressure-induced phase transitions

Many ternary oxides have crystal structures which are isomorphic to that of $AVO_4$ orthovanadates. For instance, there are phosphates, arsenates, selenates, and sulphates which crystallize in the $CrVO_4$-type structure [68]. Numerous compounds are isostructural with zircon [194]; including not only vanadates, but also silicates, phosphates, arsenates, and chromates among others; and a large number of $AXO_4$ oxides share the monazite structure with $LaVO_4$ [195]. Part of these ternary oxides has been

studied under compression and the crystal structure of HP phases has been determined. Analogies have been found in their HP behaviour. Three examples are: $ScVO_4$ and $ScPO_4$ which undergo a transformation from zircon to scheelite [25, 196]; $LaPO_4$ and $LaVO_4$ transform from the monazite structure to denser polymorphs with a larger cation coordination [40, 197]; and the fact that a wolframite-type structure is known to be the HP polymorph of $CrVO_4$ and $CrPO_4$ [15, 867].

In order to systematize the structural behaviour of $AXO_4$ compounds we will make use again of crystal-chemistry arguments. In Fig. 24 we have built a phase diagram including many different oxides with structures of interest for this review. Each of the symbols in the figure corresponds to a different compound. The compounds have been placed in the diagram using the Shannon radii [198] of each of its cationic elements. Horizontal lines have been drawn connecting compounds of a same family. The phase diagram has been built after an extensive data mining, however, there could be compounds missing on it. In the figure, it can be seen that compounds that share the same crystal structure are grouped in isolated regions of the phase diagram, which are divided by solid lines. The observed phase transitions in different oxides can be understood assuming that the crystal structure of the HP phase will be related to that of a compound with larger cations [5, 172]. In particular, phosphates [197 - 199] and vanadates [41], with structures related to the $CrVO_4$-type one transform to a structure related to wolframite. In the low-pressure phases of these $AXO_4$ compounds the $A$ cations have an octahedral coordination and the X cations are in a tetrahedral coordination. The same structural sequence is expected for arsenates having similar crystal structures. An archetypal case of this family is $InVO_4$ [41] which under compression takes the crystal structure of $InNbO_4$ [200] and $InTaO_4$ [201]. It is known that these two compounds also undergo phase transitions which increase the oxygen

anion coordination number around In and Nb (Ta) cations from six to eight. Such structure is likely to exist as post-wolframite structure in $CrVO_4$-type oxides; its search being one of the open tasks for the future. Among compounds structurally related to the $CrVO_4$-type structure, only the sulphates are those not transforming to a wolframite-type structure. Indeed, $MgSO_4$ [202] has been proposed to transform to a zircon-type structure in which Mg is eight coordinated by oxygen atoms and S is four coordinated by oxygen atoms. The distinctive behaviour of the sulphates is a consequence of the very small ionic radius of sulphur, which places sulphates at the very bottom of the phase diagram of Fig. 24, and very close to the boundary between $CrVO_4$-type and zircon. The small size of sulphur makes it very difficult for $A$SO$_4$ compounds to accommodate six oxygen atoms around S. Therefore, they prefer to increase the packing efficiency under compression by increasing the coordination of the *A* cation, thus taking the zircon structure near 17.5 GPa [202]. Interestingly, scheelite has been predicted to be a post-zircon polymorph of $MgSO_4$ [202], which is in total agreement with the structural sequence found in zircon-type oxides. The structural sequence found in $MgSO_4$ [202] has also been predicted for $CrVO_4$-type $TiSiO_4$ [79].

We will compare now the behaviour of different zircon-type oxides. As we described in section 6.3, for the vanadates with a small trivalent cation, it has been found that at high-pressure they transform to the scheelite structure. In contrast, when the trivalent cation is Nd or a cation with a large radius, the vanadate transforms from zircon to monazite. This is due to the fact, that the first group of compounds is in the upper region of zircon stability region, close to the zircon-scheelite boundary. Thus pressurizing the vanadates, which as a first approximation is equivalent to applying a chemical pressure by substituting the cations by larger atoms (i.e. moving diagonal towards the upper right corner of Fig. 24), will favour their phase transition from zircon

to scheelite. However, the second group of compounds, those with a large trivalent cation, are close to the zircon-monazite boundary, which will favour the transformation to monazite under compression, as experimentally observed.

Silicates [152, 203] and germanates [204, 205] crystallizing in the zircon structures at ambient conditions have also been found to transform to the scheelite structure under HP. An example of it is $ZrSiO_4$ (the mineral zircon) which transforms under pressure to a phase isomorphic with scheelite, named reidite [206]. The same behaviour has been observed in chromates [154, 155, 207] and arsenates [155, 208]. In the arsenates, the behaviour is completely analogous to that of vanadates, with the members of the family with large lanthanides transforming to monazite instead of transforming to scheelite. In some chromates, the phase transformation from zircon to scheelite proceeds via an intermediate monazite phase, i.e. zircon → monazite → scheelite [207]. The family of the phosphates shows a slightly different behaviour, which is also consistent with crystal chemistry arguments [153, 196, 209 – 212]. In particular, the transition to scheelite has been found in compounds like $YbPO_4$, $LuPO_4$, and $ScPO_4$. In $YPO_4$ the zircon → monazite → scheelite structural sequence has been found [196]. The same structural sequence has been determined for $TbPO_4$ [176], with calculations predicting a structure with a six-coordinated phosphor as a post-scheelite structure. The appearance of monazite as an intermediate phase between zircon and scheelite is consistent with the location of phosphates in the phase diagram of Fig. 24 and the fact that all $APO_4$ phosphates with a trivalent cation larger than Tb crystallize in the monazite structure at ambient conditions. Consequently, the radius limit determining a zircon-scheelite or zircon-monazite transition is shifted to a trivalent cation with a smaller radius in the phosphates. The same arguments can be used to

understand theoretical predictions on the phase transitions of $Y_{1-x}La_xPO_4$ [213]; with the transition pressure depending strongly on the lanthanum concentration.

A good example of the systematic discussed here are Y*X*$O_4$ oxides. Fig. 25 shows the structural sequence of zircon-structured $YAsO_4$, $YCrO_4$, $YPO_4$, and $YVO_4$ [155]. They have been ordered in the vertical axis from bottom to top by increasing the ionic radii of the *X* atom. Most of these oxides, excluding $YPO_4$, transform to scheelite under compression. $YPO_4$ transform to scheelite structure through monazite structure. The results can be rationalized as a function of the ratio of Y/*X* cation sizes [214]. $YPO_4$, has anY/P ratio (5.99) more similar to monazite $CePO_4$ (6.71) than to scheelite $CaWO_4$ (2.66), which explains the occurrence of the zircon-monazite transition. In the cases of $YAsO_4$, $YCrO_4$, and $YVO_4$, the Y/*X* ratios are 3.04, 2.95, and 2.87, respectively, justifying the zircon-scheelite transition. These results are also consistent with the fact that $YMoO_4$ (with Y/Mo=2.22) crystallizes at ambient conditions in the scheelite structure [215]. Notice that $YNbO_4$ and $YTaO_4$, with Nb and Ta having much larger radii than Mo, have a fergusonite structure [216, 217], which is the HP polymorph of scheelite [218]. This suggests that fergusonite can be the HP phase of $YMoO_4$. It also explains the zircon-scheelite-fergusonite that occurs in $YVO_4$ [26] and predicts the existence of structures with high cation-coordination numbers under extreme compression in all Y*X*$O_4$ oxides.

Several HP phase transitions have been reported in monazites [89]. In lanthanide phosphates, the monazite structure is stable up to 30 GPa [153, 197]. The HP phase is structurally related to barite [197] and has been observed as the post-monazite phase in $CeVO_4$ [33]. Since $LaPO_4$ is close to the monazite-barite boundary in Fig. 24 the reported transition is fully compatible with the crystal chemistry arguments sketch out in this section. Another example where a barite-related phase has

been found under HP is $CaSO_4$ [219]. This compound transforms under pressure from the orthorhombic anhydrite structure (*Cmcm*) to monazite at11.8 GPa. The post-monazite structure has the $AgMnO_4$-type structure (space group $P2_1/n$) and is structurally related to barite. The location of the $CaSO_4$-type sulphates in Fig. 24, makes quite reasonable their transformation into monazite and the subsequent transformation to the barite-related post-monazite structure.

Regarding the monazites located in the upper part of the stability region of monazite in Fig. 24, close to the monazite-scheelite boundary, they are expected to transform to scheelite and not the barite-related structures. This is the case of $CaSeO_4$ [220], which in fact is dimorphic, and can be prepared at ambient conditions either as scheelite or monazite. It is also the case of $SrCrO_4$ [221], which has been found to transform into the scheelite structure at 10 GPa and subsequently to the barite-related $AgMnO_4$-type structure at 13 GPa. In $LaVO_4$, the expected transition to scheelite has not been experimentally found [39]. In contrast, a transition was detected at 12 GPa to the post-scheelite phase $BaWO_4$-II-type [103]. One reason for the not observation of the scheelite structure is the presence of kinetical barriers, which in $BiPO_4$ are known to hinder the transition to scheelite, which is only obtained under the combined application of pressure and temperature [222]. To end the discussion on monazites, we will discuss the behaviour of $PbCrO_4$ [223]. In this compound, an isostructural phase transition occurs at 3 GPa. The nearly unnoticeable modification of the crystal structure produces important changes in the physical properties of $PbCrO_4$ [123]. At higher pressures, $PbCrO_4$ behaves like $SrCrO_4$, transforming subsequently to a scheelite-type structure at 11 GPa and to a post-scheelite phase at 21 GPa.

In summary, crystal-chemistry arguments allow a systematic understanding of all the results available in the literature on compounds related to $AVO_4$ vanadates. The

examples of this does not limit to the compounds discussed here but also include other oxides like $AgClO_4$ [224] and $BaSO_4$ [225]. On top of that, crystal-chemistry hypotheses have been used to explain the fact that the crystal structure of $BiSbO_4$ remains stable up to at least 70 GPa [67]. Consequently, the arguments presented in this section can be used as tool to make back-of-the-envelope predictions of HP phases in unstudied compounds. For instance, $CePO_4$ should follow the same structural sequence as $LaPO_4$. $PbSO_4$ can be anticipated to transform to a barite-related structure. High-coordination structures are expected to occur as post-monazite or post-scheelite structure in $A$$VO_4$ vanadates. Finally, $LaCrO_4$ and $LaAsO_4$ should transform into scheelite-related structures. In fact, recently monazite-to-scheelite transition in lanthanide oxo-arsenates at 11 GPa and 1373–1573 K has been reported [226].

### 9.2 Pressure-induced coordination changes

Pressure-induced coordination changes have been reported for transition-metal ions in silicate melts [227]. Such changes usually strongly affect the mechanical, vibrational, and electrical properties of a material. For the materials of interest in this review, the first compounds in which coordination changes have been reported are $A$$PO_4$ phosphates [199]. In particular, a phase with six-fold-coordinated phosphorus by oxygen was found in $AlPO_4$. However, pressures close to megabar are needed to get crystal structures based on $PO_6$ octahedral units, which have been proposed to be analogous to six-fold-coordinated silicates. As we have discussed in previous sections, a given phase transition (e.g. zircon-scheelite) in vanadates takes place at a much lower pressure than in phosphates. Therefore, $A$$VO_4$ vanadates are good candidates for studying pressure-driven coordination changes at low pressure. Thus, they can be used as models for the HP behaviour of other compounds like phosphates and silicates under extreme compression. Indeed, coordination changes have been reported to occur in orthovanadates like $MgV_2O_6$ at only 4 GPa [228]. In this compound, the HP structural change only produces a coordination change of vanadium atoms from 5 + 1 to 6. However, this change is enough to cause important changes in the electrical transport behaviour of $MgV_2O_6$.

Larger coordination changes have been found to occur in $InVO_4$ [41]. In this compound the coordination increase of vanadium is from four to six. It occurs due to a structural phase transition that takes place at 7.2 GPa. The observed new HP phase is the first one to be obtained under compression at room temperature with vanadium atoms in six coordination. A structure similar to that of the HP phase of $InVO_4$ also exist in $FeVO_4$ [229], but it requires the simultaneous application of high-pressure and high-temperature. The coordination changes induced by pressure in $InVO_4$ make the HP phase to be incompressible and to have a smaller electronic band gap than the other polytypic forms of this vanadate. This can make the new HP phase useful for photocatalytic hydrogen production [230]. Unfortunately, the HP phase cannot be recovered as a pure phase at ambient conditions. The new HP polymorph has been found to coexist with the ambient-pressure polymorph when $InVO_4$ is decompressed. The possibility of preparing $InVO_4$ nanoparticles at low calcination temperatures by using an amorphous precursor [231] can open the door to obtaining six-coordinated vanadium $InVO_4$ by pressure cycling the nanoparticles at relative low pressures [183]; which could lead to many interesting applications of $InVO_4$.

$InVO_4$ is not the only compound in which six-coordinated vanadium has been reported under compression. In $LaVO_4$ a HP phase with $VO_6$ octahedral units [39] has been found near 12 GPa. In the zircon-structured vanadates evidence of coordination changes are found only beyond 30 GPa [81], this is because of the zircon-scheelite-fergusonite structural sequence where the three structures are composed of $VO_4$ tetrahedral units. This indicates that $CrVO_4$-type vanadates are the best candidates for searching six-coordinated vanadium structures. The conclusions extracted from the study of these compounds can be of interest for the hunting of highly coordinated structures in arsenates, chromates, phosphates, selenates, and other related oxides. In this respect, the contribution of computation tools, in particular structural prediction methods [232], can be quite useful for guiding the experimental search of six-coordinated vanadium structures at HP. The performance of extended x-ray absorption fine structure (EXAFS) measurements under compression can also be an interesting tool to explore vanadium coordination changes in $AVO_4$ compounds. Such measurements can be performed at the K-edge of vanadium in a transmission mode using perforated diamond anvils [233].

**9.3 Equation of states, axial, bulk, and polyhedral compressibility**

The bulk modulus is a measure of how incompressible a substance is. It is a relevant physical parameter for evaluating the mechanical properties of materials. In particular, the bulk modulus is one of the magnitudes that can be used to estimate hardness in indirect way [234, 235]. Indeed, materials with a high bulk modulus are regularly considered as potential superhard materials [235]. As a result, significant efforts have been made to find low-compressibility materials with a bulk modulus comparable to that of diamond [236, 237]. The compressibility of $AVO_4$ orthovanadates has been extensively studied [5, 20, 24, 25, 27, 28, 29, 30, 32, 33, 36, 37, 38, 40, 41, 44, 50, 58, 59, 63, 81, 82, 95, 96, 97, 102, 123, 238, 239, 240]. In particular, room-temperature equations of state (EOS) have been determined for most of them. In Table XVI we summarized the EOS parameters corresponding to a Birch-Murnaghan EOS for different vanadates [241]. The unit-cell volume at ambient pressure ($V_0$), the bulk modulus ($B_0$), and its pressure derivate ($B_0'$) are given in the table. The references from where the values were obtained and indications on how the EOS was determined are given. Unless indicated the results were determined from XRD experiments (for which the pressure-transmitted medium used is also indicated). Among the zircon-type compounds, an increase in the bulk modulus is observed as the unit-cell volume decreases. This behaviour is analogous to that found in isomorphic phosphates [5, 153, 240]. It can also be seen that in a given compound, the HP scheelite phase has a bulk modulus, which is considerably larger than that of the zircon phase in the same compound. This is a consequence of the decrease of the volume (around 10 % [97]) and the increase of packing efficiency associated to the zircon-scheelite transition. In contrast, HP monazite phases are as compressible as the low-pressure zircon phases. At the zircon-monazite transition there is a decrease of the volume, which as a first approximation should lead to an increase of the bulk modulus. However, the flexibility of the nine-fold coordinated polyhedron of the trivalent metal (e.g. $CeO_9$ and $PrO_9$) favours the volume contraction under pressure counter balancing the effects of the volume reduction. In monazite $LaVO_4$, the phase transition to post-monazite causes a large increase of the bulk modulus. The same happen in the $CrVO_4$-type compounds, whose ambient-pressure polymorphs are the most compressible among all studied materials. However, the HP polymorphs [41, 44] have considerable larger bulk modulus (see Table XVI).

In order to understand the systematic observed, it is useful to correlate volume compression with changes induced by pressure in the compressibility of interatomic bonds. It has been reported that $VO_4$ tetrahedral units are more rigid and incompressible [27, 40, 41] than the $AO_6$, $AO_8$, or $AO_9$ polyhedra, which account for most of the volume reduction induced under compression. This behaviour is shown for the low- and high-pressure phases of $HoVO_4$ in Fig. 26. An equivalent conduct has been determined for related vanadates. For instance, in zircon-type $EuVO_4$, a polyhedral bulk modulus of 250(10) GPa has been reported for the $VO_4$ tetrahedron while a polyhedral bulk modulus of 140(10) GPa is obtained for the $EuO_8$ dodecahedron [59]. The same qualitative behaviour is found for different polymorphs of $A$VO$_4$ compounds. Such conduct has also been found in many other $AXO_4$ compounds [5, 35, 240 - 242]. Therefore, the bulk compressibility of most of these compounds, and particularly vanadates, can be described in term of the polyhedral compressibility [243]. This is because, as a first approximation, in oxides the bulk modulus is related to the stretching of its constituent chemical bonds. This fact was first used by Hazen *et al.* to describe the bulk modulus of silicates [244]. Other phenomenological approaches have been developed during the last few years [239 – 245], being an empirical equation that correlated the bulk modulus of $A$XO$_4$ compound with the $A$-O distance proposed by Errandonea and Manjon [5]. This equation properly describes the bulk modulus of $A$VO$_4$ vanadates and has been used for making correct predictions of unknown bulk moduli [5]. Thus, even within the crudeness of this phenomenological model developed in Ref. 5, it can be used to constrain the bulk modulus using only the interatomic distances and its stiffness at ambient pressure.

The differential compressibility of different polyhedral units described above has also relevant consequences in the axial compressibility, which is highly anisotropic in zircon [24, 28, 29, 58, 59, 97], monazite [40, 102], $CrVO_4$-type [41], scheelite [24, 28, 29, 58, 59, 97], and wolframite [41] vanadate polymorphs. This behaviour in the case of zircon- and scheelite-type $TbVO_4$ can be seen in Fig. 7. As a consequence of it, the axial ratio c/a is enlarged approximately by 2 % from ambient pressure to the transition pressure. Exactly the same behaviour has been found in all the studied zircon-type vanadates, phosphates, silicates, and germanates. In the scheelite-type polymorph, the opposite pressure evolution is observed. In this case, the most compressible axis is the *c*-axis. Because of it, the *c/a* axial ratio decreases under pressure nearly 2 % from the transition pressure to 20 GPa. The same anisotropic behaviour has been observed for the scheelite-type phases in related compounds [5]. The explanation to the anisotropic compression of scheelite and zircon is related to the way in which the different polyhedral are connected in each structure and with the circumstance that the $AO_8$ units are more compressible than the $VO_4$ tetrahedral units. In particular, because of the reduced compressibility of $VO_4$, those crystallographic axes in which $VO_4$ tetrahedral units are directly aligned are less compressible than those axes in which there is an *A* trivalent cation connecting two $VO_4$ tetrahedral units. Typical values of the linear compressibilities for different compounds are exemplified by $\kappa_a = 2.5 \times 10^{-3}$ $GPa^{-1}$ and $\kappa_c = 1.3 \times 10^{-3}$ $GPa^{-1}$ for zircon-type $EuVO_4$ [59] and $\kappa_a = 1.3 \times 10^{-3}$ $GPa^{-1}$ and $\kappa_c = 1.9 \times 10^{-3}$ $GPa^{-1}$ for scheelite-type $EuVO_4$ [59]. This anisotropic behaviour is enhanced in the post-scheelite fergusonite-type phase [37, 59].

In monazite-type and $CrVO_4$-type vanadates the same qualitative behaviour has been reported [39, 40, 102]. Particularly, in monazite-type $LaVO_4$, the following linear compressibilities can be obtained for the different axes: $\kappa_a = 3.2 \times 10^{-3}$ GPa, $\kappa_b = 2.5 \times 10^{-4}$ GPa, and $\kappa_c = 2.0 \times 10^{-3}$ GPa [102]. The difference between the compressibility of the most compressible and less compressible axes is approximately 50 %. In the case of $CrVO_4$-type $InVO_4$, the anisotropy is even larger. For this polymorph, the obtained linear compressibilities for describing the compression of each axis are: $\kappa_a = 7.8 \times 10^{-4}$ GPa, $\kappa_b = 8.5 \times 10^{-3}$ GPa, and $\kappa_c = 3.6 \times 10^{-4}$ GPa [102]. Notice that the *b*-axis is very soft and the *a*-axis is as uncompressible as diamond. A highly anisotropic behaviour has also been found in the HP phases of monazite- and $CrVO_4$-type vanadates [41, 44].

## 10. Technological and geophysical implications

$ZrSiO_4$ is a principal accessory mineral in terrestrial rocks [246]. The existence of zircon in rocks is useful for dating them [247] and the presence of zircon in the continental crust and oceanic ridges make it a very important material to test geological models [248]. In addition, the properties of this mineral are used to impose constraints on the origin and tectonic evolution of the metamorphic rocks [249] and to study the formation of granites by melting of the igneous crust [250]. For all these purposes, it is quite important to understand the stability of zircon at very high pressures. There are contradicting results on the behaviour of $ZrSiO_4$ under compression. Some authors report that scheelite-type $ZrSiO_4$ decomposed to $ZrO_2$ and $SiO_2$ at 22 GPa [251]. However, others [152] found that the HP phase remains stable beyond 30 GPa. On the other hand, a partial amorphization has been found in other zircon-type silicates, like coffinite at pressures close to 15 GPa [252]. The clarification of these apparent discrepancies is important to understand properly the behaviour of $ZrSiO_4$ in subducting plates; which has many geological implications [253]. One possible reason for them is the influence of non-hydrostatic conditions in the experiments, which can affect the transition pressure and trigger both chemical decomposition and amorphization [100].

Since phase transitions occur at lower pressures in vanadates than in silicates the influence of non-hydrostatic conditions in the structural stability of zircon-type vanadates is considerably reduced. Thus, vanadates can be used to model the HP behaviour of silicates and therefore the conclusions summarized in this work on the HP behaviour of vanadates are useful for geological studies. In particular, it is of particular relevance the fact that pressure-induced amorphization has not been found in any of the vanadates up to a pressure close to 50 GPa. The same can be stated for the fact that zircon-type vanadates do not decompose in the pressure range covered by the different studies and for the discovery of several new HP phases.

Monazites also exist in nature and information of their HP properties is extremely relevant for mineral physics, chemistry as well as for petrology [254]. One of the reasons for it is the fact that monazites record chemical evidence of their involvement in major geological events, for which information may be collected from the study of monazites [255]. In particular, monazite contains information on the pressure-temperature history of the minerals coexisting with it [254]. Obviously, information on pressure-induced phase transitions is relevant for these applications. The existence of phases transitions reported in $LaVO_4$ [39] and $LaPO_4$ [197] and the knowledge of transition pressure will impose a clear constrain to the pressure history of monazites. Additional information about the geophysical relevance of HP studies on monazites can be found in the review by Clavier *et al.* [195]. The information here reported on the HP behaviour of monazite $LaVO_4$ and zircon-type $AVO_4$ compounds is also relevant to understand the HP behaviour of cheralite and similar minerals containing tetravalent actinides [197] giving their structural relationship. The same can be stated for many trivalent oxides, like anhydrite, barite, and scheelite among other as become evident after the discussion is section 9.1.

Regarding technical applications, HP studies on vanadates give relevant information for the use of isomorphic phosphates as solid-state repository for radioactive waste [257]. The main reason for it is the chemical similarity between the lanthanide and actinide elements. Obviously, the knowledge of the structural stability and mechanical properties is fundamental for this use. This information is also valuable for the application of $AVO_4$ and $APO_4$ oxides as phosphors, lasers, and light emitters when doped [258]. This comes from the fact that light emission properties are directly linked to local environment of the doping element which change from one crystal structure to the other. The formation of new structures under HP, which can be recovered at ambient pressure as metastable phases might open therefore the door for developing specific materials with tailored properties for specific applications.

The understanding of the pressure stability limits of monazite and zircon structures is also for the use of vanadates and chromates as catalysts [259, 260] as well as of several oxides as ionic conductors [261] and oxidation-resistant ceramic toughening components [262]. A less obvious application on the knowledge gained from the HP study of monazites, in particular of the optical properties of chromates [67, 263], is the explanation of instabilities in paintings [264]. In particular, in those known as chrome yellow, a group of synthetic inorganic pigments that were broadly used as an artistic painting from the 19$^{th}$ century.

Additional applications derived from HP studies on vanadates comes from the possibility of creating phases, that can be metastable at ambient conditions, which have a reduced electronic band gap in comparison with the stable polymorph. The transitions from zircon-monazite [34], zircon-scheelite [34], monazite-post-monazite [40], and $CrVO_4$-type-wolframite [41] produce a collapse of the band gap, moving it from the near ultraviolet to the visible. This open the possibility for discovering and synthesizing novel metastable phases, which contribute to the developing of materials to cover a variable band-gap region, in particular in the 2 - 3 eV range, meeting different present needs in sunlight-driven photocatalysis [265]. The possibility of creating novel phases of orthovanadates by the application of moderate pressure with band gap close to 2.4 eV make them good candidates to be used for pollutant decomposition, thanks to their photocatalytic response to the visible-light region and relatively strong oxidation properties for water purification [266]. Another interesting application for them is photo-electrochemical (PEC) splitting, a green technology, for obtaining hydrogen fuel [178].

One interesting detail to consider is that the pressure-customized vanadates can be developed either in bulk form or as nanoparticles. Nanoparticles can be of particular interest because the catalysis efficiency is known to be higher for nanomaterials. Furthermore, if the new phases are prepared from doped nanoparticles of vanadates, their emissions can be tuned, which can be of special relevance for biomedical applications [267], in particular due to the chemical inertness and stability of vanadates [268]. An additional interesting fact is that the pressures required to obtain new phases in compounds like $InVO_4$ [41] are low enough to suggest that these phases could be obtained by ball milling methods [269], which will favour their production in industrial scales.

## 11. Future targets for the high-pressure studies of orthovanadates

The systematic understanding achieved on the HP behaviour of vanadates can be used to predict phase transitions in vanadates and related compounds. The more interesting aspect is towards the search of novel phases with the structures having high-coordinated vanadium (or phosphor, chromium, etc.) and/or with band gaps close to 2.4 eV. In the search of novel phases, *ab-initio* calculations and evolutionary algorithms are invoked to play an important role. Recently these methods have enabled efficient and reliable prediction of high-pressure structures [270]. In this respect, studies on $CrVO_4$-type compounds including $AlVO_4$, similar to those carried out in $InVO_4$ [41, 43, 44], are expected to produce the most interesting novel phases at relatively low-pressure (of special interests is the search of post-wolframite phases [271]). In fact, theoretical predictions already made for these compounds are waiting to be corroborated by experimental results. In particular, the existence of two post-wolframite phases at pressures between 30 and 45 GPa.

Another aspect that is expected to receive attention in future studies is pressure-induced amorphization [272], which has not yet been found in vanadates or related compounds. Such phenomenon has been observed at pressures as low as 15 GPa in double tungstates and molybdates. It will be interesting to explore if amorphization occurs in orthovanadates at higher pressures. The same can be said about the chemical decomposition at high pressure. The performance of these studies is a key to establish new concepts for understanding the physics behind such phenomena and their possible relation [273]. In the study of amorphization, EXAFS measurements can play an important role. In particular, they will be relevant for exploring the short-range order in the amorphous phases. Raman and XRD measurement will also help providing insight into the origin of lattice instabilities, which may lead to amorphization under high-pressure. Another important area to be explored is the influence of non-hydrostatic conditions in amorphization and chemical decomposition. For studying chemical decomposition, it will also be important to perform experiments under simultaneous high-pressure and high-temperature conditions. In particular, the decomposition of $ZrSiO_4$ has been observed near 90 GPa and 1200 ºC [274]. Such experimental conditions can be routinely achieved nowadays, using external heated diamond-anvil cells [275].

An interesting subject, which remains unexplored, is the behaviour under compression of solid solutions like those formed by $CrVO_4$ and $FeVO_4$ [276]. Literature survey shows that solid solutions of general formula $Fe_{1-x}Cr_xVO_4$ have different crystal structures depending upon the Cr content. In particular, for concentrations of approximately 0.85, a polymorph that differs essentially from the stable polymorphs of $FeVO_4$ and $CrVO_4$ is formed. This polymorph has the α-$MnMoO_4$ structure [277]. Other solid solutions of orthovanadates can be prepared by soft-chemical methods. They can also be prepared as nanoparticles [278]. In particular, the study of the high-pressure behaviour of solid-solution vanadates will be useful to test the general principles recently elucidated for end-member compositions. Of particular interest will be the studies of compounds like $Sm_{1-x}Nd_xVO_4$, since $SmVO_4$ [31] and $NdVO_4$ [96] undergo different phase transition under pressure; zircon-scheelite and zircon-monazite, respectively. Such studies could lead to unexpected results like recently found in $Tb_{0.5}Gd_{0.5}PO_4$ [279], which has a zircon-type structure but is at the border between this structure and monazite. This solid solution does not transform directly to monazite as $TbPO_4$ [173] but through an intermediate anhydrite-type structure. It would not be surprising that a similar phenomenon could occur in $Sm_{1-x}Nd_xVO_4$.

As we mentioned in Section 8, high-pressure structural studies on nanostructured vanadates have been performed only for $YVO_4$ and $LaVO_4$ [50, 51, 52, 99]. High-pressure studies on zircon-type $Eu^{3+}$ doped $GdVO_4$ nanostructures have been reported in which the influence of particle size in the properties of the nanoparticles was interpreted as a result of the interactions of surface defect dipoles which create an increased negative pressure with the crystallite size reduction [280]. The same phenomenon has been proposed to occur in related tungstates [281]. On the other hand, HP studies on zircon-structured chromate nanocrystals have shown that the structural phase transformation from zircon to scheelite phase observed in bulk materials [207] occurs via an intermediate monoclinic phase in nanocrystals. In addition, the transition pressure was found to be higher in the nanoparticles than in the bulk. Presently, clear picture does not exist which can explain all these phenomena in a consistent manner. Therefore, it is obvious that there is a lack of a systematic investigation of the response to high pressure of vanadate nanoparticles. Such studies should be the target on future studies. The main issue of these studies should be the effect of particle size and morphology on compressibility, transition pressure, and structural sequences, as well as the influence of the lack of hydrostaticity and the interaction between the pressure medium and sample on the high-pressure behaviour of nanoparticles. To clarify all these issues, systematic studies under identical experimental conditions should be performed on nanoparticles of a given compound covering a wide range of particle sizes.

Pressure-induced metallization has been found to occur in many oxides [282] including vanadium dioxide [283]. The insulator-metal transition observed in transition metal oxides has been associated to a breakdown of the electronic d−d correlation which leads to a Mott transition [284]. Significant attention has been given to this topic [285]. In the case of the orthovanadates, from optical-absorption measurements it was proposed that at 30 GPa, $DyVO_4$ and $TbVO_4$ were expected to become metallic by band overlap [22]. However, this metallization could not be confirmed for more recent studies in zircon-type vanadates [32, 34, 82] which were carried out up to pressures higher than 35 GPa. In addition, according to *ab-initio* calculations, the different polymorphs found in these compounds up to 50 GPa are not metallic. Therefore, pressure-induced metallization in $AVO_4$ orthovanadates, will have a lower bound of 50 GPa. Additional efforts are needed to study this challenging subject. Electrical resistivity and infrared spectroscopy measurements can significantly contribute towards this topic. The best candidates to undergo a pressure-driven metallization are $CeVO_4$, $BiVO_4$, $FeVO_4$, $CrVO_4$, $InVO_4$, and $TlVO_4$, compounds in which the effect of pressure on d electrons of the trivalent metal can favour metallization.

Zircon-type rare-earth vanadates are paramagnets [286]. Their ground state is split by the crystal field into three doublets and seven singlets. At low temperature (around 30 K), as a consequence of a cooperative Jahn-Teller interaction, a structural change is produced with extensive modifications in the electronic energy levels. The low-temperature phase is orthorhombic (space group *Fddd*) [71]. Neutron-diffraction measurements determined an antiferromagnetic state for low-temperatures. Magnetic properties and the Néel temperature ($T_N$) are strongly influenced by compression [287] and the increase of $T_N$ under compression can be possible [288]. On the other hand, neutron-diffraction measurements have shown an antiferromagnetic ordering in scheelite-type $TbCrO_4$ [154]. This suggests that a similar ordering can be obtained in scheelite-type vanadates at low temperature. Since the scheelite structure of different vanadates can be quenched as metastable phase, the study of their magnetic order can be performed by neutron diffraction experiments. Additionally, *in-situ* magnetic neutron diffraction under high pressure can be carried out [289], which opens the possibility to explore the potentially interesting magnetic behaviour of $AVO_4$ vanadates and related compounds. The magnetic order can also be indirectly studied under compression by measuring the temperature dependence of coherent phonons by ultrafast optical spectroscopy [290].

Finally, an important subject to be explored in the near future is the behaviour of elastic constants under pressure. As we mentioned in Section 5.4, there are already theoretical studies on the pressure dependence of elastic constants in zircon-type vanadates. Such studies can be compared with the characterization of the elastic properties of single-crystals under compression by ultrasonic measurements in a multi-anvil apparatus up to 10 GPa [291] and by combined Brillouin spectroscopy and x-ray diffraction [292] at least up to a pressure close to 50 GPa. Therefore, information on the elasticity of the different phases of $AVO_4$ compounds should be available soon. This information is fundamental for understanding transition mechanisms.

## 12. Summary

In this review, we have systematically reviewed the progress made during the last years on the study of high-pressure effects on the structural, vibrational, and electronic properties of $AVO_4$ orthovanadates. We have focused on the occurrence of pressure-induced phase transitions in zircon-, monazite-, and $CrVO_4$-type compounds, which have been discussed exhaustively. Results obtained from XRD, Raman spectroscopy, optical-absorption, resistivity measurements, and other experimental techniques are presented. We also describe recent results obtained from *ab-initio* calculations.

A systematics for the HP behaviour of $AVO_4$ vanadates, based upon crystal chemistry arguments, is proposed and discussed. In addition, we present a comparative analysis with the HP behaviour of related oxides. From the results available in the literature, it has been concluded that zircon-type orthovanadates follow two different structural sequences. 1) zircon→ scheelite → fergusonite → denser phases with six-coordinated V, for compounds with a rare-earth with an ionic radius equal or smaller than Sm, or 2) zircon → monazite → denser phases with six-coordinated V, for compounds with a rare-earth with an ionic radius equal or larger than Nd. Monazite-type $LaVO_4$ also transforms under compression to a denser phase with six-coordinated V. On the other hand, $CrVO_4$-type compounds also follow an "analogous" structural sequence, with the HP structure having a crystal structure related to wolframite in which V is also six-coordinated.

Possible phase transitions expected at pressures higher than those achieved in the experiments until today have been proposed based upon the discussion of the already existing results. The consequences of the observed structural changes in phonons and electronic properties have also been commented. The implications of the reviewed results for technological applications of vanadates and in geophysics have also been examined and possible directions for the future research on $AVO_4$ vanadates described. Two issues of particular interest for the future are the study of pressure effects on $AVO_4$ nanoparticles and the HP synthesis of metastable phases with tailor-made properties for specific applications.

**Acknowledgments**

This study was possible through the financial support from the Spanish Ministerio de Economía y Competitividad (MINECO), the Spanish Research Agency (AEI), and the European Fund for Regional Development (FEDER) under Grants No. MAT2016-75586-C4-1-P, MAT2013-46649-C4-1-P, and No. MAT2015-71070-REDC (MALTA Consolider). The authors are grateful to all the collaborators who participated in their research on the reviewed subject.

**Table I:** Unit-cell parameters and coordinates of the oxygen atoms in different vanadates crystallizing in the zircon structure [53 - 65]. They are listed according to decreasing ionic radii from top to bottom.

| Compound | *a* (Å) | *c* (Å) | *y* | *z* |
|---|---|---|---|---|
| $LaVO_4$ | 7.4569 | 6.5483 | 0.4269 | 0.2072 |
| $CeVO_4$ | 7.4004 | 6.4972 | 0.4279 | 0.2067 |
| $BiVO_4$ | 7.2999 | 6.4573 | 0.4280 | 0.2080 |
| $PrVO_4$ | 7.3631 | 6.4650 | 0.4288 | 0.2057 |
| $NdVO_4$ | 7.3308 | 6.4356 | 0.4295 | 0.2050 |
| $PmVO_4$ | 7.2963 | 6.4096 | 0.4303 | 0.2044 |
| $SmVO_4$ | 7.2618 | 6.3837 | 0.4350 | 0.207 |
| $EuVO_4$ | 7.2133 | 6.3559 | 0.4303 | 0.2095 |
| $GdVO_4$ | 7.1986 | 6.3353 | 0.4036 | 0.2271 |
| $TbVO_4$ | 7.1774 | 6.3264 | 0.4330 | 0.2021 |
| $DyVO_4$ | 7.1481 | 6.3082 | 0.4350 | 0.1990 |
| $HoVO_4$ | 7.1227 | 6.2891 | 0.4341 | 0.2010 |
| $YVO_4$ | 7.1183 | 6.2893 | 0.4342 | 0.2008 |
| $ErVO_4$ | 7.0957 | 6.2729 | 0.4349 | 0.2004 |
| $TmVO_4$ | 7.0682 | 6.2593 | 0.4355 | 0.2001 |
| $YbVO_4$ | 7.0427 | 6.2472 | 0.4362 | 0.2000 |
| $LuVO_4$ | 7.0254 | 6.2347 | 0.4364 | 0.1995 |
| $ScVO_4$ | 6.7804 | 6.1345 | 0.4422 | 0.1967 |

**Table II:** Structural information of the crystal structure of $CrVO_4$-type $InVO_4$. The parameters correspond to the crystal structure determined at 0.8 GPa [41].

| Space group *Cmcm*, Z = 4 | | | | |
|---|---|---|---|---|
| *a* = 5.738(5) Å, *b* = 8.492(8) Å, *c* = 6.582(6) Å, V = 320.72(2) Å$^3$ | | | | |
| Atom | Site | *x* | *Y* | *z* |
| In | 4a | 0 | 0 | 0 |
| V | 4c | 0 | 0.3617(5) | 0.25 |
| $O_1$ | 8g | 0.2568(4) | 0.4831(6) | 0.25 |
| $O_2$ | 8f | 0 | 0.7492(8) | 0.9573(9) |

**Table III:** Unit-cell parameters of $AVO_4$ compounds crystallizing in $CrVO_4$-type structure [41, 68, 69].

| Compound | *a* (Å) | *b* (Å) | *c* (Å) |
|---|---|---|---|
| $CrVO_4$ | 5.589 | 8.252 | 5.993 |
| $FeVO_4$-II | 5.628 | 8.272 | 6.112 |
| $InVO_4$ | 5.765 | 8.542 | 6.592 |
| $TlVO_4$ | 5.839 | 8.687 | 6.800 |

**Table IV:** Frequencies of the Raman modes (ω) observed in zircon-type $NdVO_4$ and $TbVO_4$ at ambient conditions [81, 82]. Theoretical (a) and experimental (b) frequencies are compared. The relative difference between measured ($\omega_E$) and calculated ($\omega_T$) frequencies is also given ($R_\omega = (\omega_T - \omega_E)/\omega_E$).

| Raman Mode Symmetry | $NdVO_4$ | | | $TbVO_4$ | | |
|---|---|---|---|---|---|---|
| | $\omega_E$ [$cm^{-1}$][b] | $\omega_T$ [$cm^{-1}$][a] | $R_\omega$ | $\omega_E$ [$cm^{-1}$][b] | $\omega_T$ [$cm^{-1}$][a] | $R_\omega$ |
| $T(E_g)$ | 113.0 | 112.1 | -0.0080 | ----- | 108.9 | ----- |
| $T(B_{1g})$ | 123.8 | 122.3 | -0.0121 | 120 | 117.7 | -0.0192 |
| $T(E_g)$ | 151.1 | 149.7 | -0.0093 | 154 | 151.5 | -0.0162 |
| $R(E_g)$ | ----- | 224.6 | ----- | 246 | 230.9 | -0.0614 |
| $T(B_{1g})$ | 243.1 | 233.9 | -0.0378 | ----- | 246.0 | ----- |
| $\nu_2(B_{2g})$ | 260.9 | 258.2 | -0.0103 | 260 | 255.9 | -0.0158 |
| $\nu_4(E_g)$ | 373.4 | 363.9 | -0.0254 | ----- | 371.7 | ----- |
| $\nu_2(A_{1g})$ | 381.9 | 365.5 | -0.0429 | 381 | 364.4 | -0.0436 |
| $\nu_4(B_{1g})$ | 472.2 | 453.0 | -0.0407 | 483 | 464.9 | -0.0375 |
| $\nu_3(B_{1g})$ | 794.6 | 797.0 | 0.0030 | 809 | 814.6 | 0.0069 |
| $\nu_3(E_g)$ | 808.1 | 801.9 | -0.0077 | 826 | 822.16 | -0.0046 |
| $\nu_1(A_{1g})$ | 871.1 | 859.0 | -0.0139 | 885 | 878.4 | -0.0075 |

**Table V:** Raman and IR modes of $CrVO_4$-type $InVO_4$ [41, 44]. For Raman, theoretical (a) and experimental (b) frequencies ($\omega_E$ and $\omega_T$) are compared and $R_\omega$ is given.

| Raman | | | | IR | |
|---|---|---|---|---|---|
| Mode assignment | $\omega_E$ [$cm^{-1}$][b] | $\omega_T$ [$cm^{-1}$][a] | $R_\omega$ | Mode assignment | $\omega_T$ [$cm^{-1}$][b] |
| $T(B_{3g})$ | 135 | 128.36 | -0.0492 | $T(B_{1u})$ | 102.64 |
| $T(B_{1g})$ | 191 | 153.78 | -0.1949 | $T(B_{3u})$ | 150.91 |
| $T(A_g)$ | 218 | 193.24 | -0.1136 | $T(B_{1u})$ | 152.31 |
| $R(B_{1g})$ | 252 | 208.32 | -0.1733 | $T(B_{2u})$ | 214.39 |
| $R(B_{2g})$ | 342 | 237.68 | -0.3050 | $R(B_{3u})$ | 253.72 |
| $\nu_2(A_g)$ | 348 | 334.48 | -0.0389 | $T(B_{2u})$ | 253.85 |
| $R(B_{3g})$ | 377 | 352.16 | -0.0659 | $\nu_4(B_{3u})$ | 328.31 |
| $\nu_4(B_{1g})$ | 389 | 361.53 | -0.0706 | $R(B_{1u})$ | 346.36 |
| $\nu_2(B_{2g})$ | 390 | 370.74 | -0.0494 | $\nu_2(B_{2u})$ | 358.37 |
| $\nu_4(A_g)$ | 456 | 380.68 | -0.1652 | $\nu_4(B_{2u})$ | 396.86 |
| $\nu_4(B_{3g})$ | 637 | 422.85 | -0.3362 | $\nu_4(B_{1u})$ | 423.21 |
| $\nu_3(B_{1g})$ | 755 | 657.99 | -0.1285 | $\nu_3(B_{3u})$ | 681.81 |
| $\nu_3(A_g)$ | 847 | 752.49 | -0.1116 | $\nu_1(B_{2u})$ | 758.20 |
| $\nu_1(A_g)$ | 914 | 920.05 | 0.0066 | $\nu_3(B_{1u})$ | 893.20 |
| $\nu_3(B_{3g})$ | 918 | 925.12 | 0.0078 | $\nu_3(B_{2u})$ | 936.59 |

**Table VI:** Raman and IR modes of monazite-type $LaVO_4$ [40]. For Raman, theoretical (a) and experimental (b) frequencies ($\omega_E$ and $\omega_T$) are compared and $R_\omega$ is given.

| Raman mode | $\omega_T$ ($cm^{-1}$)[a] | $\omega_E$ ($cm^{-1}$)[b] | $R_\omega$ | Infrared mode | $\omega_T$ ($cm^{-1}$)[a] |
|---|---|---|---|---|---|
| $B_g$ | 70 | 64 | 0.0938 | $A_u$ | 81 |
| $A_g$ | 72 | 61 | 0.1803 | $B_u$ | 90 |
| $A_g$ | 91 | 88 | 0.0341 | $A_u$ | 101 |
| $B_g$ | 92 | 102 | -0.0980 | $A_u$ | 118 |
| $A_g$ | 102 | 102 | 0 | $B_u$ | 128 |
| $B_g$ | 115 | 115 | 0 | $A_u$ | 150 |
| $B_g$ | 127 | 127 | 0 | $B_u$ | 154 |
| $A_g$ | 134 | 137 | -0.0219 | $B_u$ | 164 |
| $A_g$ | 143 | 146 | -0.0205 | $A_u$ | 165 |
| $A_g$ | 154 | 160 | -0.0375 | $B_u$ | 171 |
| $B_g$ | 158 | 158 | 0 | $A_u$ | 205 |
| $B_g$ | 183 | 189 | -0.0317 | $B_u$ | 221 |
| $A_g$ | 188 | 193 | -0.0259 | $A_u$ | 223 |
| $B_g$ | 204 | 209 | -0.0239 | $A_u$ | 246 |
| $B_g$ | 224 | 232 | -0.0345 | $B_u$ | 252 |
| $A_g$ | 230 | 235 | -0.0213 | $A_u$ | 279 |
| $B_g$ | 242 | 241 | 0.0041 | $B_u$ | 290 |
| $A_g$ | 252 | 252 | 0 | $A_u$ | 290 |
| $B_g$ | 297 | 309 | -0.0388 | $B_u$ | 342 |
| $B_g$ | 316 | 331 | -0.0453 | $A_u$ | 355 |
| $A_g$ | 317 | 326 | -0.0276 | $A_u$ | 372 |
| $A_g$ | 336 | 349 | -0.0372 | $B_u$ | 378 |
| $A_g$ | 355 | 373 | -0.0483 | $B_u$ | 416 |
| $A_g$ | 380 | 397 | -0.0428 | $B_u$ | 419 |
| $B_g$ | 389 | 400 | -0.0275 | $A_u$ | 466 |
| $B_g$ | 410 | 426 | -0.0376 | $A_u$ | 782 |
| $A_g$ | 423 | 439 | -0.0364 | $B_u$ | 790 |
| $B_g$ | 427 | 440 | -0.0295 | $B_u$ | 806 |
| $A_g$ | 784 | 768 | 0.0208 | $A_u$ | 821 |
| $B_g$ | 799 | 790 | 0.0114 | $B_u$ | 842 |
| $A_g$ | 806 | 794 | 0.0151 | $A_u$ | 843 |
| $A_g$ | 836 | 819 | 0.0208 | $B_u$ | 854 |
| $B_g$ | 850 | 843 | 0.0083 | $A_u$ | 878 |
| $B_g$ | 861 | 855 | 0.0070 | | |
| $A_g$ | 870 | 859 | 0.0128 | | |
| $B_g$ | 892 | 882 | 0.0113 | | |

**Table VII:** Unit-cell parameters and atomic coordinates of the oxygen atoms in different $RVO_4$ vanadates in the metastable scheelite structure (space group $I4_1/a$) at ambient conditions [24, 28, 31, 73, 96]. The *R* atom is at the 4b Wyckoff position (0, 1/4, 5/8) and the V atom at the 4a Wyckoff position (0, 1/4, 1/8). The oxygen positions (16f) have been transformed to an equivalent position in all the compounds to facilitate comparison.

| Compound | *a* (Å) | *c* (Å) | *x* | *y* | *z* |
|---|---|---|---|---|---|
| $YVO_4$ | 5.0320(3) | 11.233(1) | 0.2454(11) | 0.6090(11) | 0.5441(5) |
| $HoVO_4$ | 5.021(5) | 11.209(9) | 0.2362(5) | 0.5975(8) | 0.5860(8) |
| $SmVO_4$ | 5.1090(1) | 11.4113(3) | 0.2493(13) | 0.6101(13) | 0.5464(5) |
| $ErVO_4$ | 5.003(1) | 11.143(2) | 0.2460(6) | 0.6040(6) | 0.5446(3) |
| $DyVO_4$ | 5.0410(1) | 11.2689(3) | 0.2487(7) | 0.5951(8) | 0.0465(3) |

**Table VIII:** Fergusonite-type $EuVO_4$ (space group $I2_1/a$) at 25.6 GPa and room temperature. $a$ = 5.036 (4) Å, $b$ = 11.077(10) Å, $c$ = 4.675(6) Å, and $\beta$ = 92.79(6)° [59].

| Atom | site | *x* | *y* | *z* |
|---|---|---|---|---|
| Eu | 4e | 1/4 | 0.6327(10) | 0 |
| V | 4e | 1/4 | 0.1202(29) | 0 |
| $O_1$ | 8f | 0.871(9) | 0.906(5) | 0.254(15) |
| $O_2$ | 8f | 0.448(4) | 0.219(6) | 0.866(17) |

**Table IX:** Monazite-type $PrVO_4$ (space group: $P2_1/n$) at ambient conditions $a$ = 6.9550(1) Å, $b$ = 7.1652(1) Å, c = 6.6851(1) Å, $\beta$ =105.224(1)° [31].

| Atom | site | $x$ | $y$ | $z$ |
|---|---|---|---|---|
| Pr | 4e | 0.2833(2) | 0.1585(2) | 0.1017(2) |
| V | 4e | 0.3069(2) | 0.1634(2) | 0.6132(2) |
| $O_1$ | 4e | 0.2510(15) | −0.0010(5) | 0.4237(6) |
| $O_2$ | 4e | 0.3904(6) | 0.3449(4) | 0.4959(6) |
| $O_3$ | 4e | 0.4863(7) | 0.1122(9) | 0.8266(4) |
| $O_4$ | 4e | 0.1230(7) | 0.2212(16) | 0.7184(6) |

**Table X:** $BaWO_4$-II-type polymorph of $LaVO_4$ (space group: $P2_1/n$) at 16.4 GPa and room temperature. $a$ = 12.289(9) Å, $b$ = 6.492(5) Å, $c$ = 6.836(5) Å, and $\beta$ = 95.6(1) [40].

| Atom | site | $x$ | $y$ | $z$ |
|---|---|---|---|---|
| $La_1$ | 4e | 0.89138 | 0.34364 | 0.11856 |
| $La_2$ | 4e | 0.87725 | 0.05433 | 0.63987 |
| $V_1$ | 4e | 0.86896 | 0.83095 | 0.17888 |
| $V_2$ | 4e | 0.83191 | 0.55824 | 0.60671 |
| $O_1$ | 4e | 0.93820 | 0.04308 | 0.31838 |
| $O_2$ | 4e | 0.79212 | 0.35971 | 0.77788 |
| $O_3$ | 4e | 0.90611 | 0.40060 | 0.47845 |
| $O_4$ | 4e | 0.77509 | 0.63949 | 0.09695 |
| $O_5$ | 4e | 0.91628 | 0.69045 | 0.87024 |
| $O_6$ | 4e | 0.81457 | 0.77366 | 0.42413 |
| $O_7$ | 4e | 0.99659 | 0.70000 | 0.15006 |
| $O_8$ | 4e | 0.82884 | 0.01541 | 0.98855 |

**Table XI:** Experimental Raman frequencies (ω) and pressure coefficients (dω/dP) in zircon (top), scheelite (center), and fergusonite (bottom) phases of $YVO_4$ at ambient pressure (at 25 GPa for fergusonite). Calculated values using LDA (a) and GGA (b) are also given [26].

| Mode | ω ($cm^{-1}$) | dω/dP ($cm^{-1}$/GPa) | $ω^a$ ($cm^{-1}$) | $dω/dP^a$ ($cm^{-1}$/GPa) | $ω^b$ ($cm^{-1}$) | $dω/dP^b$ ($cm^{-1}$/GPa) |
|---|---|---|---|---|---|---|
| $E_g$ | | | 134 | −0.30 | 131 | 0.38 |
| $B_g$ | 156.8 | 1.33 | 163 | 0.57 | 150 | 1.50 |
| $E_g$ | 163.2 | 0.05 | 168 | 0.31 | 157 | −0.81 |
| $B_{2g}$ | 259.6 | −1.30 | 256 | -1.53 | 255 | −1.54 |
| $B_g$ | | | 274 | 3.25 | 252 | 3.11 |
| $E_g$ | 260.5 | 5.62 | 276 | 5.01 | 238 | 6.62 |
| $E_g$ | | | 376 | 0.82 | 371 | 0.68 |
| $A_g$ | 378.4 | 2.34 | 380 | 3.33 | 360 | 1.94 |
| $B_g$ | 489.3 | 2.77 | 480 | 2.90 | 461 | 2.79 |
| $B_g$ | 816.0 | 5.93 | 862 | 6.03 | 811 | 6.11 |
| $E_g$ | 838.8 | 5.46 | 874 | 4.91 | 825 | 5.68 |
| $A_g$ | 891.1 | 5.99 | 924 | 3.45 | 875 | 6.31 |
| $E_g$ | 148.9 | 1.00 | 152 | 0.27 | 143 | 1.37 |
| $B_g$ | 166.2 | −0.16 | 157 | −1.51 | 162 | −0.33 |
| $B_g$ | 194.9 | 3.90 | 220 | 3.50 | 189 | 4.26 |
| $E_g$ | 202.2 | 3.49 | 209 | 2.94 | 190 | 3.67 |
| $A_g$ | 243.9 | 1.19 | 244 | −1.62 | 232 | 1.22 |
| $E_g$ | 320.2 | 2.62 | 324 | 2.86 | 300 | 3.84 |
| $A_g$ | 349.1 | 2.53 | 348 | 3.24 | 327 | 3.07 |
| $B_g$ | 371.8 | 3.20 | 374 | 3.05 | 349 | 3.59 |
| $B_g$ | 414.3 | 3.02 | 411 | 1.94 | 397 | 3.24 |
| $E_g$ | 437.3 | 3.67 | 435 | 3.25 | 413 | 3.01 |
| $E_g$ | 747.8 | 4.94 | 784 | 4.34 | 741 | 4.75 |
| $B_g$ | 812.7 | 3.49 | 833 | 4.06 | 794 | 6.48 |
| $A_g$ | 829.4 | 3.94 | 845 | 7.60 | 814 | 3.11 |
| $B_g$ | 154 | 0.38 | 162 | 1.19 | 155 | −0.22 |
| $A_g$ | 169 | 0.81 | 170 | 3.98 | 151 | 0.31 |
| $B_g$ | 180 | 2.07 | 175 | 2.75 | 157 | 0.44 |
| $A_g$ | 241 | 1.44 | 258 | 1.73 | 255 | 0.76 |
| $B_g$ | 251 | 1.65 | 267 | 2.70 | 252 | 1.87 |
| $A_g$ | | | 277 | −0.03 | 260 | 1.80 |
| $B_g$ | 278 | 1.53 | 296 | 2.73 | 254 | 2.30 |
| $B_g$ | 374 | 1.12 | 374 | 0.78 | 360 | 1.74 |
| $A_g$ | 411 | 1.07 | 412 | 1.13 | 390 | 2.13 |
| $A_g$ | 427 | 1.92 | 429 | 1.87 | 409 | 1.87 |
| $B_g$ | | | 432 | 3.48 | 363 | 2.09 |
| $B_g$ | 489 | 0.27 | 509 | 1.05 | 491 | 3.19 |
| $B_g$ | 511 | 0.84 | 537 | 2.00 | 493 | 4.47 |
| $A_g$ | | | 546 | 5.57 | 464 | 3.02 |
| $B_g$ | 778 | 0.45 | 796 | −2.53 | 830 | 2.82 |
| $A_g$ | 804 | 0.23 | 796 | −0.21 | 869 | 2.30 |
| $B_g$ | 831 | 1.24 | 852 | 0.98 | 834 | 3.02 |
| $A_g$ | 924 | 3.14 | 940 | 2.83 | 889 | 2.39 |

**Table XII:** Elastic constants (given in GPa) of different zircon-type vanadates [58, 124 – 131].

| Elastic constant | $TbVO_4$ | $DyVO_4$ | $HoVO_4$ | $ErVO_4$ | $YVO_4$ | $CeVO_4$ | $SmVO_4$ |
|---|---|---|---|---|---|---|---|
| $C_{11}$ | 240(2) | 242(2) | 246.4 | 256.6(5.1) | 244.5 | 233.2 | 212.8 |
| $C_{12}$ | 55(3) | 50(4) | | 53(3) | 48.9 | 58.5 | 42.3 |
| $C_{13}$ | 93(2) | 96(2) | 79.2 | 79(6) | 81.2 | 84.8 | 76.8 |
| $C_{33}$ | | | 310.5 | 313(6) | 313.7 | 233.8 | 287.4 |
| $C_{44}$ | | | 48.5 | 50.1(1.0) | 48.3 | 38.98 | 37.4 |
| $C_{66}$ | 13.1(2) | 15.0(0.5) | 16.07 | 17.7(0.9) | 16.2 | 44.38 | 14.3 |

**Table XIII:** Crystal structure of orthorhombic HP phase of $TbVO_4$ at 35.4 GPa. Space group *Cmca*, Z = 8; *a* = 7.30480 Å, *b* =12.12854 Å, and *c* = 4.89861 Å [81].

| | Site | x | y | z |
|---|---|---|---|---|
| Tb | 8e | 0.25 | 0.84115 | 0.25 |
| V | 8f | 0.5 | 0.58939 | 0.28068 |
| $O_1$ | 8f | 0.5 | 0.41326 | 0.07580 |
| $O_2$ | 8d | 0.15369 | 0 | 0.5 |
| $O_3$ | 8f | 0.5 | 0.79106 | 0.99398 |
| $O_4$ | 8e | 0.25 | 0.15306 | 0.25 |

**Table XIV:** Crystal structure of $BaWO_4$-II-type $NdVO_4$ calculated at 15.8 GPa (space group $P2_1/n$, Z = 8). The Wyckoff positions of different atoms are indicated between brackets [82].

| | |
|---|---|
| *a* | 12.3630 Å |
| *b* | 6.4442Å |
| *c* | 6.8174Å |
| β | 96.16º |
| $Nd_1$ (4e) | (0..89292, 0.83968, 0.11759) |
| $Nd_2$ (4e) | (0.87508, 0.55366, 0.63608) |
| $V_1$ (4e) | (0.87160, 0.32781, 0.17440) |
| $V_2$(4e) | (0.83104, 0.05367, 0.60301) |
| $O_1$ (4e) | (0.93827, 0.53816, 0.32154) |
| $O_2$ (4e) | (0.79256, 0.86204, 0.77774) |
| $O_3$ (4e) | (0.90259, 0.89948, 0.46801) |
| $O_4$ (4e) | (0.77872, 0.13698, 0.08896) |
| $O_5$(4e) | (0.91798, 0.19443, 0.75993) |
| $O_6$(4e) | (0.80881, 0.27289, 0.41770) |
| $O_7$(4e) | (0.98972, 0.19608, 0.14742) |
| $O_8$(4e) | (0.83469, 0.51500, 0.98392) |

**Table XV:** Summary of transition pressures found in the literature [21 – 38, 40, 58, 59, 62, 81, 95 – 97, 101, 165 – 170] for rare-earth vanadates, $ScVO_4$, and $YVO_4$.

| Compound | zircon - scheelite | scheelite - fergusonite | post-fergusonite | zircon - monazite | post-monazite |
|---|---|---|---|---|---|
| $LaVO_4$ | | | | | 12.9 GPa |
| $CeVO_4$ | | | | 3.8 – 5.6 GPa | 12.8 – 14.7 GPa |
| $PrVO_4$ | | | | 6 GPa | |
| $NdVO_4$ | | | | 5.9 – 6.5 GPa | 11.4 – 18.1 GPa |
| $SmVO_4$ | 6.5 GPa | | | | |
| $EuVO_4$ | 6.8 – 7.8 GPa | 23.4 GPa | | | |
| $GdVO_4$ | 5 – 7.4 GPa | 23.1 GPa | | | |
| $TbVO_4$ | 6.4 – 6.7 GPa | 26.7 – 33.9 GPa | 34.4 – 48 GPa | | |
| $DyVO_4$ | 6.5 GPa | | | | |
| $HoVO_4$ | 4.5 – 8.2 GPa | 20.4 GPa | | | |
| $ErVO_4$ | 8 GPa | | | | |
| $TmVO_4$ | 7.3 GPa | | | | |
| $YbVO_4$ | 5.9 – 7.3 GPa | 15.8 GPa | | | |
| $LuVO_4$ | 8 GPa | 16 GPa | | | |
| $ScVO_4$ | 8.2 – 8.7 GPa | 18.2 GPa | | | |
| $YVO_4$ | 7.5 – 8.5 GPa | 20 GPa | | | |

**Table XVI:** Bulk modulus, pressure-derivative of the bulk modulus, and ambient pressure volume of different structures in *A*$VO_4$ compounds taken from the literature. References are indicated in the right-hand column. It is also indicated if the results were obtained from experiments or calculations. The pressure-transmitting medium (PTM) used in the experiments is also indicated. ME = methanol-ethanol. MEW = methanol-ethanol-water.

| Compound | Structure | PTM | $V_0$ [Å$^3$] | $B_0$ [GPa] | $B_0'$ | Reference |
|---|---|---|---|---|---|---|
| $LaVO_4$ Nano rods | Zircon | ME | 364.0(2) | 93(2) | 4 | 50 |
| $LaVO_4$ | Monazite | MEW | 333.2 (1) | 106(1) | 4 | 96 |
| | Monazite | MEW | 334.2(5) | 99(5) | 4 | 96 |
| | Monazite | Theory | 328.2 | 105.2 | 4.3 | 40 |
| | $BaWO_4$-II | Theory | 609.2 | 154 | 4.2 | 40 |
| $CeVO_4$ | Zircon | ME | 356.53 | 112(3) | 4 | 32 |
| | Zircon | Ne | 352.9 | 125(9) | 4 | 33 |
| | Zircon | ME | 354.86 | 118.9 | 4 | 30 |
| | Monazite | ME | 328.23 | 98(3) | 4 | 32 |
| | Monazite | Ne | 326.2(8) | 133(5) | 4.4(6) | 33 |
| | Monazite | ME | 327.21 | 142 | 4.4 | 30 |
| $PrVO_4$ | Zircon | Theory | 350.5 | 142 | 4 | 5 |
| $NdVO_4$ | Zircon | MEW | 350.8(9) | 148(4) | 3.72(2) | 96 |
| | Monazite | Theory | 316.62 | 138.8 | 3.79 | 82 |
| | *Cmca* | Theory | 596.35 | 146.8 | 3.42 | 82 |
| $SmVO_4$ | Zircon | MEW | 336.5(9) | 129(4) | 4 | 58 |
| | Zircon | ME | 338.6(9) | 131(7) | 4 | 58 |
| | Scheelite | MEW | 299.7(1.2) | 133(5) | 4 | 58 |
| | Scheelite | ME | 298.9(1.5), | 256(12) | 4 | 58 |
| $EuVO_4$ | Zircon | Silicon oil | 333.2(9) | 149(6) | 5.6(6) | 25 |
| | Zircon | MEW | 333.9(3) | 119(3) | 4 | 29 |
| | Zircon | ME | 333.84 (4) | 150 (7) | 5.3 (6) | 59 |
| | Scheelite | Silicon oil | 299.4(9) | 199(9) | 4.1(9) | 25 |
| | Scheelite | MEW | 296.2(6) | 135(7) | 4 | 29 |
| | Scheelite | ME | 296 (2) | 195 (8) | 5.5 (9) | 59 |
| Er:$GdVO_4$ | Zircon | silicon oil | 326.5(2.2) | 102.4(14.0) | 4 | 36 |
| | Scheelite | silicon oil | 288.6(1.4) | 136.6(15.4) | 4 | 36 |
| $GdVO_4$ | Zircon | No medium | 327.35 (2.4) | 185(28) | 4 | 38 |
| | Zircon | From ultrasonic Measurements | | 114.6 | | 63 |
| | Scheelite | No medium | 283.64(0.9) | 379(32) | 4 | 38 |
| $TbVO_4$ | Zircon | Ne | 324.4(9) | 122(5) | 6.2(5) | 33 |
| | Scheelite | Ne | 288.2(6) | 163(9) | 5.8(6) | 33 |
| | *Cmca* | Theory | 524.66 | 134.3 | 3.38 | 81 |
| $DyVO_4$ | Zircon | Estimated from elastic constants | 323.3 | 140(5) | 4 | 124 |
| | Zircon | Theory | 325.9 | 149.4 | 4 | 238 |
| | Zircon | Boron Nitride | 322.37(9) | 118(4) | 4.6(1.7) | 97 |
| | Scheelite | Boron Nitride | 286.5(1) | 153(6) | 4.2(2.7) | 97 |
| $HoVO_4$ | Zircon | ME | 320.5(9) | 160(15) | 5(2) | 28 |
| | Zircon | Ar | 319(1) | 143(3) | 4.7(9) | 28 |
| | Scheelite | ME | 282.5 | 180(15) | 5.5(9) | 28 |
| | Scheelite | Ar | 282.5(2) | 160(3) | 4.3(9) | 28 |
| $ErVO_4$ | Zircon | Estimated from elastic constants | 315.9 | 136(9) | 4 | 124 |
| | Zircon | Theory | 315.9 | 154.2 | 4 | 238 |
| $TmVO_4$ | Zircon | Theory | 312.7 | 137.9 | 4 | 239 |
| $YbVO_4$ | Zircon | ME | 310.7 | 141 | 4 | 27 |
| | Scheelite | ME | N.A. | 159 | 4 | 27 |
| | Fergusonite | ME | 291.0 | 116 | 4 | 27 |
| $LuVO_4$ | Zircon | silicon oil | 307.9(9) | 166(7) | 5.6(6) | 25 |
| | Zircon | ME | 307.73 | 147 | 4.3 | 95 |
| | Scheelite | silicon oil | 271.4(9) | 195(9) | 4.9(9) | 25 |
| | Scheelite | ME | NA, | 194 | 5.3 | 95 |
| $ScVO_4$ | Zircon | silicon oil | 281.1(9) | 178(9) | 5.9(9) | 25 |
| | Scheelite | silicon oil | 256.9(9) | 210(12 | 5.3(8) | 25 |
| | Fergusonite | No medium | 261 | 187.78 | 5.93 | 37 |
| $YVO_4$ | Zircon | $N_2$ | 319.15(1) | 130(3) | 4.4(10) | 24 |
| | Scheelite | $N_2$ | 284.48(3) | 138(8) | 4.5(8) | 24 |
| $InVO_4$ | $CrVO_4$-type | MEW | 324.0(9) | 69(1) | 4 | 41 |
| | Woframite | MEW | 131.6(6) | 168(9) | 4 | 41 |
| | $\alpha$-$MnMoO_4$-type | Theory | 671.6 | 120.9 | 4.52 | 44 |
| $BiVO_4$ | Fergusonite | Theory | 310.3 | 109.93 | 4 | 240 |
| | Scheelite | ME | 310.5 | 150(5) | 4 | 20 |
| | Scheelite | Theory | 310.5 | 132.13 | 4 | 237 |
| | pucherite | Theory | 351.15 | 102.88 | 4 | 237 |
| | zircon | Theory | 324.3 | 108.97 | 4 | 237 |

**Figure Captions**

**Figure 1:** Most common crystal structures of *A*$VO_4$ compounds. The coordination polyhedra of V and the trivalent cation *A* are shown in red and blue. The oxygen atoms are shown as small red spheres. (a) zircon, (b) $CrVO_4$-type, (c) $FeVO_4$-I, (d) fergusonite, (e) pucherite, and (f) monazite.

**Figure 2:** Plot of ambient pressure structural parameters for all *R*$VO_4$ compounds. The compounds are identified by the *R* cation in the horizontal axis (e.g. La is used for $LaVO_4$). The tetragonal distortion increases with the decrease in ionic radii of the rare-earth which is reflected in the *c/a* ratio. The overall volume of the unit cell decreases, which is the consequence of the rare-earth contraction effect. The meaning of the different symbols is indicated in the plot. Different scales (indicated also in the plot) have been used for different magnitudes in the vertical axis.

**Figure 3:** X-ray powder diffraction diagrams of $YVO_4$ at different pressures. The top most diagram was measured after a pressure cycle up to 26.6 GPa. Pressures are indicated in the plot and the (112) Bragg peak of scheelite is identified. X-ray wavelength = 0.4176 Å. Figure reproduced from Ref. 24.

**Figure 4:** Observed, calculated, and difference x-ray powder patterns for the zircon ($I4_1/amd$, Z = 4) and scheelite ($I4_1/a$, Z = 4) phases of $YVO_4$ at 7.8 GPa (top) and 16.4 GPa (bottom), respectively. Vertical markers indicate Bragg reflections and the residuals shown. The diffraction pattern in the top frame was refined assuming an admixture of the high-pressure scheelite phase. X-ray wavelength = 0.4176 Å. Figure reproduced from Ref. 24.

**Figure 5:** Pressure evolution of x-ray diffraction patterns of $TbVO_4$. (r) denotes the patterns collected while pressure release. In the pattern collected at 8.8 GPa for the HP scheelite phase, the experimental data (dots), the refined pattern (solid line), and the residual of the refinement

(dotted line) are also shown. X-ray wavelength = 0.4246 Å. Pressures and structure are indicated in the figure. Figure reproduced from Ref. 33.

**Figure 6:** Schematic representation of different HP polymorphs. (a) Scheelite, (b)isomorphic to $BaWO_4$−II(c)post-fergusonite phase orthorhombic structure (space group *Cmca*), and (d) wolframite. Most of these structures imply a coordination number increase for both cations. In the HP orthorhombic polymorph, Tb and V atoms are coordinated to eleven and seven oxygen atoms, respectively.

**Figure 7:** Pressure dependence of the lattice parameters and unit-cell volume for the zircon and scheelite phases of $TbVO_4$ (squares and circles, respectively). In order to facilitate the comparison, for the scheelite phase, we plotted $c/2$ instead of $c$. Lines: quadratic fits. Symbols: experiments. Solid squares: zircon, solid circles: scheelite, empty circles: scheelite upon decompression, and empty squares: zircon data from Ref. 62. The vertical dashed line indicates the transition pressure. Results taken from Ref. 33.

**Figure 8:** Selection of x-ray diffraction patterns measured in $CeVO_4$ at different pressures. The pressures as well as the structure assigned to each pattern are indicated. Bragg peaks from Ne (the pressure medium) are identified. There is a XRD pattern collected under pressure release, which is denoted by (r). The vertical ticks indicate the calculated position for Bragg reflections. In the pattern collected at 5.6 GPa for the HP phase, the measurement (dots), the refined pattern (solid line), and the residual of the refinement (dotted line) are shown. X-ray wavelength = 0.4246 Å. Figure reproduced from Ref. 33.

**Figure 9:** XRD patterns for $LaVO_4$ at two different pressures illustrating the phase transition. Rietveld refinements are shown. Dots represent the experiments and solid lines the calculated profiles. The residuals are shown as solid lines. Vertical ticks indicate the Bragg reflections position. Pressures and structures are labelled in the figure. Peaks from Ne (the pressure medium) are also shown. X-ray wavelength = 0.5340 Å. Figure reproduced from Ref. 40.

**Figure 10:** (Top) Measured high-pressure Raman spectra for $PrVO_4$. (Bottom) Pressure evolution of the Raman modes of the different phases. Circles: low-pressure phase. Squares: high-pressure phase. The Raman modes of the zircon phase are labelled. Figure reproduced from Ref. 31.

**Figure 11:** (Top) Measured high-pressure Raman spectra for $SmVO_4$. (Bottom) Pressure evolution of the Raman modes of the different phases. Circles: low-pressure phase. Squares: high-pressure phase. The Raman modes of the zircon phase are labelled. Figure reproduced from Ref. 31.

**Figure 12:** Raman spectra of $TbVO_4$ zircon phase up to 6.7 GPa (left) and scheelite phase from 8.3 GPa to 26.3 GPa (center). We also show a selection of spectra showing the three phase transitions observed under compression (right). On the top of the right-hand side panel a Raman spectra measured after decompression is shown. Figure reproduced from Ref. 81.

**Figure 13:** Pressure-dependent sequence of selected high-pressure Raman spectra of $LaVO_4$. "r" indicates a spectrum measured after decompression. The onset of the transition is indicated. Figure reproduced from Ref. 40.

**Figure 14:** Pressure dependence of the band-gap energy for $YVO_4$, $YbVO_4$, $LuVO_4$ and $NdVO_4$. Sold lines are the linear fits to the experimental data shown by circles. Figure reproduced from Ref. 36.

**Figure 15:** Luminescence spectra of $Eu^{3+}$ doped $GdVO_4$ at selected pressures. Pressures are shown and the signal from ruby is identified. Figure reproduced from Ref. 108.

**Figure 16:** Observed variation of electrical resistance (top) and volume (bottom) of $CeVO_4$ with pressure. Results taken from Refs. 32 and 33.

**Figure 17:** Calculated energy-volume curves for $TbVO_4$. Since the total-energy curves of scheelite and fergusonite structures overlap, we remark the stability range of scheelite phase

(fergusonite phase) using a solid (dashed) line. The inset shows differences in the Gibbs free energy with respect to the zircon phase. Figure reproduced from Ref. 81.

**Figure 18:** Enthalpy as a function of pressure for $NdVO_4$. The zircon structure is taken as a reference. Figure reproduced from Ref. 82.

**Figure 19:** Calculated frequency of the $B_{1u}$ silent mode of the zircon-type structure as a function of pressure for $TbVO_4$. Figure reproduced from Ref. 81.

**Figure 20:** Comparison of the zircon, scheelite, and fergusonite structures seen from different projections. $VO_4$ tetrahedra are in red and $AO_8$ dodecahedra in blue. Oxygen atoms are the small red spheres.

**Figure 21:** Phase diagram of lanthanide vanadates. The dotted and solid lines are a schematic representation of different phase boundaries. Post-fergusonite phases appear only at very HP and have only been reported for $TbVO_4$. The region of stability of these new phases is tentatively indicated in the figure. The compounds are plotted using the ionic radii for the horizontal axis and are identified by the lanthanide cation.

**Figure 22:** Powder XRD patterns of $InVO_4$ at selected pressures. Cu (pressure marker) and gasket peaks are identified. Ticks indicate the position of Bragg peaks of phases III and V. Figure reproduced from Ref. 41.

**Figure 23:** Volume curves as a function of pressure for the most stable polymorphs of $InVO_4$. Symbols are explained in the inset. Experimental results from Ref. 41 and calculations from Ref. 44. Figure reproduced from Ref. 44.

**Figure 24:** Schematic structural phase diagram for $AXO_4$ compounds. The different crystal structures are identified using different symbols. The dashed lines connect compounds of the same family, which are indicated on right-hand side of the figure.

**Figure 25:** Structural sequence observed in different zircon-type oxides and related oxides. Figure reproduced from Ref. 155. The higher coordination phases refer either to the $SrUO_4$-

type structure or any of the high-coordination phases foundin related oxides, e.g., orthorhombic (*Cmca*) found in $TbVO_4$ [81].

**Figure 26:** Pressure dependence of Ho–O and V–O distances in the zircon and scheelite structures of $HoVO_4$. Figure reproduced from Ref. 28.

**Figure 1**

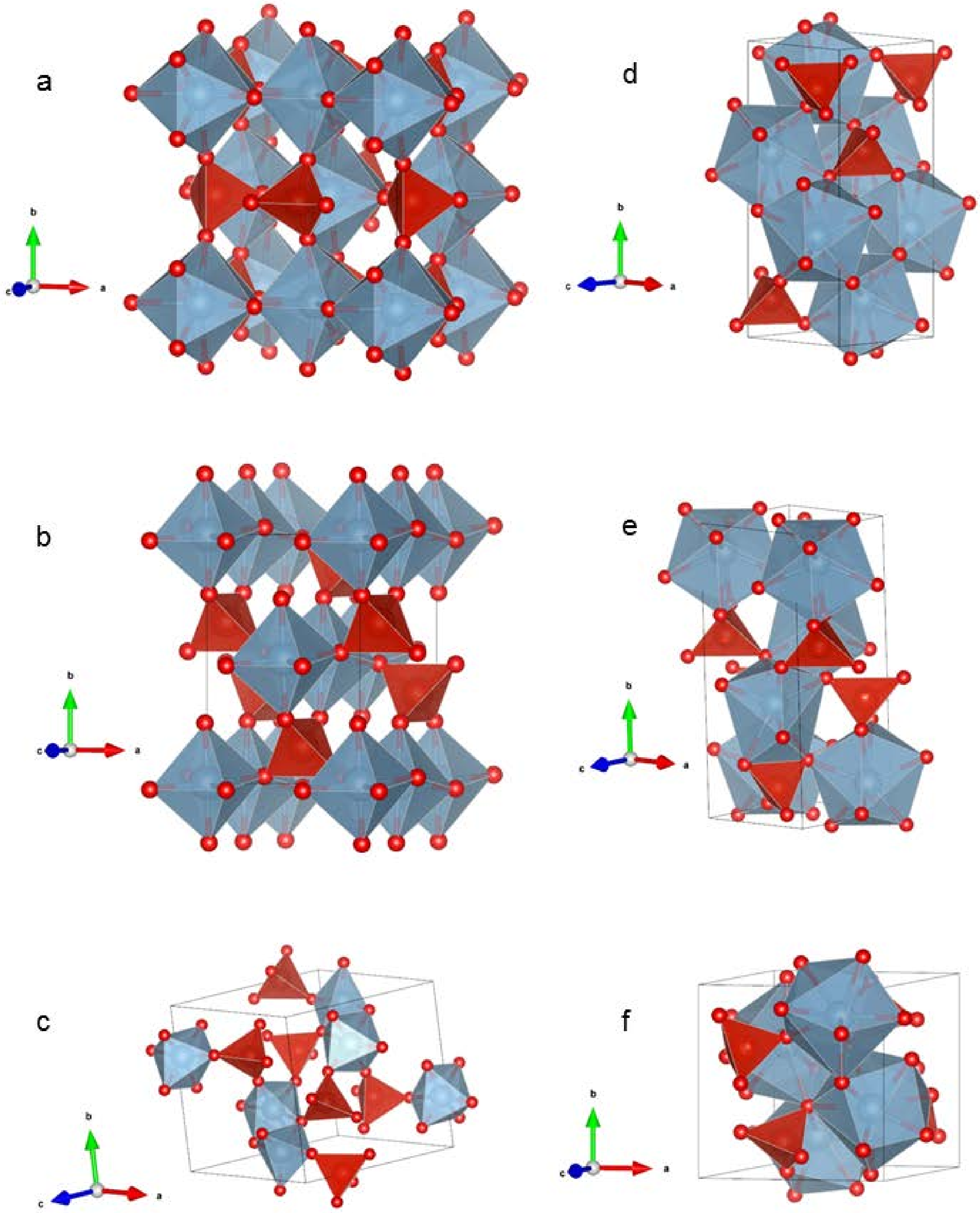

**Figure 2**

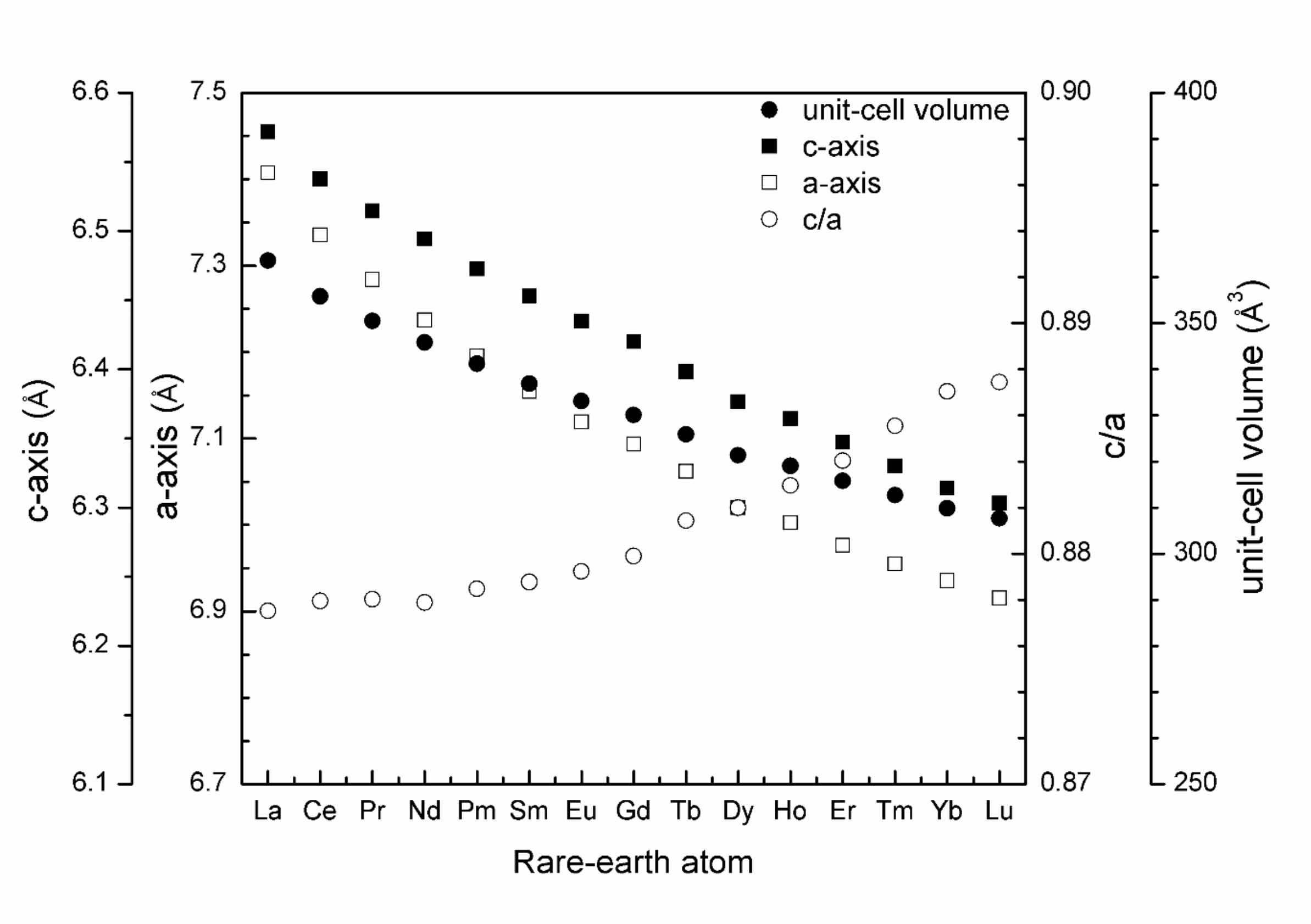
unit-cell volume
c-axis
a-axis
c/a
c-axis (Å)
a-axis (Å)
c/a
unit-cell volume (Å$^3$)
6.6
6.5
6.4
6.3
6.2
6.1
7.5
7.3
7.1
6.9
6.7
0.90
0.89
0.88
0.87
400
350
300
250
La Ce Pr Nd Pm Sm Eu Gd Tb Dy Ho Er Tm Yb Lu
Rare-earth atom

**Figure 3**

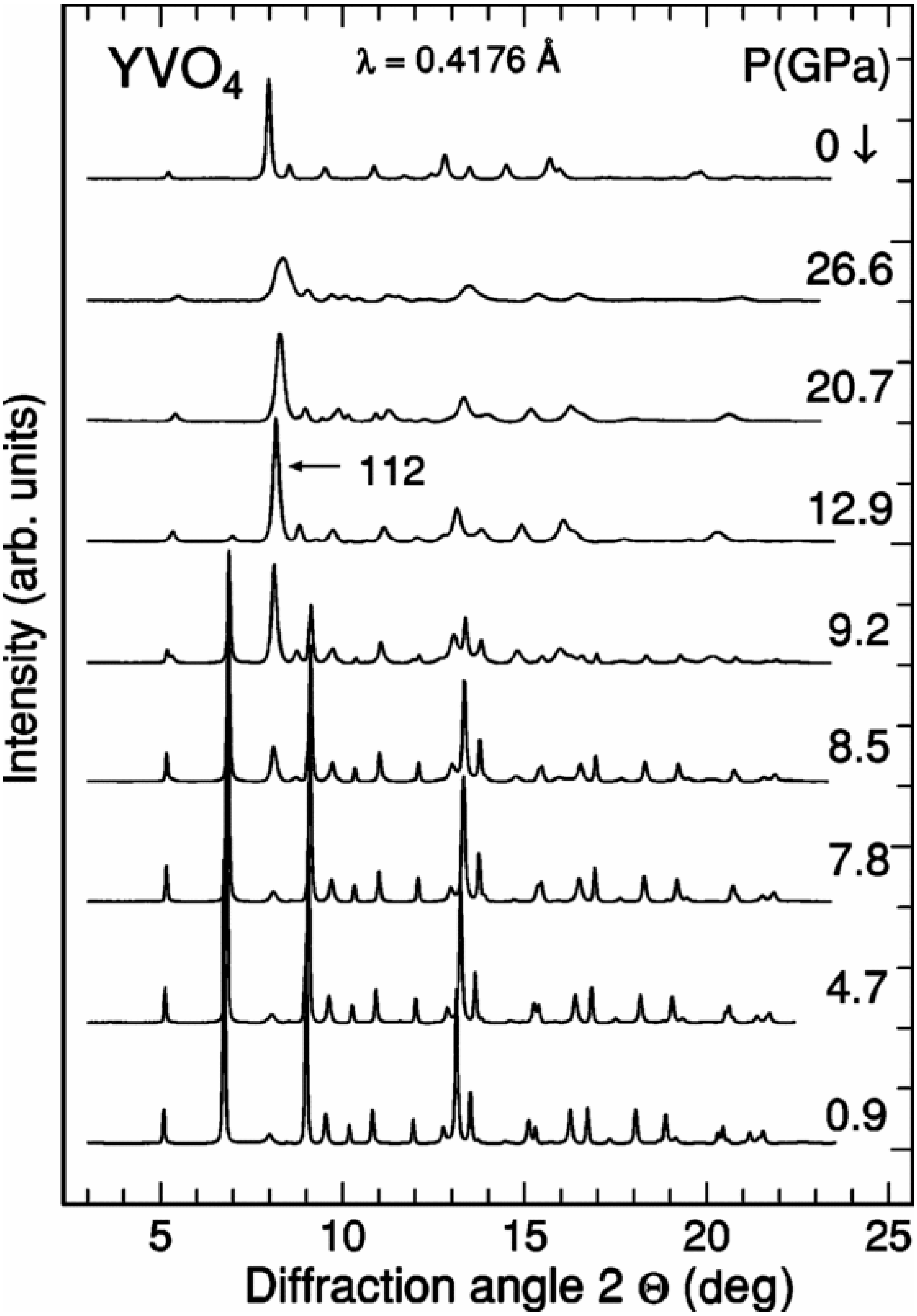

**Figure 4**

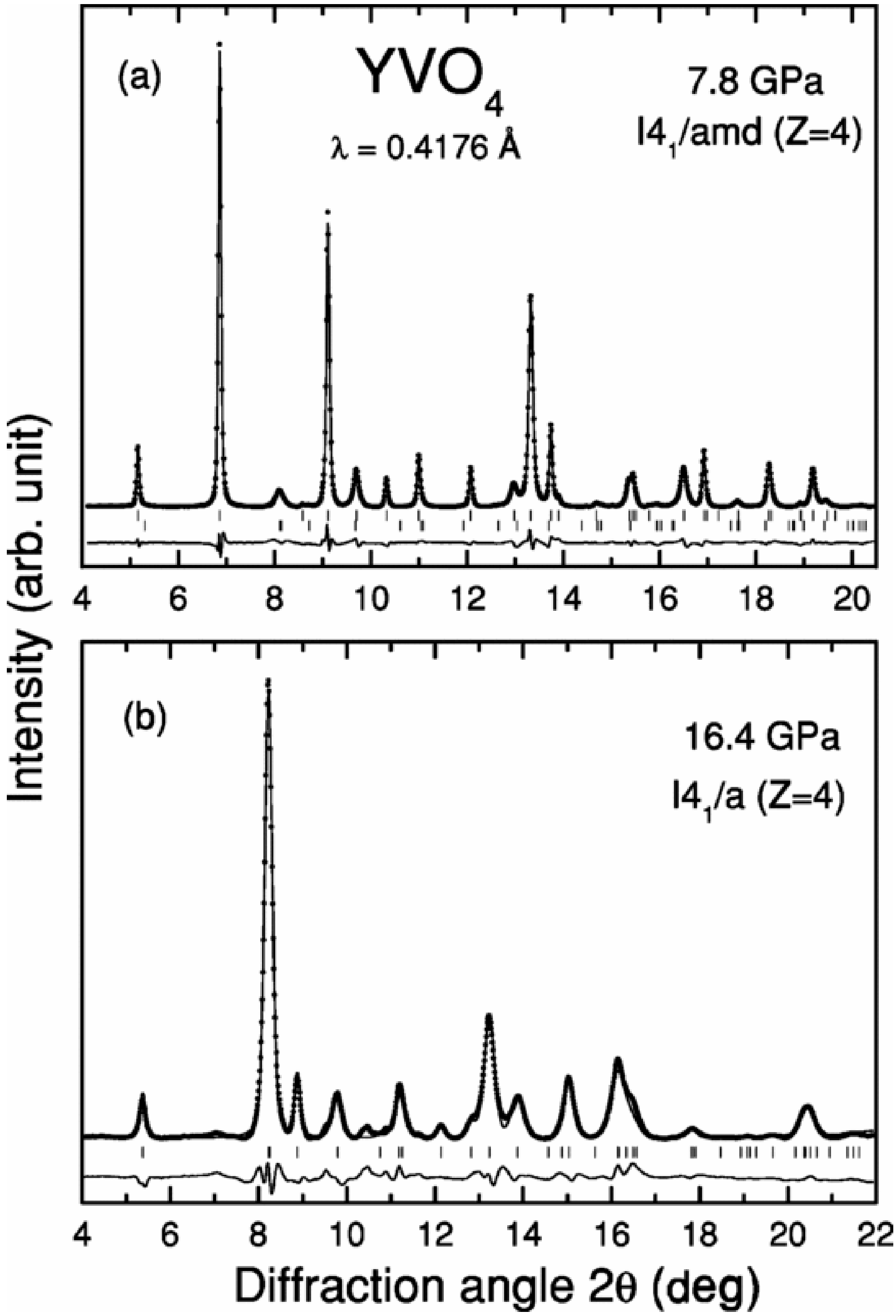

**Figure 5**

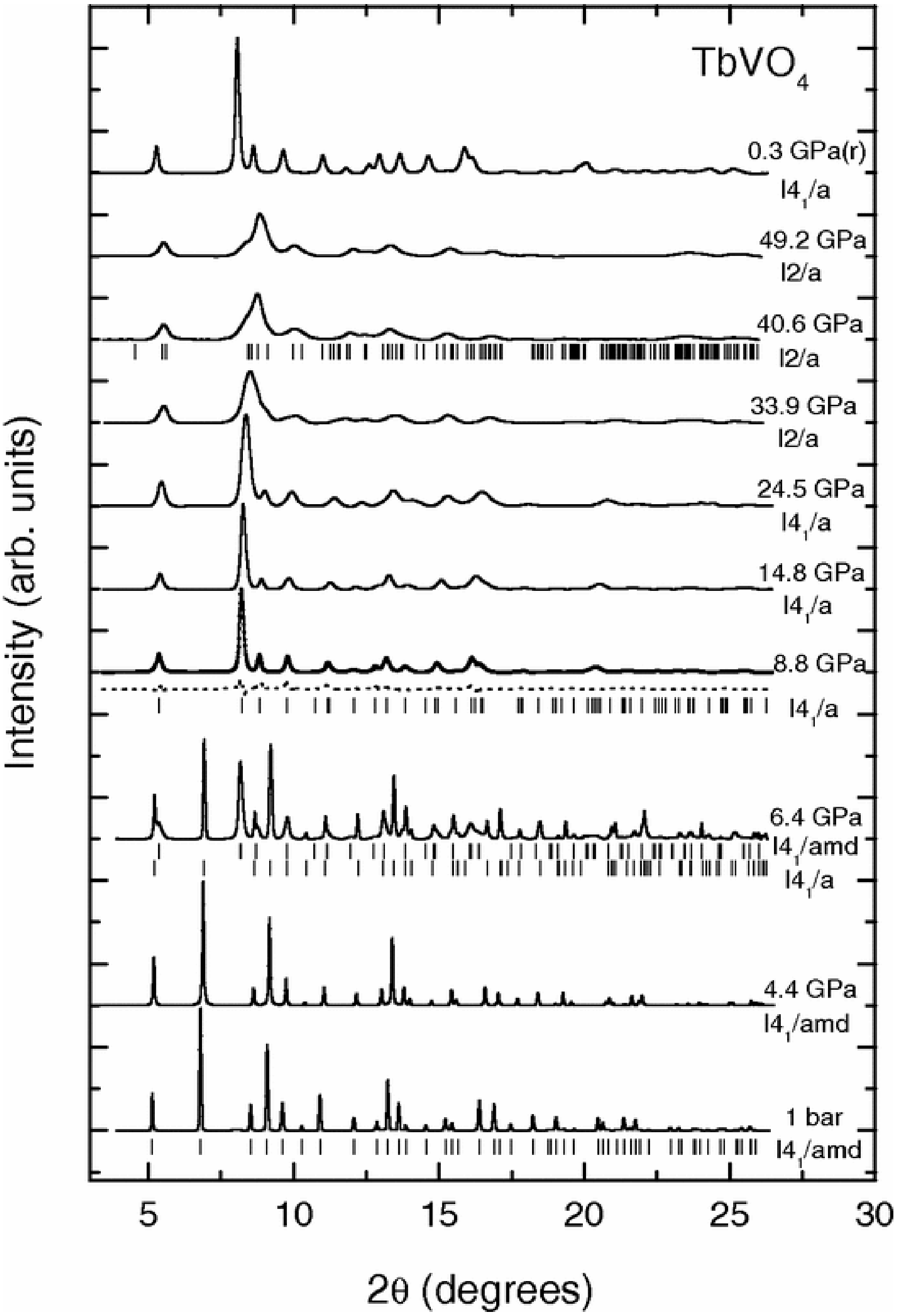

Figure 6

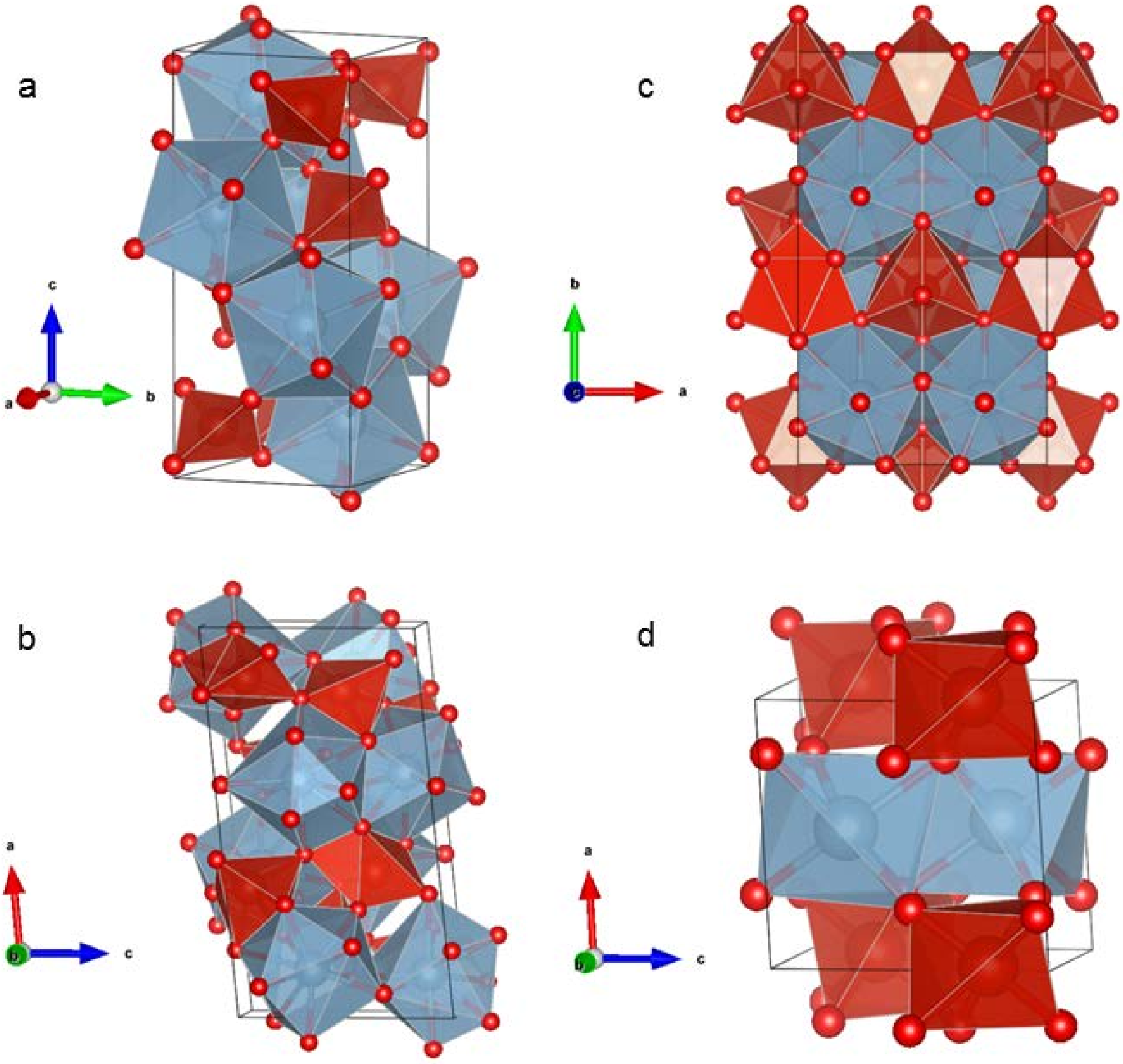

**Figure 7**

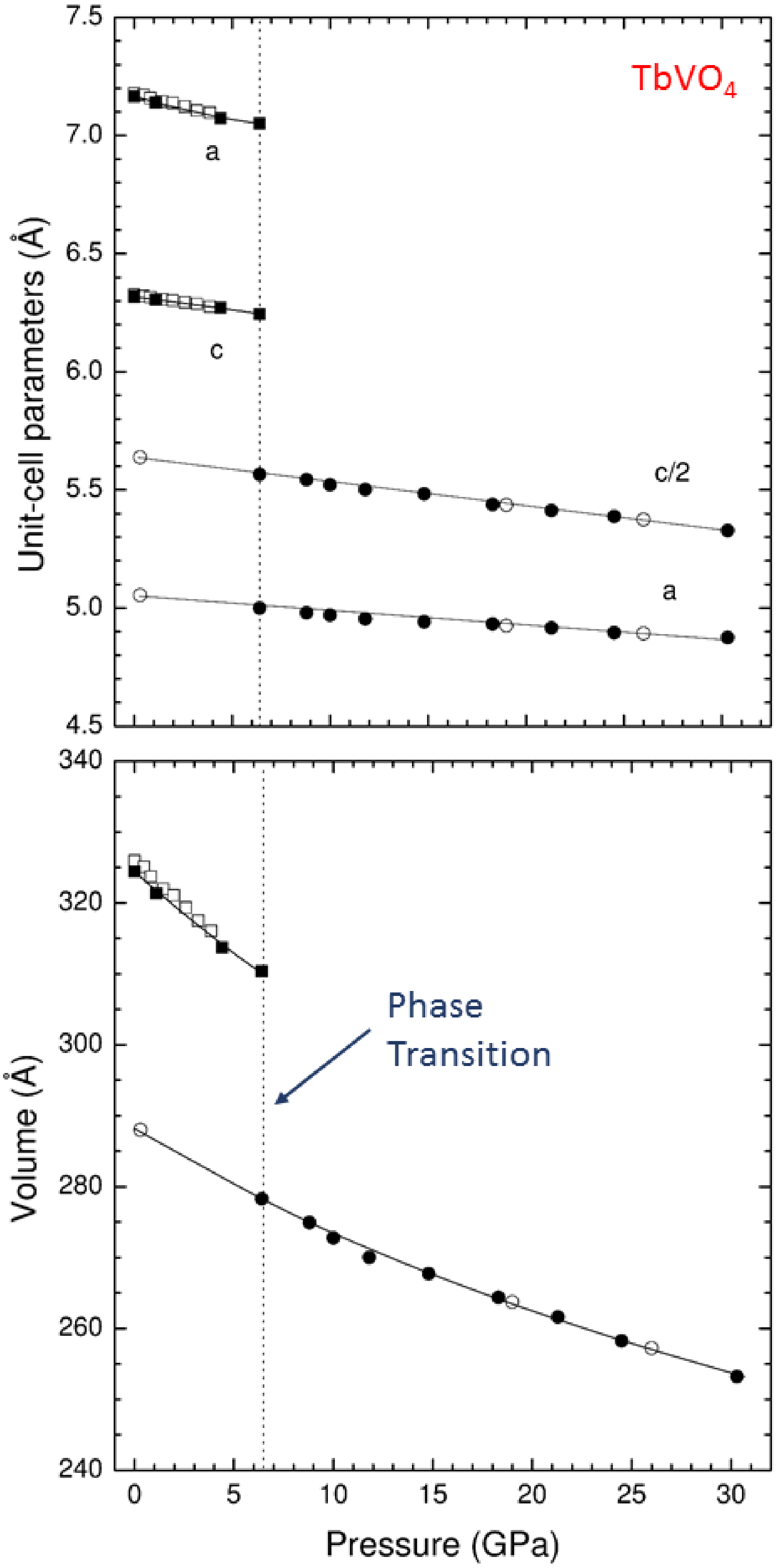

**Figure 8**

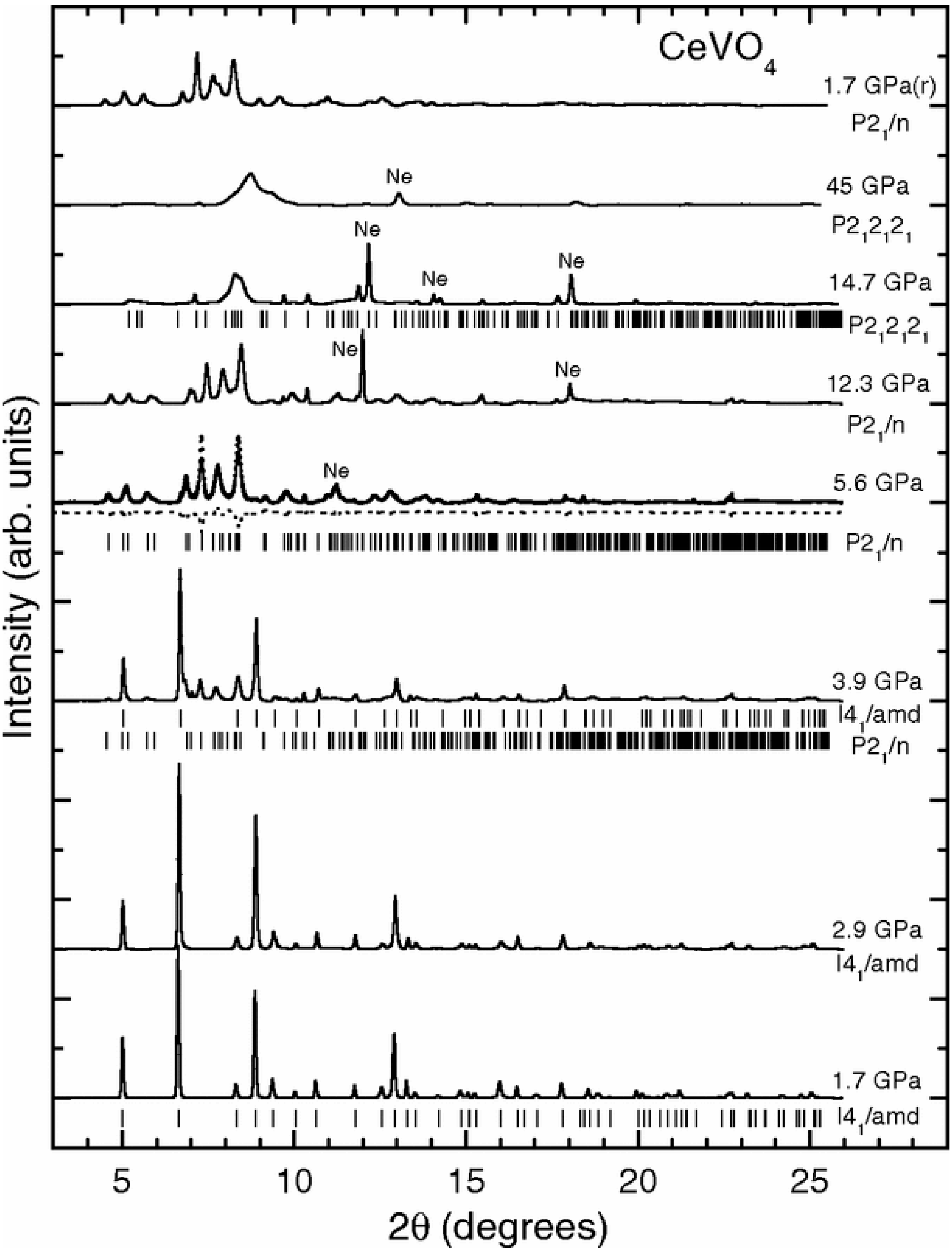

**Figure 9**

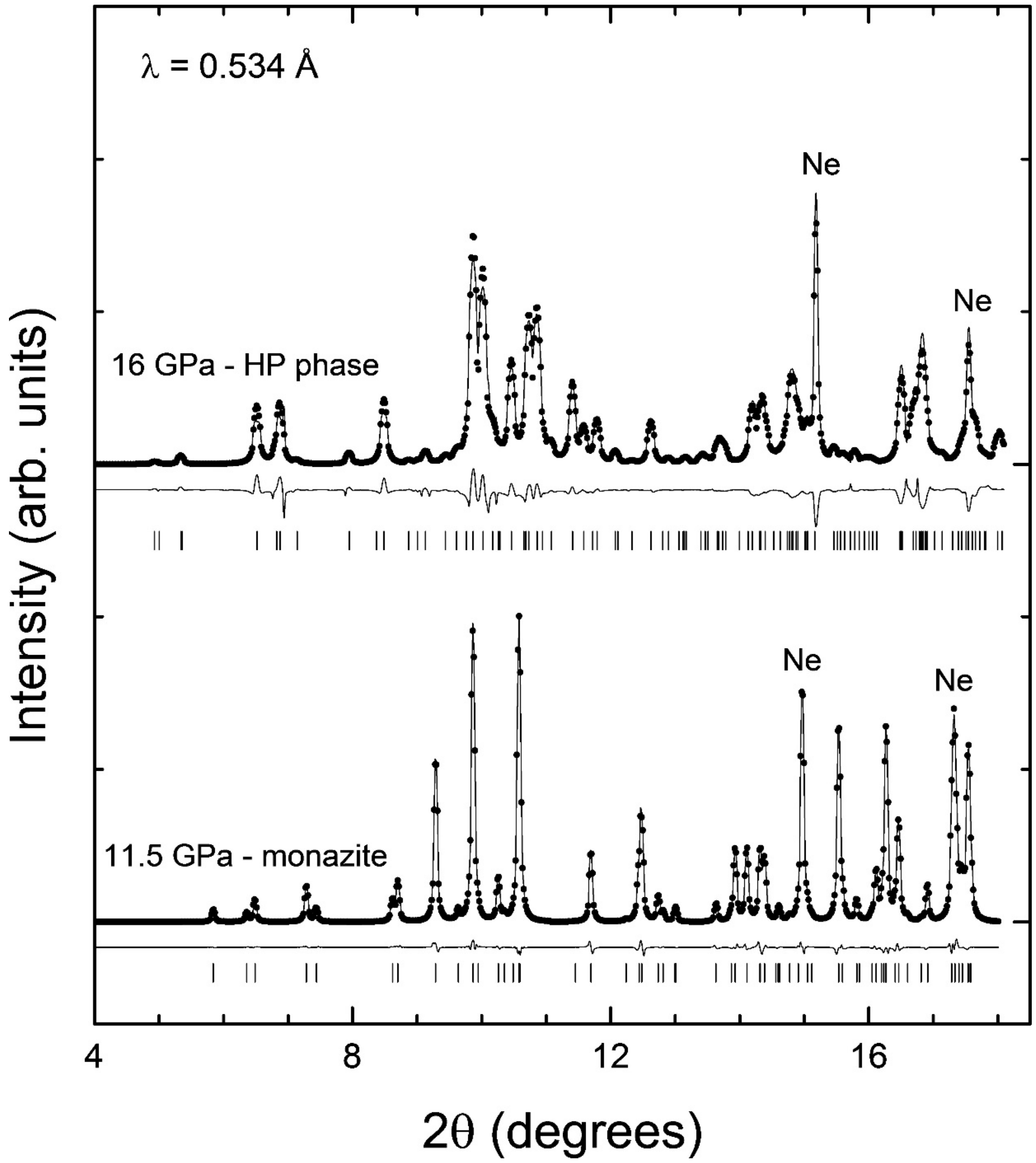
λ = 0.534 Å
Ne
Ne
16 GPa - HP phase
Ne
Ne
11.5 GPa - monazite
Intensity (arb. units)
4
8
12
16
2θ (degrees)

**Figure 10**

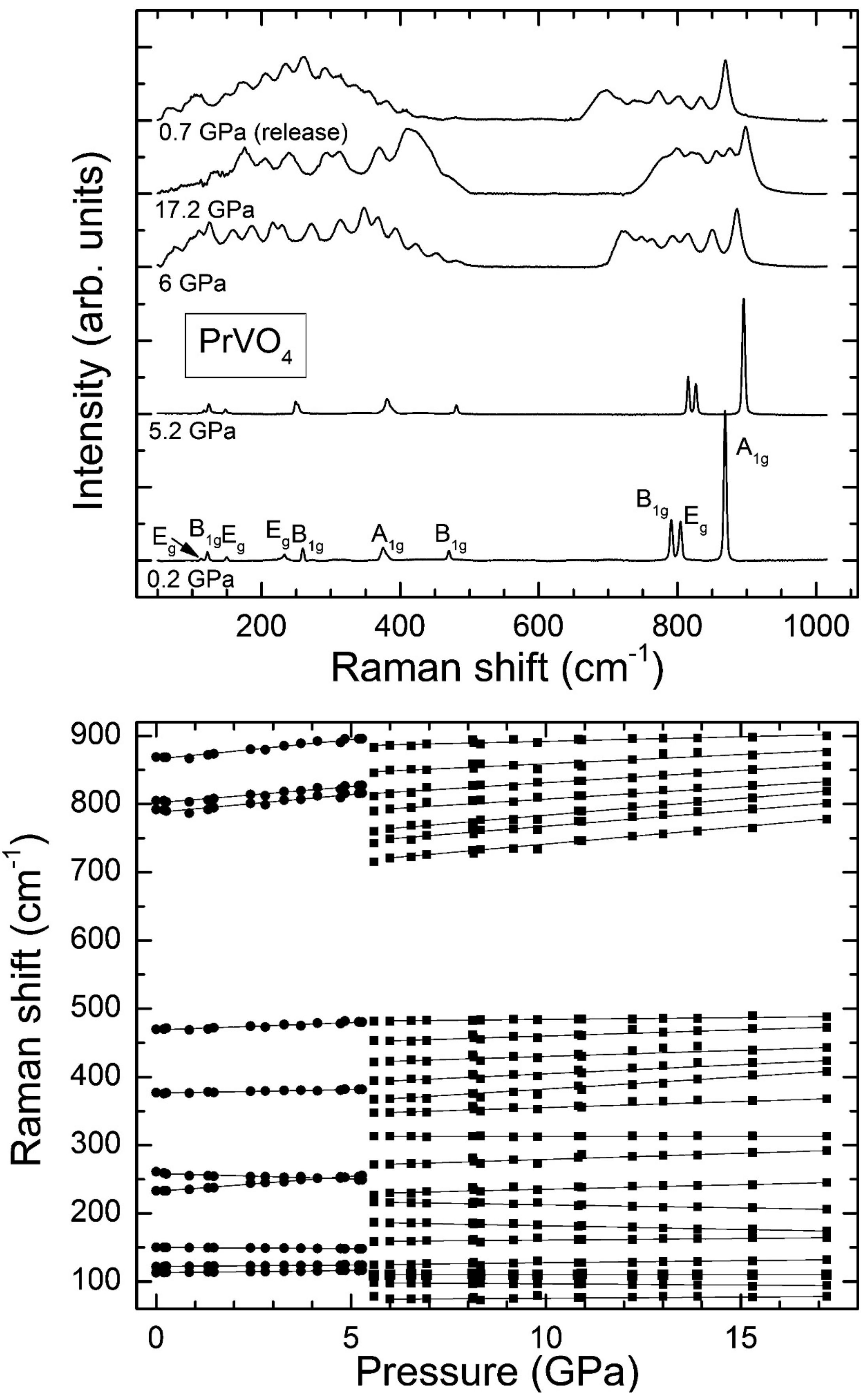

0.7 GPa (release)
17.2 GPa
6 GPa
PrVO$_4$
5.2 GPa
0.2 GPa
E$_g$ B$_{1g}$ E$_g$ E$_g$ B$_{1g}$ A$_{1g}$ B$_{1g}$
B$_{1g}$ E$_g$ A$_{1g}$
Intensity (arb. units)
200 400 600 800 1000
Raman shift (cm$^{-1}$)
900 800 700 600 500 400 300 200 100
Raman shift (cm$^{-1}$)
0 5 10 15
Pressure (GPa)

**Figure 11**

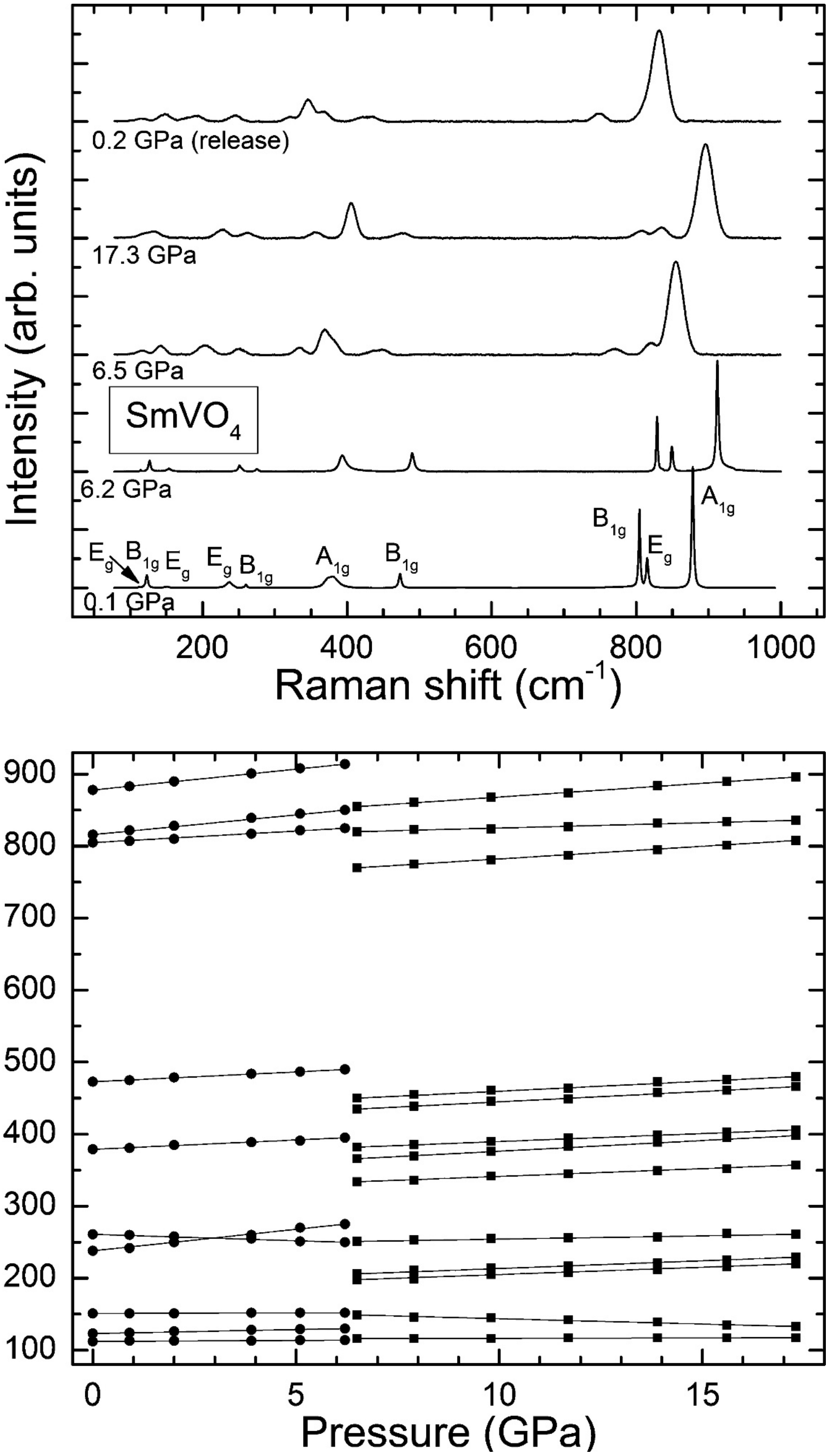
0.2 GPa (release)
17.3 GPa
6.5 GPa
SmVO4
6.2 GPa
0.1 GPa
Eg B1g Eg Eg B1g A1g B1g B1g Eg A1g
Intensity (arb. units)
200 400 600 800 1000
Raman shift (cm-1)
900 800 700 600 500 400 300 200 100
Raman shift (cm-1)
0 5 10 15
Pressure (GPa)

**Figure 12**

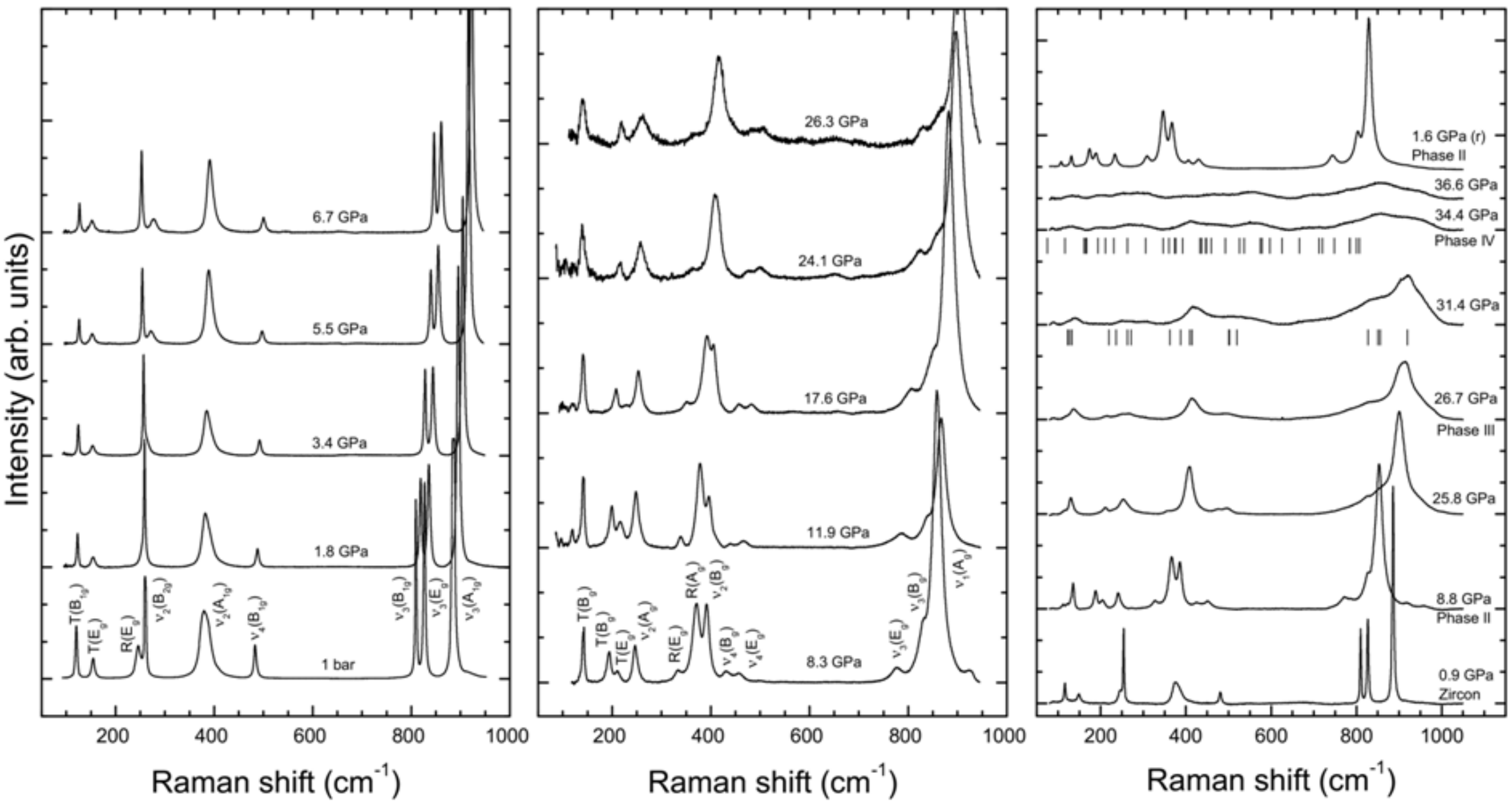

**Figure 13**

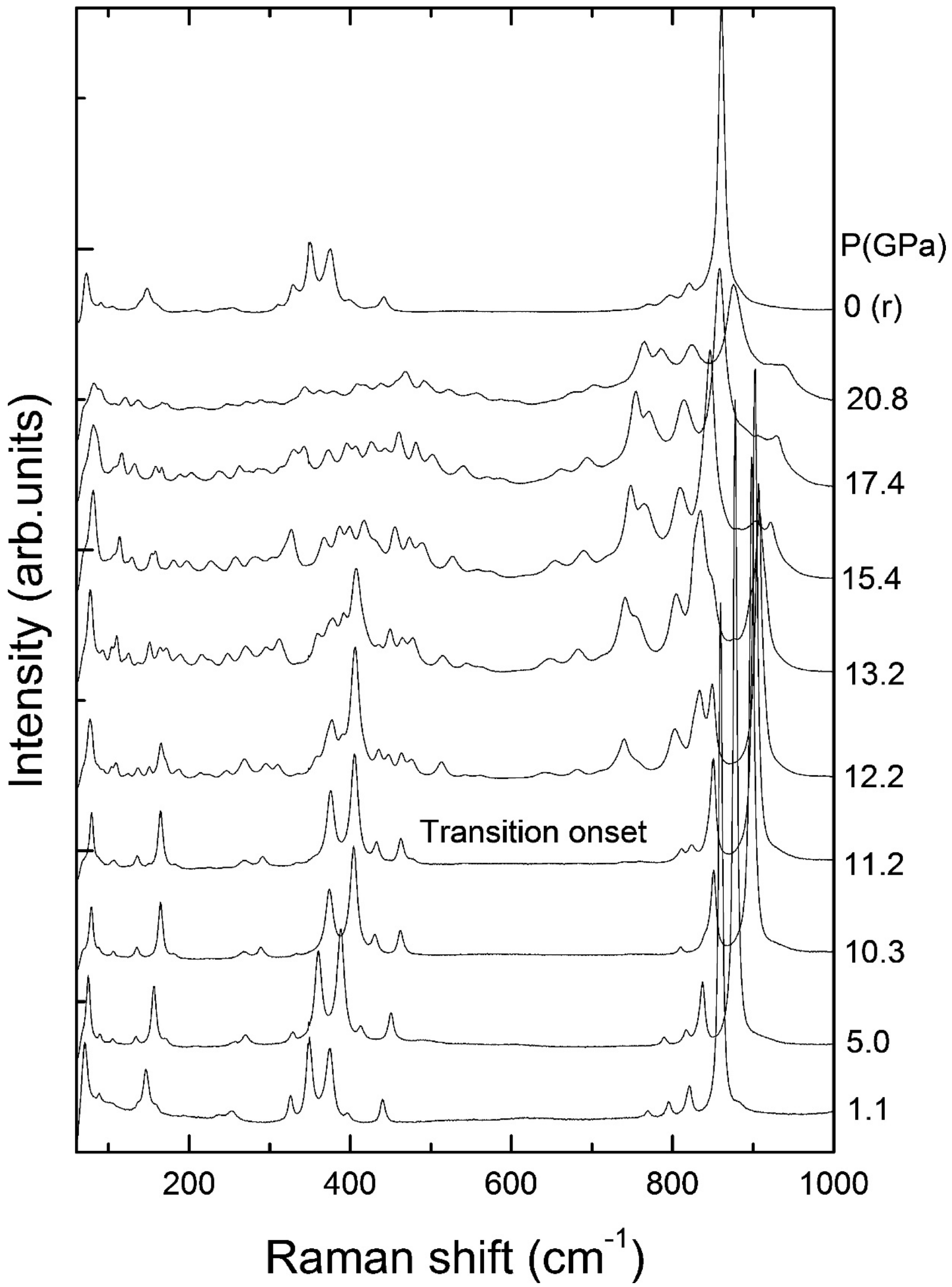

**Figure 14**

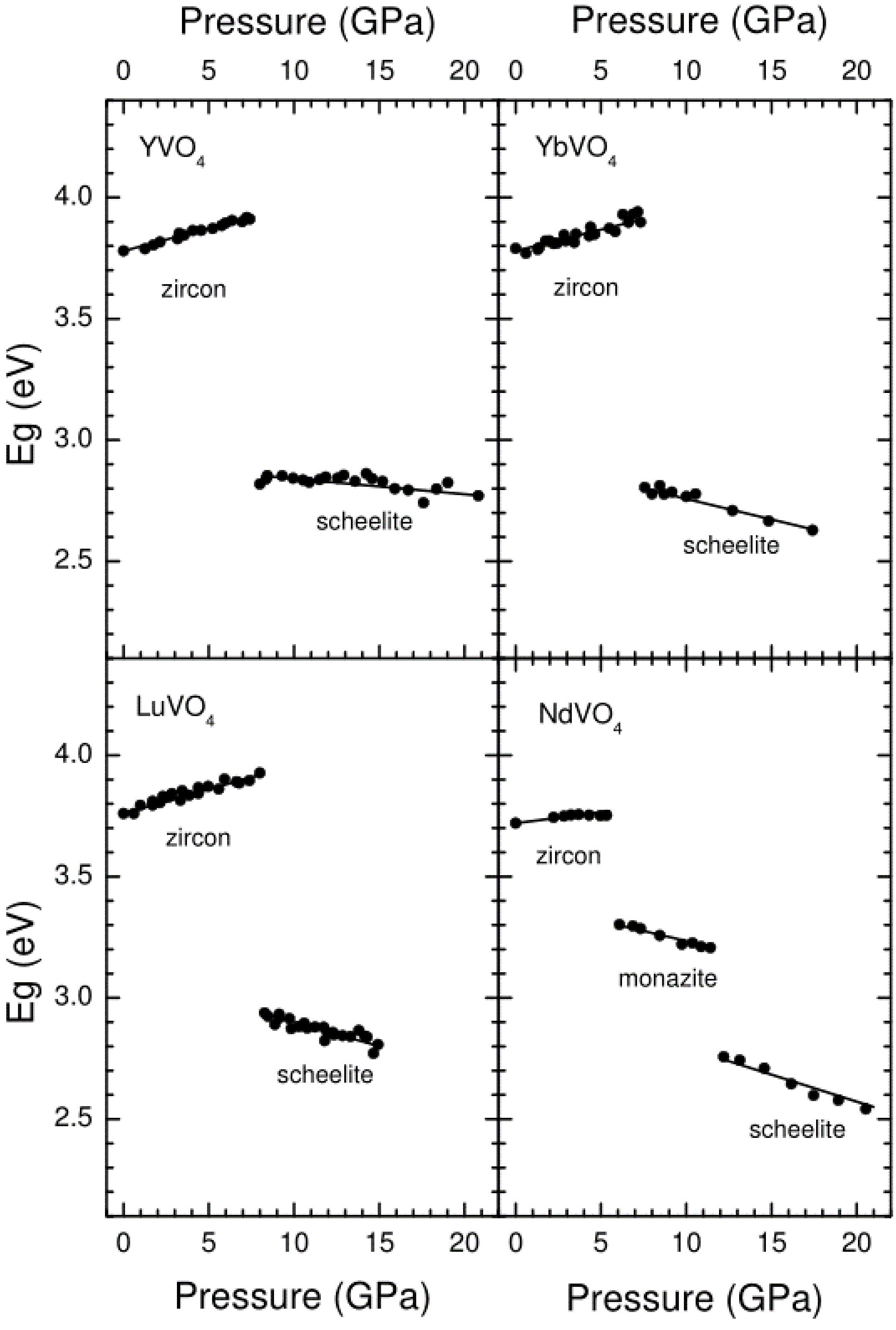

**Figure 15**

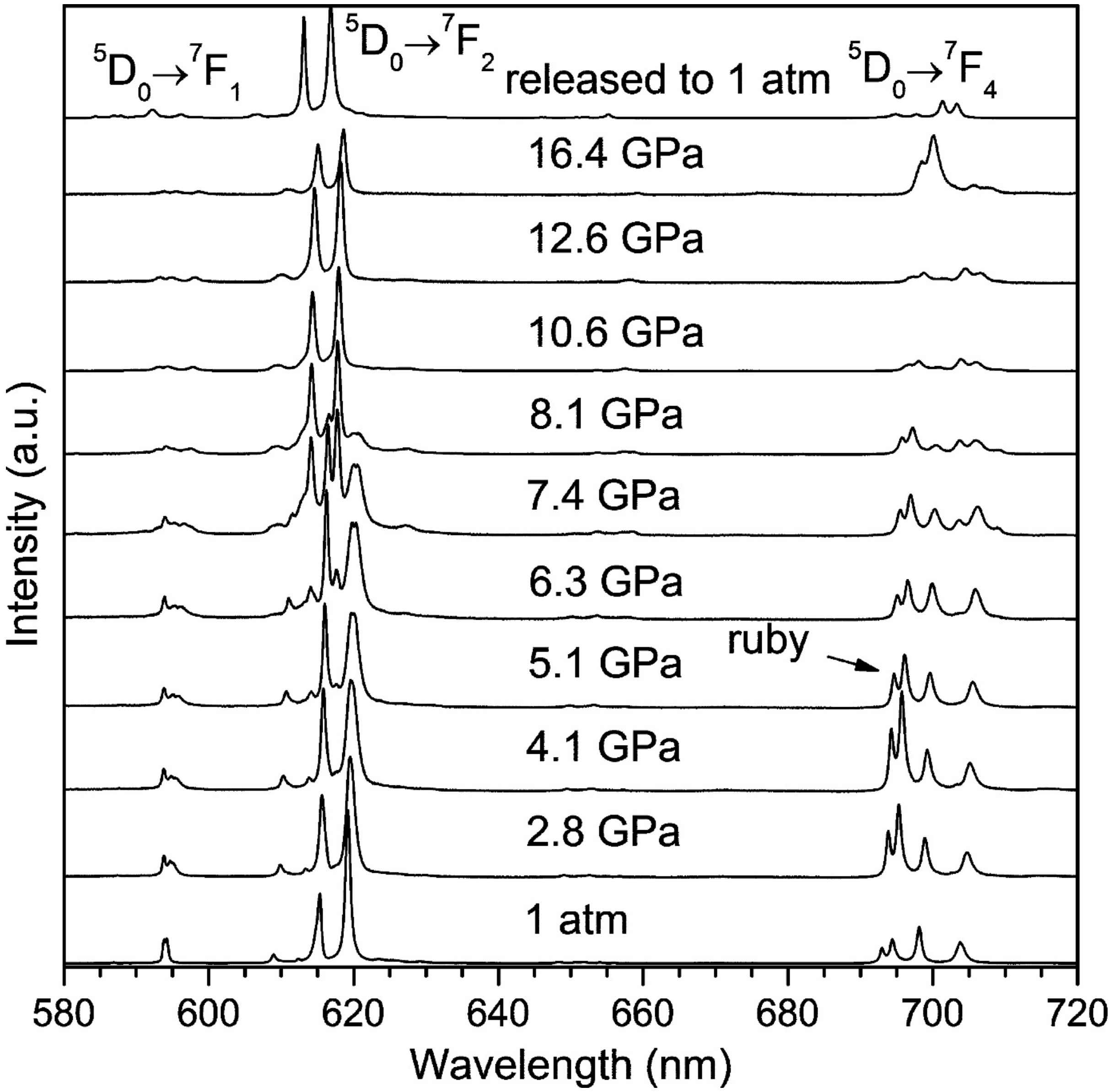

**Figure 16**

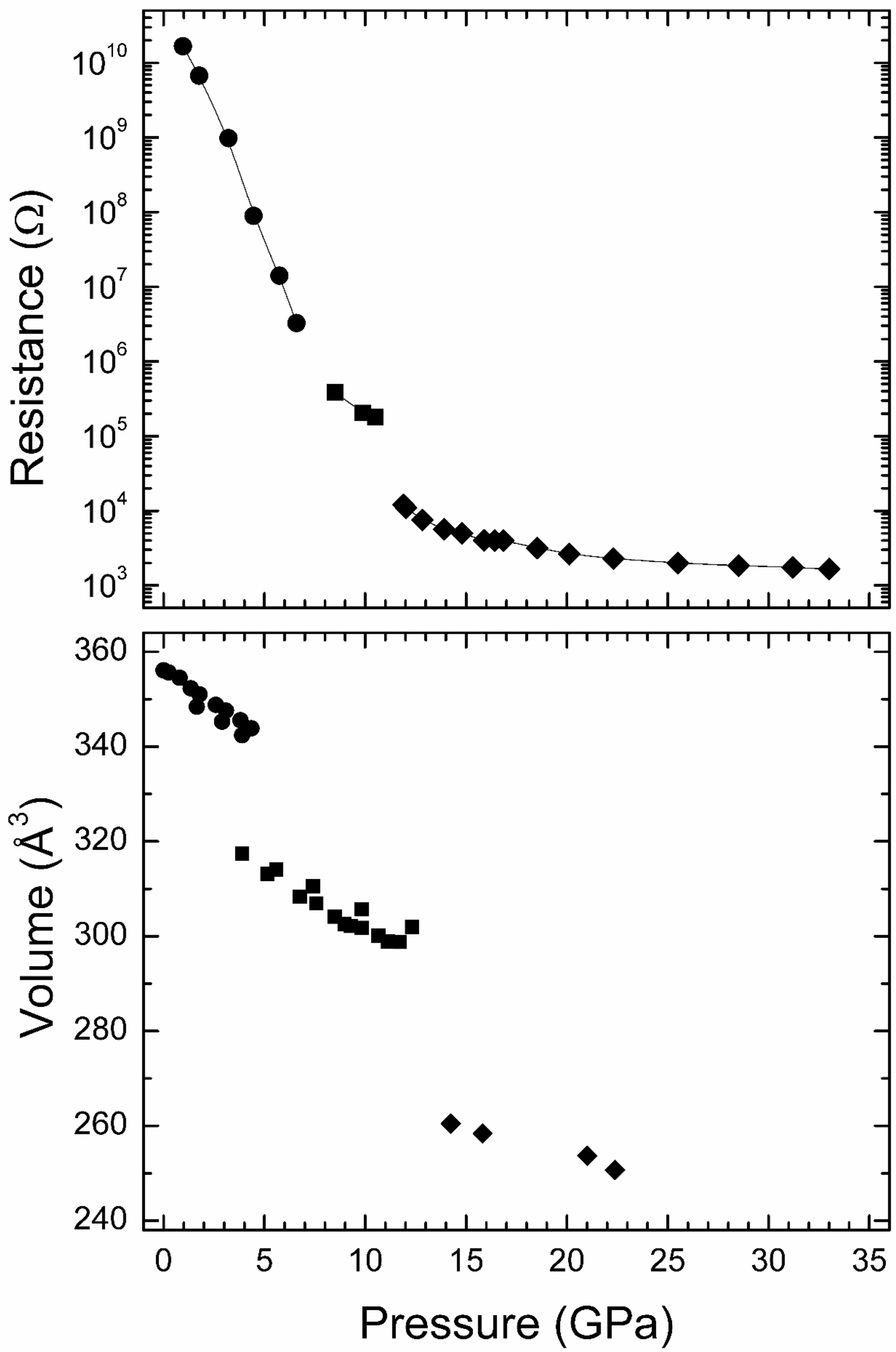

**Figure 17**

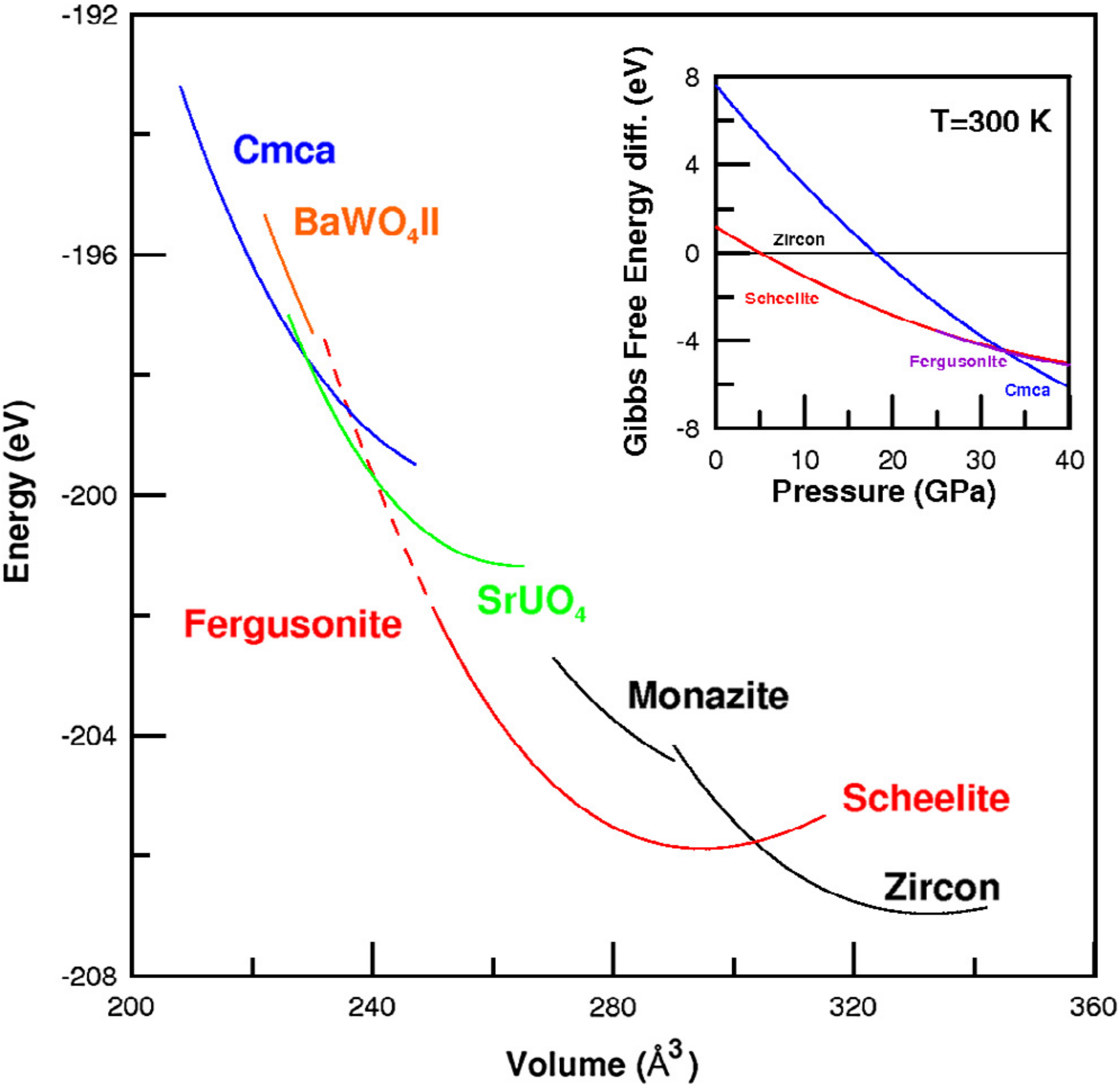

**Figure 18**

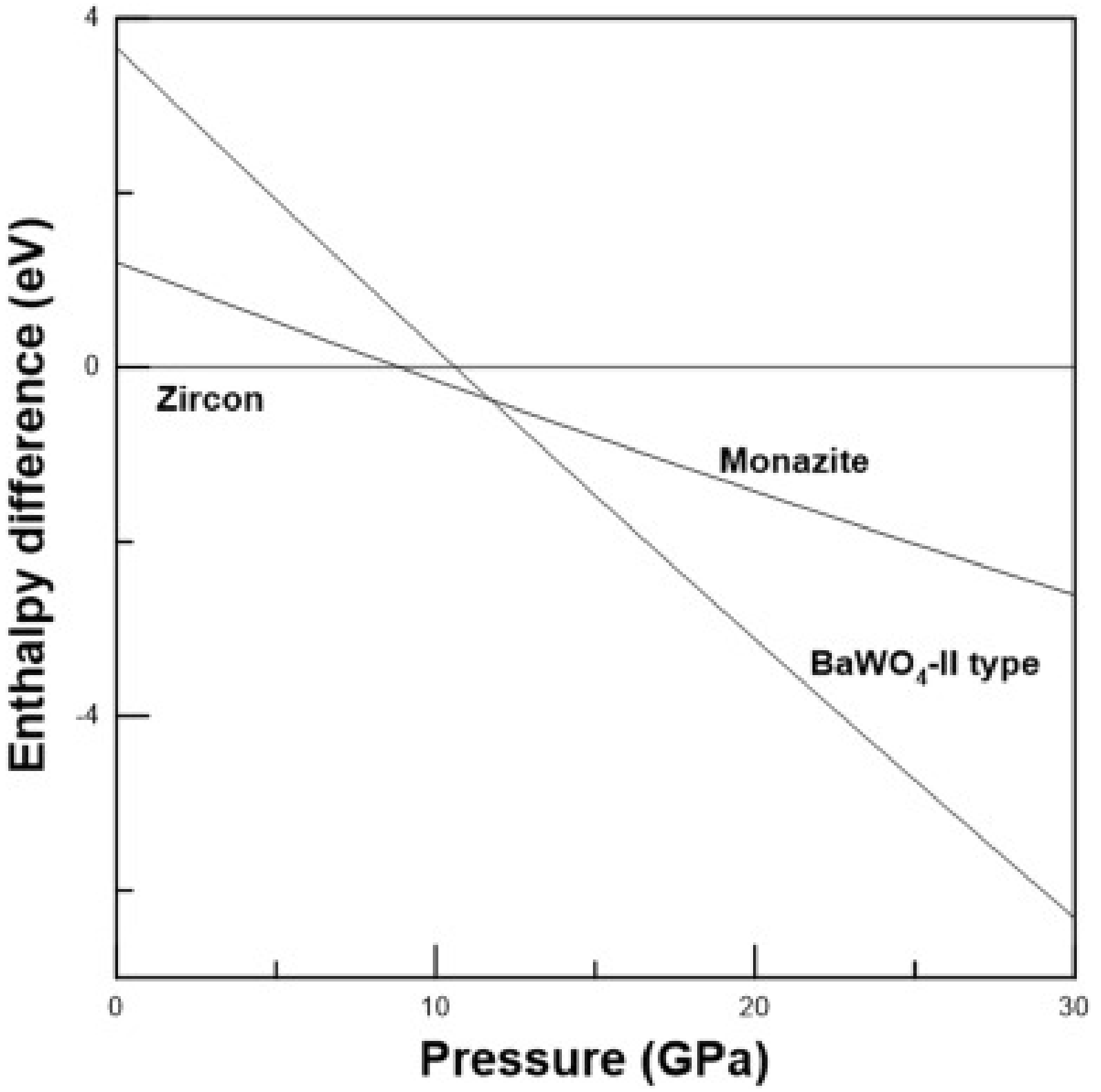

**Figure 19**

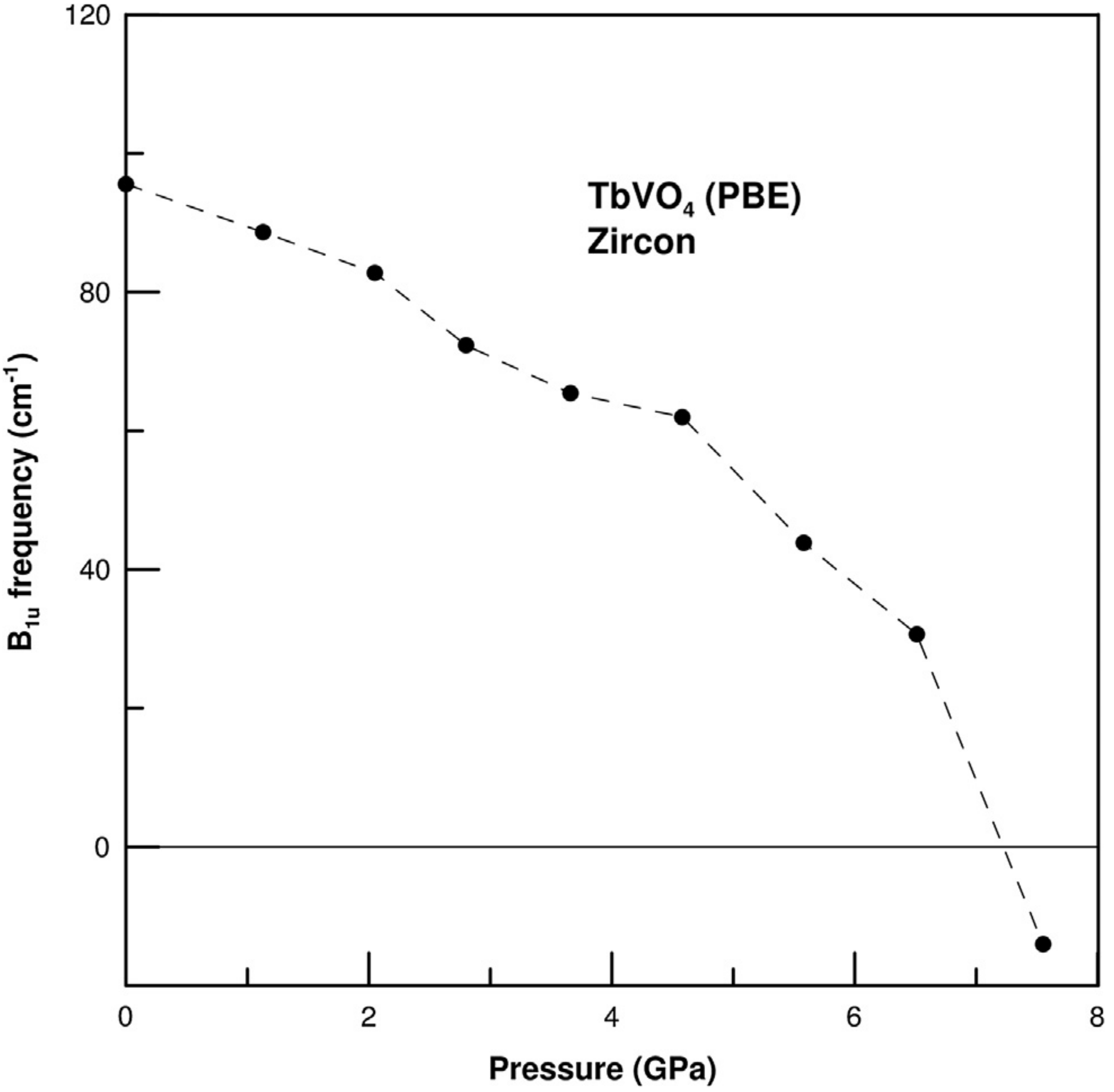

**Figure 20**

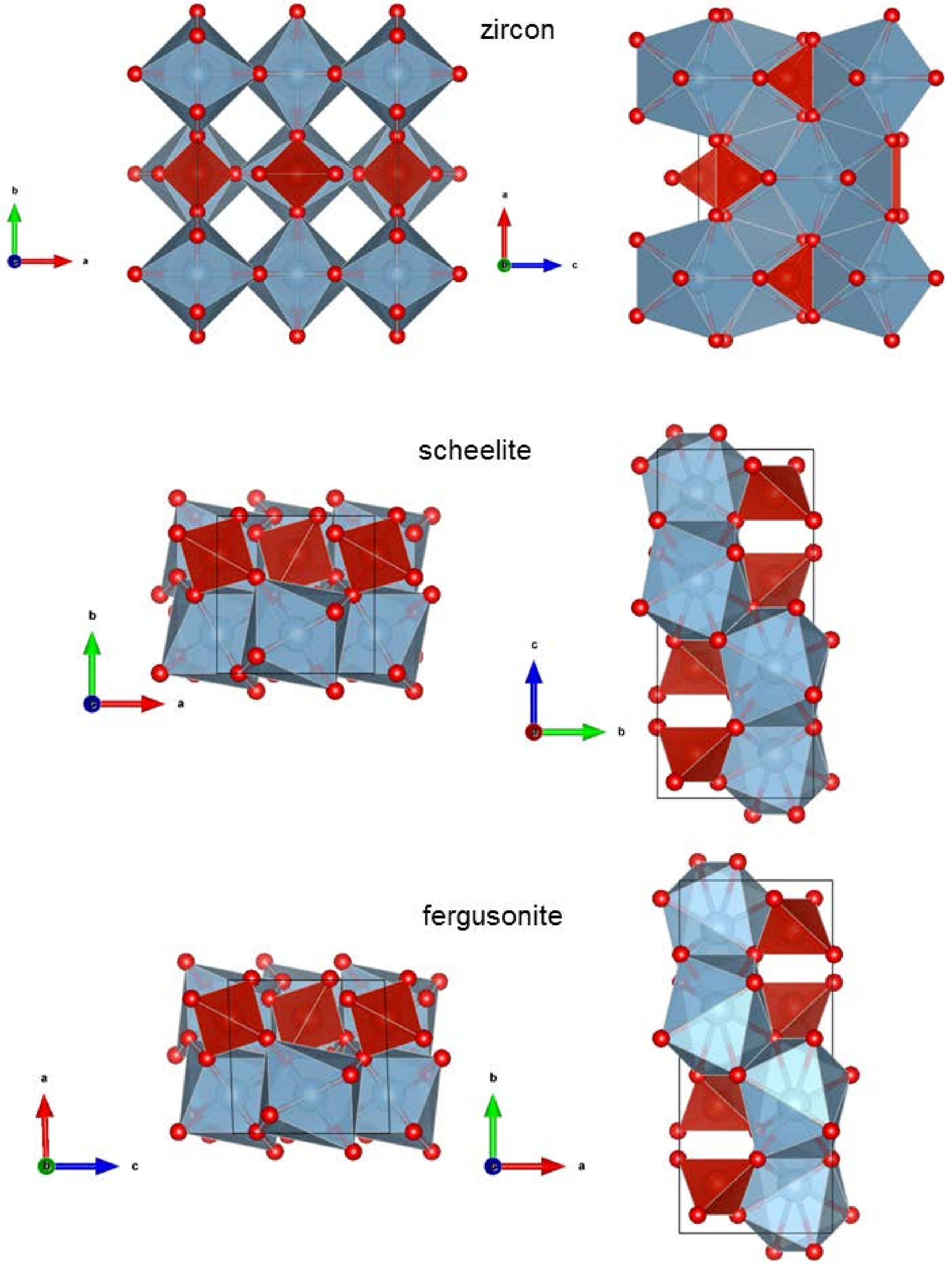

**Figure 21**

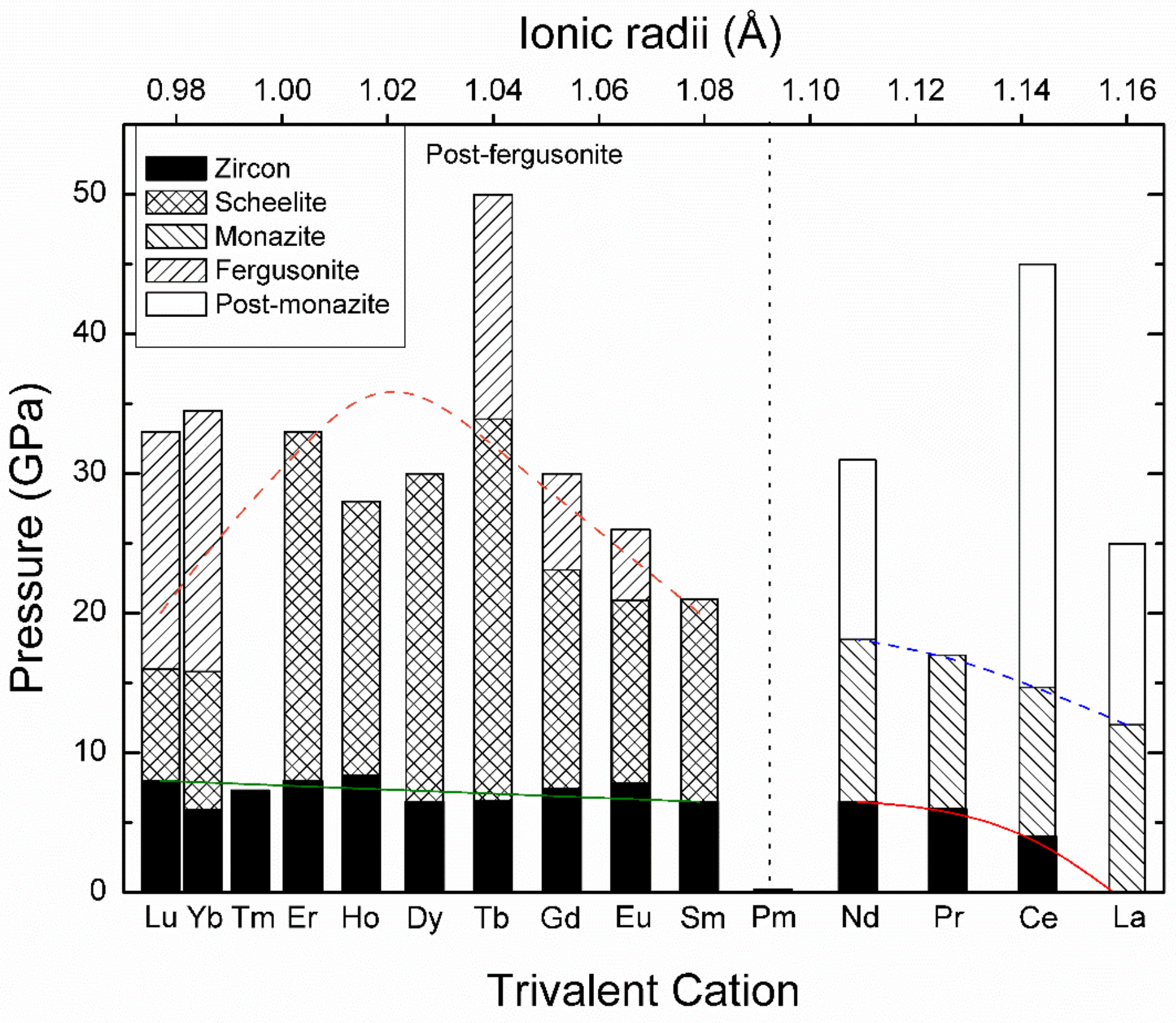
Ionic radii (Å)
0.98
1.00
1.02
1.04
1.06
1.08
1.10
1.12
1.14
1.16
Zircon
Scheelite
Monazite
Fergusonite
Post-monazite
Post-fergusonite
Pressure (GPa)
0
10
20
30
40
50
Lu Yb Tm Er Ho Dy Tb Gd Eu Sm Pm Nd Pr Ce La
Trivalent Cation

**Figure 22**

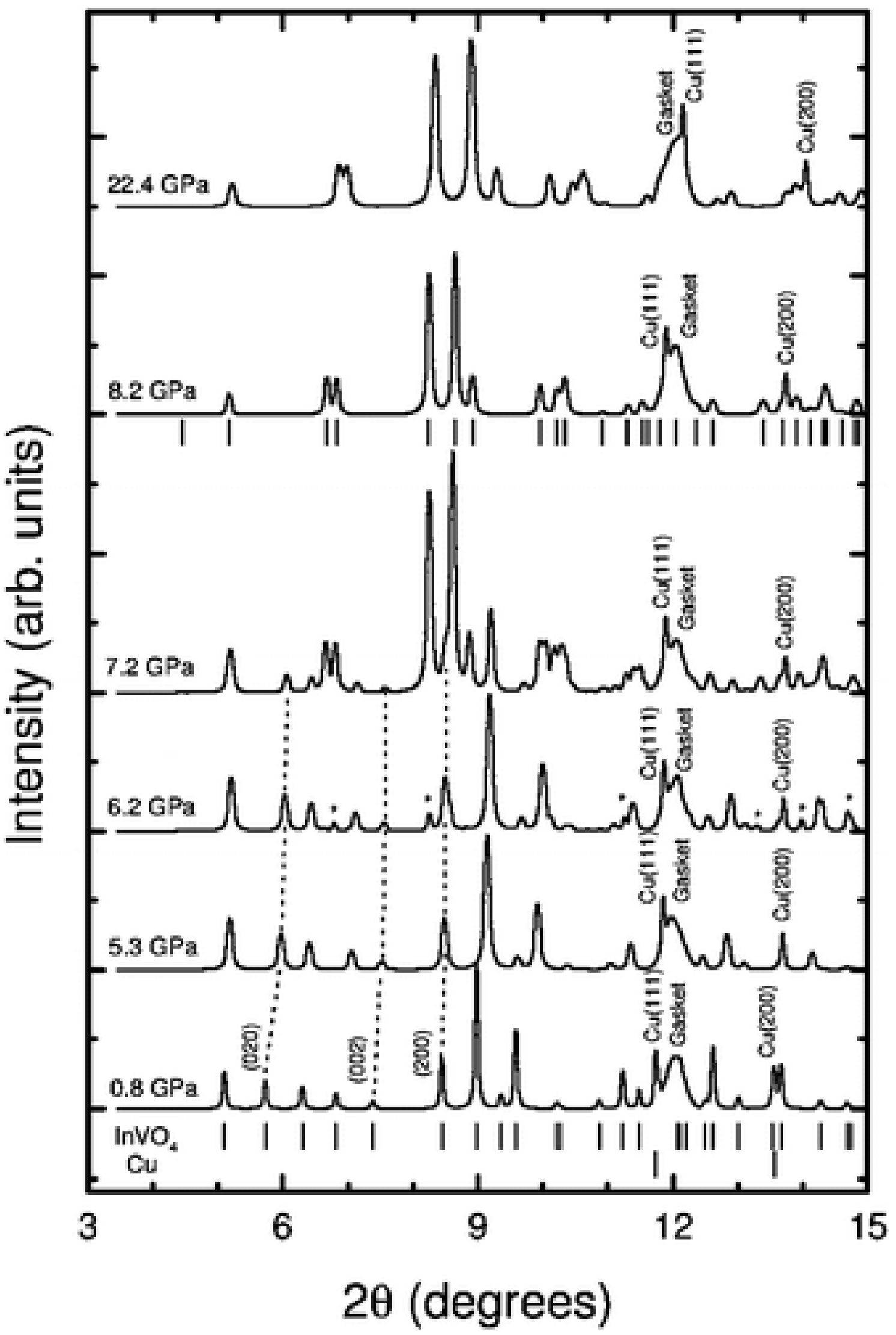

**Figure 23**

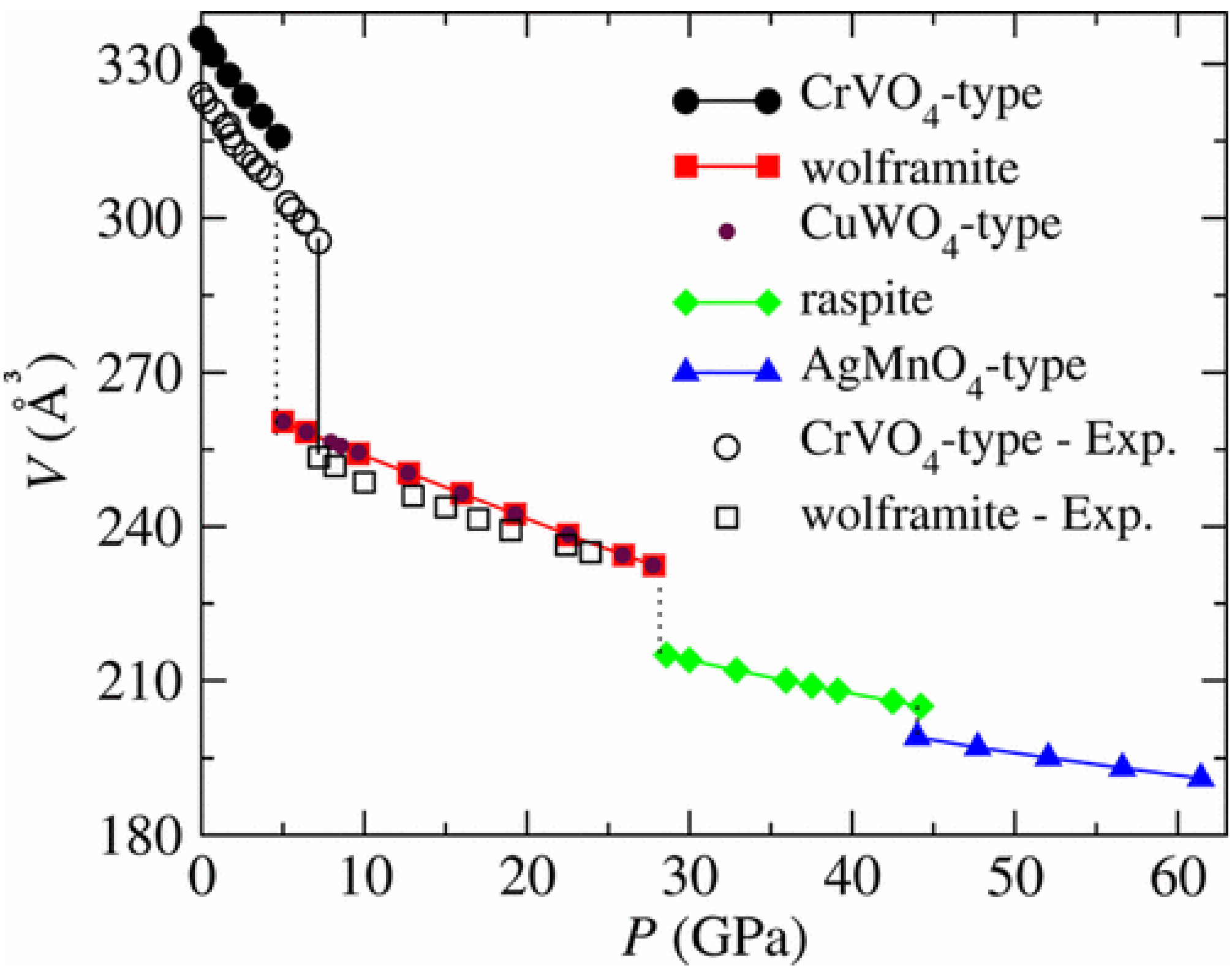

**Figure 24**

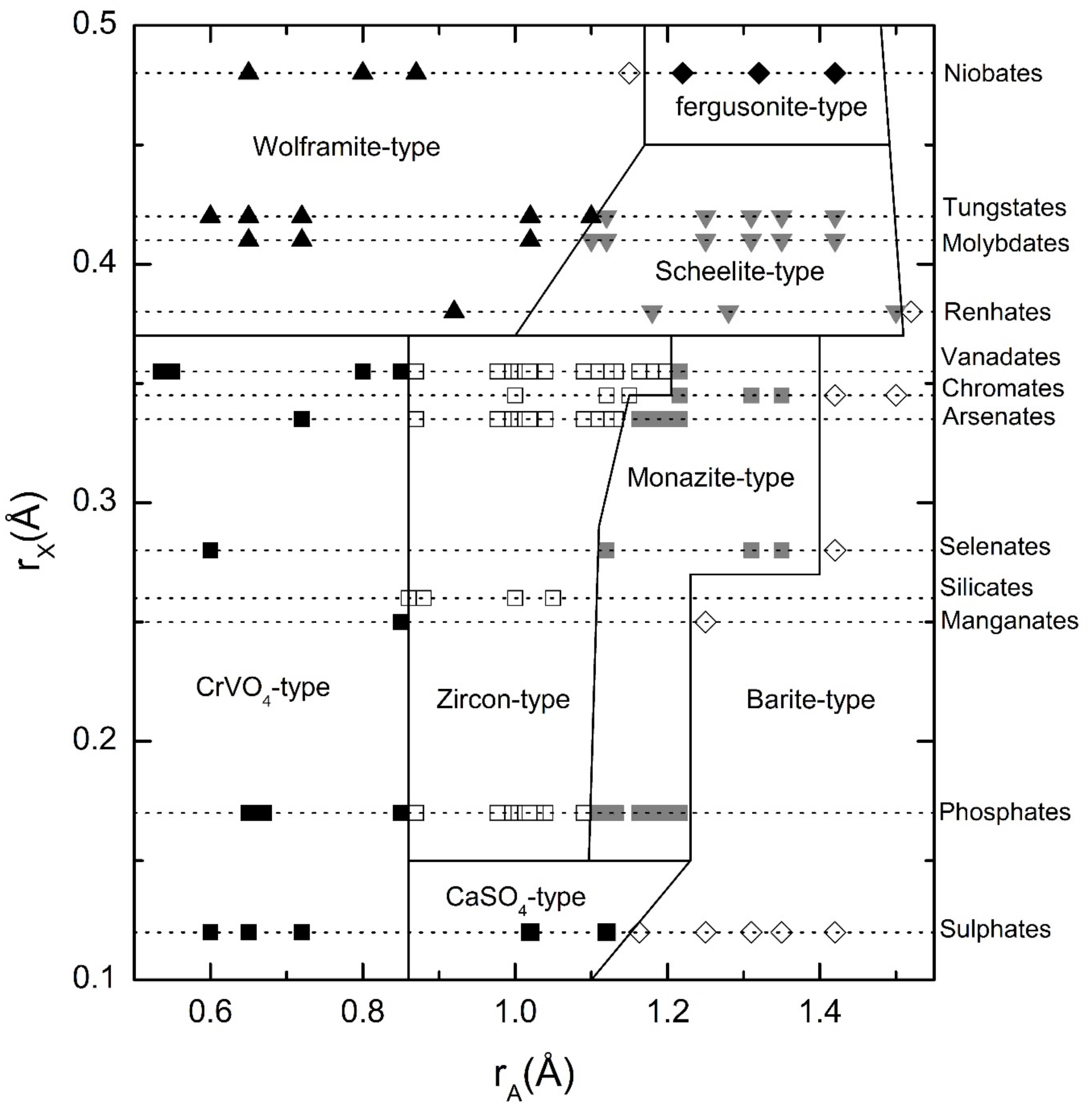

**Figure 25**

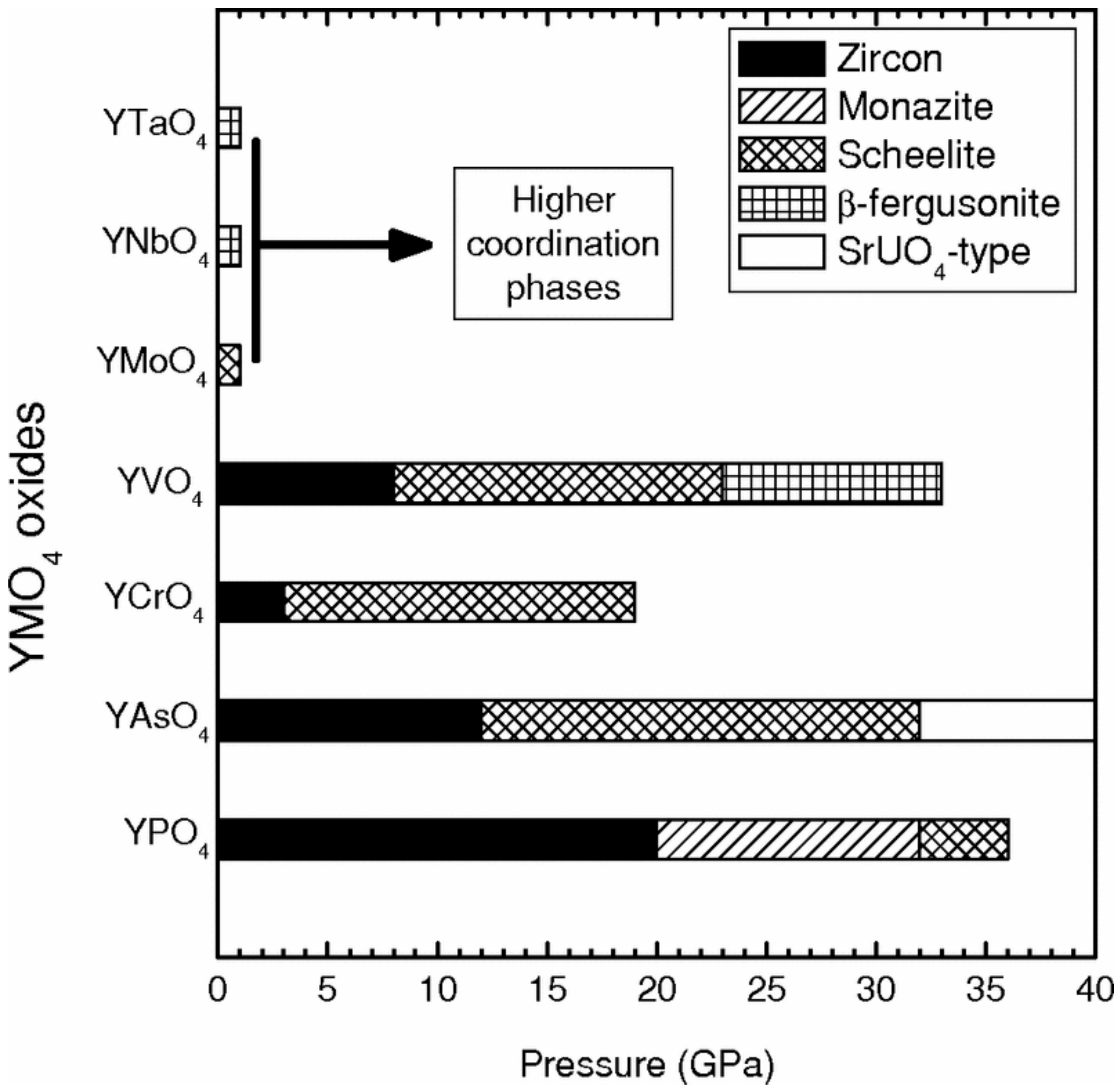

**Figure 26**

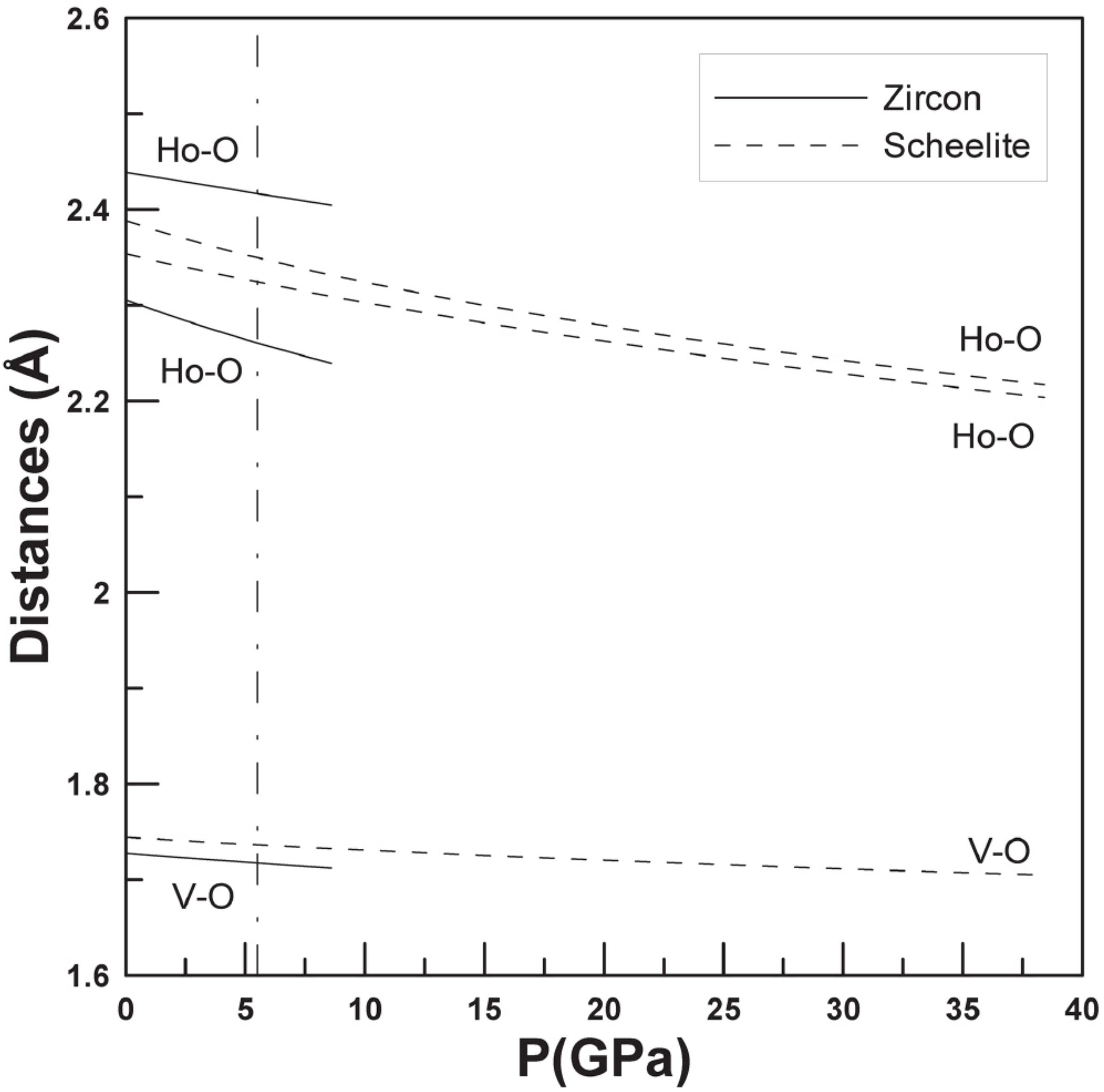